\documentclass[prb,amsmath,amssymb,twocolumn,lettersize]{revtex4}
\usepackage{slashed}
\usepackage{graphicx}
\usepackage{dcolumn}
\usepackage{bm}
\usepackage{graphics}
\usepackage{subfigure}
\usepackage{epsfig}
\usepackage{float}
\usepackage{epstopdf}
\usepackage{bbm,color,ulem}
\usepackage{comment}

\newcommand{\bk}{{\bf k}}

\DeclareMathAlphabet{\mathpzc}{OT1}{pzc}{m}{it} 
\begin{document}
\title{Many-body effects and ultraviolet renormalization in three-dimensional Dirac materials}
\author{Robert E. Throckmorton}
\author{Johannes Hofmann}
\author{Edwin Barnes}
\author{S. Das Sarma}
\affiliation{Condensed Matter Theory Center and Joint Quantum Institute, Department of Physics, University of Maryland, College Park, Maryland 20742-4111 USA}
\date{\today}
\begin{abstract}
We develop a theory for electron-electron interaction-induced many-body effects in three-dimensional Weyl or Dirac semimetals, including interaction corrections to the polarizability, electron self-energy, and vertex function, up to second order in the effective fine structure constant of the Dirac material. These results are used to derive the higher-order ultraviolet renormalization of the Fermi velocity, effective coupling, and quasiparticle residue, revealing that the corrections to the renormalization
group flows of both the velocity and coupling counteract the leading-order tendencies of velocity enhancement and coupling suppression at low energies. This in turn leads to the emergence of a critical coupling above which the interaction strength grows with decreasing energy scale. In addition, we identify a range of coupling strengths below the critical point in which the Fermi velocity varies non-monotonically as the low-energy, non-interacting fixed point is approached. Furthermore, we find that while the higher-order correction to the flow of the coupling is generally small compared to the leading order, the corresponding correction to the velocity flow carries an additional factor of the Dirac cone flavor number (the multiplicity of electron species, e.g. ground-state valley degeneracy arising from the band structure) relative to the leading-order result. Thus, for materials with a larger multiplicity, the regime of velocity non-monotonicity is reached for modest values of the coupling strength. This is in stark contrast to an approach based on a large-$N$ expansion or the random phase approximation (RPA), where higher-order corrections are strongly suppressed for larger values of the Dirac cone multiplicity. This suggests that perturbation theory in the coupling constant (i.e., the loop expansion) and the RPA/large-$N$ expansion are complementary in the sense that they are applicable in different parameter regimes of the theory. We show how our results for the ultraviolet renormalization of quasiparticle properties can be tested experimentally through measurements of quantities such as the optical conductivity or dielectric function (with carrier density or temperature acting as the scale being varied to induce the running coupling). Although experiments typically access the finite-density regime, we show that our zero-density results still capture clear many-body signatures that should be visible at higher temperatures even in real systems with disorder and finite doping.
\end{abstract}
\maketitle
\section{Introduction} \label{Sec:Introduction}
Three-dimensional (3D) Weyl-Dirac semimetals, materials that possess one or more Dirac points at which two bands in the bulk touch with a linear dispersion, $E_\pm(\vec{k})=\pm\hbar v_F|\vec{k}|$, have been of great interest in recent years. These materials may be thought of as
three-dimensional analogs of graphene or as solid state (and nonrelativistic, i.e., the velocity of light is taken to be infinite so the bare Coulomb interaction is unretarded although there is a static background lattice dielectric constant reducing the overall value of the Coulomb coupling) incarnations of relativistic quantum electrodynamics since they are described by quasi-relativistic fermions with Coulomb interactions at low energies. Dirac semimetals were first theorized long ago by Herring\cite{HerringPR1937} and by Abrikosov and Beneslavski\u{i}\cite{AbrikosovJETP1971}. More recently, these materials have attracted considerable attention not only because they host a system of interacting massless Dirac quasiparticles, but also due to the fact that they lie at the intersection of many interesting topological phases of matter. This observation has driven a concerted effort to identify theoretically materials that possess symmetry-protected Dirac points in the bulk\cite{WanPRB2011,BurkovPRL2011,YoungPRL2012,ZyuzinPRB2012a,YangNatCommun2014,WengPRX2015,ChiuArXiv2015,HuangArXiv2015a,HuangArXiv2015b}. Different topological states can be realized by breaking certain symmetries; for example, breaking inversion symmetry lifts the Dirac cone degeneracy, giving rise to Weyl semimetals,\cite{WanPRB2011,BurkovPRL2011,ZyuzinPRB2012a,HuangArXiv2015a,HuangArXiv2015b} materials which exhibit anomalous transport properties\cite{YangPRB2011,HosurPRL2012,ZyuzinPRB2012b,RosensteinPRB2013,MoonPRL2013,BiswasPRB2014,ParameswaranPRX2014,ZieglerArXiv2015} and topologically protected Fermi arc states on their surface\cite{WanPRB2011,PotterNatCommun2014,GaneshanPRB2015}. Over the last two years, numerous experimental works have confirmed that certain materials such as Na$_3$Bi and Cd$_3$As$_2$ are Dirac semimetals by directly observing Dirac cones in the bulk\cite{XuArXiv2013,LiuScience2014,NeupaneNatCommun2014,LiuNatMater2014,BorisenkoPRL2014,XuArXiv2015c} and Fermi arcs on the surface\cite{XuScience2015}. Many additional works observing and elucidating transport properties\cite{BanerjeePRB2012,JeonNatMater2014,LiangNatMater2015,NarayananPRL2015,Kushwaha_APLMat15,TaguchiPRB2015,FengArXiv2014,DasSarmaPRB2015,XiongArXiv2015,DasSarmaPRB2015_2} and phase transitions\cite{GoswamiPRL2011,WitczakKrempaPRB2012,IsobePRB2012,WangPRB2013b,GoswamiPRB2013,GonzalezPRB2014,ZhangPRB2015,SchoopArXiv2014,YangNatPhys2014,PixleyArXiv2015,RoyPRB2014} in Dirac semimetals have appeared. Even more recently, the first Weyl semimetals were discovered experimentally\cite{LvArXiv2015a,LvArXiv2015b,XuArxiv2015a}, with evidence for Fermi arcs\cite{XuArxiv2015a} and the unusual transport signatures they entail\cite{WangPRB2013a,ZhangArXiv2015a,ZhangArXiv2015b}. The theory developed in this work applies equally well to both Dirac and Weyl systems [i.e., chiral linearly dispersing gapless 3D systems with filled (empty) valence (conduction) bands touching at singular Dirac points], and we refer to both as Dirac systems from now on in this paper.  The number of such Dirac points (e.g., from possible valley degeneracy in the system arising from band-structure effects) or, equivalently, the number of Dirac cones defines the number of flavors or multiplicity ($N$) giving the ground-state degeneracy (excluding the spin degeneracy for a Dirac material or due to there being two separate cones per flavor in a Weyl material) which would play an important role in the theory.

A crucial feature of interacting Dirac materials is the field-theoretic many-body renormalization of the quasiparticle properties, i.e., the Fermi velocity, the quasiparticle residue, and the effective Coulomb interaction strength are no longer pinned at their noninteracting band structure values but acquire a scale dependence. The strength of the renormalization depends on the dimensionless ratio of the Coulomb interaction to the kinetic energy as given by the effective fine-structure constant:
\begin{align}
\alpha&=\frac{g^2}{4\pi v_F} = \frac{e^2}{\hbar v_F \kappa} , \label{eq:fine}
\end{align}
where $e$ is the electron charge, $v_F$ is the Fermi velocity, and $\kappa$ is the effective lattice dielectric constant of the material ($\kappa>1$ in general for solid state materials). Because of the linear band approximation, any correlation function is dominated by high-momentum excitations as is manifested by an ultraviolet divergent dependence of these correlators on a cutoff scale $\Lambda$. Through a renormalization of the quasiparticle properties, this ultraviolet cutoff dependence can be removed at the expense of introducing renormalized parameters which now depend on the scale (such as momentum, energy, temperature, or density) at which they are measured. This scale dependence is dictated by renormalization group (RG) equations. For example, the coupling~\eqref{eq:fine} obeys
\begin{align}
\frac{d\alpha}{d\ln \mu} &= \beta_\alpha , \label{eq:rgflow}
\end{align}
where $\mu$ denotes the renormalization scale. The coefficient on the right-hand side is called the beta function and is uniquely determined by the divergence structure of the theory. If $\alpha(\mu_0)$ is measured at one scale $\mu_0$, Eq.~\eqref{eq:rgflow} then predicts the value of $\alpha(\mu')$ at a different scale $\mu'$. The quasiparticle renormalization is an observable effect and has been measured most notably in the two-dimesional (2D) Dirac material graphene in a wide variety of experiments using energy, density (which determines the Fermi energy), and momentum as the running scale\cite{EliasNatPhys2011,LiNatPhys2008,ChaePRL2012,SiegelPNAS2011,YuPNAS2013}.

In principle, temperature, provided it is much higher than the Fermi energy, could also be a varying scale to study the RG flow, and may very well be the most suitable scaling variable for 3D Dirac systems in terms of experimental investigations.  We note that our approach to the RG flow of quasiparticle renormalization does not make any explicit use of the precise value of the ultraviolet cutoff $\Lambda$ since we express renormalization at one scale simply in terms of that at another value of the scaling parameter.  This completely eliminates choosing an arbitrary value of the ultraviolet cutoff which is theoretically ill defined except as an ultraviolet cutoff, making it problematic to assign its precise value.  In particular, the basic tenet of the RG approach is that physical quantities vary with the energy or momentum scale (and equivalently, with carrier density or temperature), but do not depend on the ultraviolet cutoff scale which should not show up in observable quantities.  We should, however, mention that, in solid state systems, unlike in relativistic quantum field theories [e.g. quantum electrodynamics (QED) or quantum chromodynamics (QCD)], there is indeed a true ultraviolet momentum cutoff given by the inverse lattice constant.  However, this cutoff is by no means a sharply defined unique quantity for calculating renormalized quasi-particle properties (e.g., it could be multiplied by $2\pi$ or some other constant or one could simply use the bandwidth as an ultraviolet energy cutoff), and thus any explicit calculation of quasi-particle renormalization using such a lattice (or band) cutoff momentum (or energy) is not quantitatively meaningful in systems, such as Dirac materials, manifesting ultraviolet divergences.

In this work, we present a calculation of the renormalization group (RG) equations for coupling, Fermi velocity, and quasiparticle residue up to second order in the interaction strength $\alpha$. Previous theoretical works have investigated renormalization phenomena in 3D Weyl or Dirac systems using the random phase approximation (RPA) or a large-$N$ expansion\cite{LvIJMPB2013,HofmannArXiv2014,HofmannArXiv2015,GonzalezPRB2014}, as well as first-order perturbation theory\cite{HosurPRL2012,IsobePRB2012,IsobePRB2013}. One of the primary motivations for our current work is to explore the effects of the leading corrections to RPA and to identify the parameter regime in which a perturbative approach is valid. Similar investigations played an important role in the case of 2D Dirac materials, in particular graphene, where it was demonstrated that perturbative RG results are invalid for all but very strongly screening media, and that apparent agreement between experiment and first-order theory is often completely spurious\cite{BarnesPRB2014,HofmannPRL2014}. These findings led to the development of a theory that quantitatively describes renormalization effects in graphene that applies to a broader range of experimental setups\cite{HofmannPRL2014}. It is therefore of direct experimental relevance and of considerable fundamental interest to perform a similar analysis in the context of 3D Dirac materials to investigate how physical properties of 3D Dirac materials vary under RG flows with higher-order corrections included.

Therefore, in this work, we consider the effect of a Coulomb interaction on a system with $N$ ``flavors'' of massless Dirac fermions to two-loop order. At this stage, $N$ is simply a mathematical quantity defining the number of electron flavors or species, but in reality $N$ is typically the (often large) valley degeneracy in the relevant 3D Dirac system.  After reviewing the previous one-loop results, we proceed to calculate the polarization, electron self-energy, and vertex function to two loops.  We use the term ``loop expansion'' to mean the usual order-by-order diagrammatic perturbative many-body expansion in the Coulomb interaction, which is precisely equivalent to an expansion in the effective fine-structure constant for linearly dispersing Dirac systems.  At one-loop order, we already see logarithmic divergences in both the polarization and self-energy as a function of momentum.  Therefore, at two loops, we expect to see log-squared divergences, and we confirm this with explicit analytical calculations.  We then consider the renormalization of the theory, in which we not only determine the necessary counterterms to cancel out the divergences, but also perform several useful self-consistency checks on our results.  We then make use of these results to derive RG equations for the parameters in our theory, namely, the Fermi velocity $v_F$, the effective fine-structure constant $\alpha$, and the quasiparticle residue.

While the remainder of this paper gives a detailed account of both the systematics and the calculations of the renormalization group equations, we summarize here our main findings, which are the renormalization group equations obtained up to the two-loop expansion calculation for residue, Fermi velocity, and coupling given in Eqs.~\eqref{eq:rgZ},~\eqref{eq:rgvF}, and~\eqref{eq:rgalpha}, respectively:
\begin{equation}
\frac{d\ln{Z_\psi}}{d\ln{\mu}}=\left(\frac{15+N}{3\pi^2}-\frac{1}{2}\right)\alpha^2.
\end{equation}
\begin{equation}
\frac{d\ln{v_F}}{d\ln{\mu}}=-\frac{2\alpha}{3\pi}+\frac{2N}{9\pi^2}\alpha^2.
\end{equation}
\begin{equation}
\frac{d\alpha}{d\ln{\mu}}=\frac{2(N+1)}{3\pi}\alpha^2+\frac{27C-44}{54\pi^2}N\alpha^3,
\end{equation}
with $C\approx 1.33318$ a constant that we determined numerically [cf. Eq.~\eqref{eq:C}]. We find that, in contrast to the one-loop result that implies $\alpha$ always going to zero as we go to low-momentum scales, the two-loop result reveals that there is a critical value above which $\alpha$ instead diverges, just as in the case of graphene\cite{BarnesPRB2014}. However, this critical value is much larger than in graphene; in 3D Dirac materials, it is $\alpha_c=14.1298\left (1+\frac{1}{N}\right )$, which should be compared to $\alpha_c=0.78$ for graphene. This would seem to imply that perturbation theory is reliable over a much wider range of values of $\alpha$ in the 3D case relative to the 2D case. In particular, with experimental values of $\alpha \sim 0.1$-$1$ in typical 3D Dirac materials (e.g., in Cd$_3$As$_2$, dielectric constants in the range $\kappa\sim20$-$40$ have been measured\cite{ZivitzPRB1974,JayGerinSSC1977}, while velocities are typically\cite{LiuNatMater2014,NeupaneNatCommun2014,BorisenkoPRL2014} in the range $10^{5}$-$10^{6}$m/s), we might expect perturbation theory to give quantitatively reliable results since the condition $\alpha\ll\alpha_c$ applies here. However, we see from the equation for the velocity renormalization that the second-order correction has again the opposite sign relative to the first-order result, producing a second special value of $\alpha$, $\alpha_*=3\pi/N$, where the second-order term is equal in magnitude to the leading-order term. It is generally the case that $\alpha_*<\alpha_c$, and for $\alpha_*<\alpha<\alpha_c$, we find the unusual situation in which both the velocity and effective coupling decrease as the energy scale is reduced. This trend continues until $\alpha$ reaches $\alpha_*$, at which point the velocity reverses its trend and begins to grow as the non-interacting fixed point in the infrared is approached. Moreover, we see that $\alpha_*$ decreases with increasing Dirac cone multiplicity $N$, so that for larger values of $N$, $\alpha\geq\alpha_*$ is achieved for modest values of the interaction strength. This observation may be particularly relevant for the pyrochlore iridates where $N=12$\cite{WanPRB2011}, meaning that for $\alpha\sim0.1$, the second-order term in the velocity constitutes a large 25\% correction. On the other hand, for Cd$_3$As$_2$ with $N=1$, this term yields a few-percent correction to the leading order. If we interpret the appearance of $\alpha_*$ as signifying a breakdown of perturbation theory (given that for this coupling, the second-order corrections are comparable to the leading order), then $\alpha_*$ can be used as a criterion for determining the validity of results based on the perturbation theory in $\alpha$. The fact that perturbation theory appears to work better for smaller degeneracies is particularly interesting in light of the fact that methods such as RPA or large-$N$ expansions work in exactly the opposite regime, suggesting that perturbation theory may be complementary to these approaches\cite{HofmannPRL2014,HofmannArXiv2014}.  Thus, the perturbative loop expansion may be a reasonable approximation for 3D Dirac systems for small values of $N$, whereas in graphene it is only reasonable when the effective coupling constant itself is very small (for example, for substrates with very large dielectric constants so that effective $\kappa$ values are large).

We also discuss how this predicted renormalization can be observed experimentally, namely, by measuring the plasmon frequency and the optical conductivity. While the control over the doping in experiments on a 3D Dirac material is more restricted than in experiments on graphene (since gating is not an option for 3D materials as it is for 2D graphene), the temperature of the system can serve as another energy scale.  If this scale is much greater than the Fermi energy, which is determined by the doping density, then we expect that our results, which have been derived for an intrinsic (undoped) system, are valid for the system realized in experiments.  In fact, having a finite temperature higher than the Fermi energy associated with the unintentional doping density in the system is a convenient and practical way of approaching the intrinsic Dirac point physics\cite{DasSarmaPRB2013,DasSarmaPRB2013_2,HofmannArXiv2015}.  We elaborate on this argument by an explicit calculation of the Drude weight which is independent of initial doping even at intermediate temperatures of $T > \varepsilon_F/2$, thus probing the intrinsic limit. While simple dimensional analysis predicts a linear temperature dependence of the intrinsic Drude weight, this linear scaling is violated by the logarithmic renormalization of the charge and Fermi velocity, giving rise to a superlinear temperature dependence instead. To leading order in $\alpha$, the strength of this superlinear scaling is set by a renormalization group invariant quantity known as the Landau pole $\Lambda_L$, a dimensionally transmuted scale which marks the point of divergence of the one-loop fine-structure constant at high energy. To probe the relative effects of doping, temperature, and renormalization, we provide a full calculation of the plasmon frequency (which is related to the Drude weight) to leading order in the RPA, which clearly shows the transition from a finite-density extrinsic regime at low temperature to a zero-density intrinsic regime at high temperature with superlinear logarithmic scaling violations. An experimental observation of this super-linear temperature dependence will yield direct evidence for our predicted ultra-violet renormalization corrections.

\begin{table*}
\begin{tabular}{c | c | c}
 & \textbf{Graphene}\cite{SodemannPRB2012,BarnesPRB2014} & \textbf{3D Dirac} \\
\hline
\hline
Vertex & $\log$ divergence expected at $O(\alpha^2)$ & $\log$ divergence at $O(\alpha^2)$ \\
\hline
Self-energy & $\log$ divergence at $O(\alpha)$, $\log^2$ at $O(\alpha^3)$ & $\log$ divergence at $O(\alpha)$, $\log^2$ at $O(\alpha^2)$ \\
\hline
Polarizability & $\log$ divergence at $O(\alpha^2)$ & $\log$ divergence at $O(\alpha)$, $\log^2$ at $O(\alpha^2)$ \\
\hline
Charge & No renormalization & Renormalizes \\
\hline
Fermi velocity & Renormalizes & Renormalizes \\
\hline
Optical conductivity & $\log$ divergence at $O(\alpha^2)$ & $\log$ divergence at $O(\alpha)$
\end{tabular}
\caption{Table summarizing the divergences appearing in graphene and 3D Dirac materials for various quantities. \label{Tab:Graphenevs3DDirac}}
\end{table*}
\begin{table*}
\begin{tabular}{c | c | c}
 & \textbf{3D Dirac} & \textbf{3+1D QED}\cite{PeskinSchroederBook,JostHPA1950,ItzyksonBook} \\
\hline
\hline
Vertex & $\log$ divergence at $O(\alpha^2)$ & $\log$ divergence at $O(\alpha)$ \\
\hline
Self-energy & $\log$ divergence at $O(\alpha)$, $\log^2$ at $O(\alpha^2)$ & $\log$ divergence at $O(\alpha)$, $\log^2$ expected at $O(\alpha^2)$ \\
\hline
Polarizability & $\log$ divergence at $O(\alpha)$, $\log^2$ at $O(\alpha^2)$ & $\log$ divergence at $O(\alpha)$ \\
\hline
Charge & Renormalizes & Renormalizes \\
\hline
Fermi velocity/speed of light & Renormalizes & No renormalization \\
\hline
Electron mass & N/A & Renormalizes
\end{tabular}
\caption{Table summarizing the divergences appearing in 3D Dirac materials and in 3+1D QED for various quantities. \label{Tab:3DDiracvsQED}}
\end{table*}

We summarize how our ultraviolet renormalization results compare with graphene, the 2D analog of the 3D Dirac materials studied here, in Table \ref{Tab:Graphenevs3DDirac} and with (3+1)-dimensional [(3+1)D] QED in Table \ref{Tab:3DDiracvsQED}, to which 3D Dirac materials provide a nonrelativistic condensed matter analog.  Let us first discuss the comparison to graphene, whose many-body renormalization effects were considered in detail in our earlier works\cite{BarnesPRB2014,HofmannPRL2014}.  We note that, while the Fermi velocity renormalizes in both graphene and 3D Dirac materials, the charge only renormalizes in 3D Dirac materials (but not in graphene).  This is because, in graphene (and, in fact, in all solid state 2D Dirac systems), the Coulomb interaction is given by a non-analytic term in the action, and thus we never obtain any terms in perturbation theory that renormalize the charge.  Physically, this is connected to the fact that the Coulomb interaction in graphene still has the 3D Coulomb form since graphene is not the solid state analog of (2+1)-dimensional QED and is actually a 2D physical system existing in the 3D world.  Therefore, the only renormalization of the overall interaction strength, as given by $\alpha$, comes from renormalization of the Fermi velocity in graphene.  On the other hand, the renormalization of $\alpha$ in 3D Dirac materials is due to that of both charge and Fermi velocity [as in (3+1)-dimensional QED where both charge and mass manifest ultraviolet renormalization]; the Fermi velocity may be seen in this case as an independent parameter of the theory from $\alpha$.  One very important consequence of this is that, unlike in graphene, in which the system becoming weakly interacting at low energy scales necessarily means that the Fermi velocity diverges logarithmically at the Dirac point, the Fermi velocity can (and, as it turns out, does) actually remain finite even at low energies in 3D Dirac systems, since the non-interacting limit can be achieved by the charge renormalizing to zero at low energy.  This gives us a very major distinction between the 2D and 3D Dirac systems: in 2D, we \textit{must} see a logarithmically divergent Fermi velocity at low energy because it is the only way for the system to become weakly interacting\cite{EliasNatPhys2011,YuPNAS2013,BarnesPRB2014}, but this is not necessary in 3D.  Note also that we expect to see a logarithmic divergence appearing for the vertex renormalization in graphene at $O(\alpha^2)$, even though it has not been calculated.  This is because, at second order, the self-energy acquires a temporal component\cite{BarnesPRB2014}, just as it does in 3D Dirac materials.  Due to gauge invariance, we expect that the vertex renormalization should diverge as well at orders higher than the first, and in such a way that the Coulomb field strength does not renormalize at any order, as we see happens in 3D Dirac materials. Thus, as summarized in Table~\ref{Tab:Graphenevs3DDirac}, there are major mathematical and physical differences between many-body effects induced by the ultraviolet renormalization in 3D and 2D Dirac systems arising in condensed matter physics.

We now consider the comparison between 3D Dirac materials and (3+1)D QED, as summarized in Table \ref{Tab:3DDiracvsQED}.  Here, the situation is in some sense opposite to that of graphene.  The interaction strength in 3+1D QED renormalizes only because \textit{charge} renormalizes, since the speed of light cannot renormalize due to the requirement that it be constant.  Both, however, still exhibit the well-known Landau pole (as does graphene\cite{BarnesPRB2014}), a divergence of $\alpha$ at a large, but finite, energy scale at one-loop order.  While the Fermi velocity renormalizes (and, in fact, diverges itself at the Landau pole), it does not diverge rapidly enough to cancel the effect of the diverging charge.  We also note that logarithmic divergences start appearing at lower orders in $\alpha$ in QED than in 3D Dirac materials in the vertex correction.  This is because we treat a non-relativistic Coulomb interaction here, rather than the full relativistic electromagnetic force, as is done in QED.  This means that the quasiparticle residue, corresponding to a temporal component of the self-energy, already differs from unity at $O(\alpha)$ in QED; in the 3D Dirac case, this only happens at $O(\alpha^2)$.  In both cases, however, the vertex function and the quasiparticle residue are renormalized in such a way that the strength of the electromagnetic field remains unchanged, as is required by gauge invariance.  This divergence of the vertex in QED at $O(\alpha)$ also, ironically, leads to the cancellation of a potential $\log^2$ divergence at $O(\alpha^2)$.  We will find that, for the 3D Dirac material case considered here, only the diagram in which the self-energy is inserted into one of the electron lines in the bubble has a $\log^2$ divergence.  In QED, on the other hand, the vertex correction contribution also has a $\log^2$ divergence\cite{JostHPA1950,ItzyksonBook}; this divergence exactly cancels that from the self-energy correction contribution, leaving only a simple logarithmic divergence at $O(\alpha^2)$.

One fundamental difference between solid state Dirac materials and QED is of course the large difference in the applicable \textit{effective} coupling in the two cases.  Whereas the fine-structure constant for QED is always $\frac{e^2}{\hbar c}\approx 1/137$, the corresponding effective fine-structure constant in Dirac systems (either 2D or 3D) is multiplied by a factor of $c/v_F\kappa$ [see Eq.\ \eqref{eq:fine}], which is typically very large since the Fermi velocity ($\sim 10^6\,\text{m}/\text{s}$) is much less than the speed of light ($\approx 3\times 10^8\,\text{m}/\text{s}$), although the presence of the effective background dielectric constant reduces the effective coupling in Dirac systems by a factor of order 10 typically in most materials (except for graphene suspended in vacuum where $\kappa=1$, giving a very large graphene fine-structure constant of around $\tfrac{300}{137}\approx 2.2$). The largeness of the effective fine structure constant makes Dirac systems more like a strong-coupling QED problem (albeit with nonrelativistic Coulomb interactions) rather than the regular QED where the loop expansion in powers of the fine structure constant is asymptotic up to a very high ($\gg 137$) order in the perturbation theory.  We discuss the asymptotic nature of the loop expansion for 3D Dirac systems in this paper [see, e.g., Eq.\ \eqref{Eq:f1OalphaSq}] and this issue was discussed in depth for graphene in Ref.\ \onlinecite{BarnesPRB2014}, where it was found that the loop expansion may already fail at the first order for graphene suspended in vacuum.  One may wonder, given the large coupling constant in Dirac materials, whether some aspects of its interaction physics resemble QCD rather than QED.  It turns out that one specific aspect of interaction effects in Dirac materials (for both 2D and 3D) does indeed have some superficial similarities with QCD.  This is the behavior of the Dirac system when the effective coupling is larger than the critical coupling ($\alpha>\alpha_c$) so that the running coupling increases with decreasing energy, eventually diverging at the Dirac point.  This appears similar to the QCD strong-coupling behavior (provided one starts with a sufficiently large coupling $\alpha>\alpha_c$ in the beginning for the Dirac system, whereas in QCD of course this behavior is generic because of the structure of the beta function itself).  But, this similarity is somewhat misleading because the beta functions in the two cases are fundamentally different, and the divergent running coupling low-energy behavior of the Dirac system we find (for $\alpha>\alpha_c$) may very well be just an artifact arising from the failure of the perturbation theory.  In addition, the Dirac problem neither has confinement nor true asymptotic freedom (as is obvious from the existence of the Landau pole in the theory), which are two hallmarks of QCD.

Another qualitative difference between QED and Dirac systems worth pointing out is the role of the Fermi energy (i.e., a finite chemical potential away from the Dirac point) as well as temperature in Dirac materials, which provide additional physical parameters (doping density and temperature) for the experimental implementation of the RG flow in solid state materials.  Either the finite Fermi energy (arising from any finite doping of the system) or temperature, by itself, can act as a low-energy cutoff for the RG flow just as the band width or the inverse lattice constant acts as a high-energy or high-momentum cutoff in solids.  But, in the presence of a finite Fermi energy (temperature), the other parameter, i.e., temperature (Fermi energy), can be used as a practical parameter to induce the RG flow as long as temperature (Fermi energy) is much greater than the Fermi energy (temperature).  This can be very useful in experimental measurements where the pure undoped Dirac limit (or zero temperature) can never be reached.  Since the Fermi energy (i.e., the chemical potential) depends monotonically on the doping density, one can change the doping density (at fixed low temperature) or the temperature (at fixed low doping density) as the scaling parameter to study the ultraviolet renormalization of Dirac materials.  By contrast, in purely relativistic field-theoretic problems (e.g., QED, QCD), the concepts of doping, finite chemical potential (or Fermi energy), temperature, etc., simply do not apply, and the system is always an intrinsic system (in our sense) and can never be detuned from the precise critical point.

It is also instructive to compare the many-body renormalization properties of 3D Dirac systems with normal 3D metals (or doped semiconductors) which are often the textbook systems for studying many-body effects arising from long-range Coulomb interactions in solids\cite{Nozieres1964,Abrikosov1975,Mahan1981,Fetter2003}. Three-dimensional metals are defined by the sharp existence of a finite Fermi surface and the associated Fermi liquid properties, with ultraviolet renormalization or divergences not playing any role whatsoever since the energy dispersion in these cases is parabolic, and there is no zero-density critical point because of the existence of a large band gap separating the filled valence band from the partially filled conduction band with the finite Fermi energy.  Thus, the key physics of interest in Dirac materials, namely, the many-body interaction corrections associated with ultraviolet renormalization, is not relevant for ordinary metals at all.  In fact, there exist detailed RG analyses\cite{ShankarRMP1994,PolchinskiArXiv1999} of the stability of the Fermi surface in 3D and 2D metals starting from high momentum and systematically approaching the Fermi surface, with the conclusion that, except for the possible occurrence of interaction-driven superconductivity at high orbital momentum channels (the so-called Kohn-Luttinger superconductivity\cite{KohnPRL1965}) at exponentially low temperatures, the Fermi surface is stable and there is no ultraviolet divergence (in the QED sense) anywhere for 3D (or 2D) metals.  Thus, the key physics of ultraviolet renormalization and scale-dependent RG flow with logarithmic running of the coupling are features not arising in normal metals as they do in Dirac systems.  Of course, there are strong Fermi liquid renormalization effects in normal metals, as discussed extensively in many standard many-body theory textbooks\cite{Nozieres1964,Abrikosov1975,Mahan1981,Fetter2003}, and these effects are non-perturbative since the dimensionless interaction coupling strength in 3D metals (which depends explicitly on the metallic electron density and is not a simple effective fine-structure constant as in Dirac systems) is large ($\sim 5-6$ typically), but these many-body corrections (often quantitatively large) are not indicative of any fundamental ultraviolet divergence in the system with no logarithmic cutoff-dependent renormalization corrections arising anywhere in the problem.  In fact, the divergence one worries about in metals is the infrared divergence associated with the long-range Coulomb interaction (and not an ultraviolet divergence associated with large momenta) which is ``cured'' by considering a loop expansion in the dynamically screened Coulomb interaction (rather than the loop expansion in the bare Coulomb interaction we consider for the Dirac system in the current work), which provides a controlled perturbation theory for metals at high electron density (which is equivalent to weak metallic interaction coupling). At metallic densities, such an expansion is only qualitatively valid (since there is no quantum phase transition in the problem), but a comparison with experiment (or with quantum Monte Carlo simulations) shows that metallic many-body corrections calculated using an expansion in the screened Coulomb interaction works reasonably well even at metallic densities where the system is far from being weakly coupled for reasons not very clear at this stage.\cite{ZhangPRB2005}  Density-dependent (but not cutoff-dependent) many-body corrections similar to the metallic quasiparticle renormalization effects also show up in doped extrinsic Dirac materials as non-ultraviolet non-logarithmic subleading corrections, but we ignore them completely in our current work since they are completely negligible compared with the ultraviolet logarithmic corrections as the Dirac point is approached.

The rest of this paper is organized as follows: In Sec.\ \ref{Sec:Model}, we describe the (Euclidean-time) action for our system and the associated Feynman rules.  Section \ref{Sec:Polarization} is dedicated to reviewing the one-loop results and deriving the two-loop polarization, while Sec.\ \ref{Sec:SelfEnergy_SE} provides a detailed calculation of the two-loop electron self-energy.  In Sec.\ \ref{Sec:Renormalization}, we formulate the renormalized perturbation theory, determine the counterterms that arise therein, and derive the RG equations for our model. In Sec.\ \ref{Sec:Experiment}, we show how our results can be tested in finite-temperature experiments on finite-density samples even though our results are derived at zero density. We present our conclusions in Sec.\ \ref{Sec:Conclusion}.  We provide derivations of the one-loop polarization and electron self-energy, the two-loop vertex corrections, and the optical conductivity within the Drude-Boltzmann approximation in the Appendix, along with an alternate derivation of the RG equations using the Callan-Symanzik equation.  The latter provides yet another consistency check on our results.

\section{Model} \label{Sec:Model}
The model that we will be using is that of a 3D system with $N$ ``flavors'' (e.g.,
valleys) of Dirac electrons interacting via a Coulomb interaction. We note that the
total degeneracy of the 3D system we consider is $2N$ including both spin and valleys.
We will work with the following Euclidean-time action (setting $\hbar = 1$):
\begin{eqnarray}
S&=&-\sum_{a=1}^{N}\int dt\,d^3\vec{R}\,(\bar{\psi}_a\gamma^0\partial_0\psi_a+v_F\bar{\psi}_a\gamma^i\partial_i\psi_a\cr
&+&\varphi\bar{\psi}_a\gamma^0\psi_a)+\frac{1}{2g^2}\int dt\,d^3\vec{R}\,(\partial_i\varphi)^2, \label{Eq:Action}
\end{eqnarray}
where the $\gamma$ matrices form a Clifford algebra, i.e., $\{\gamma^\mu,\gamma^\nu\}=2\delta^{\mu\nu}$
for $\mu$ and $\nu=0,1,2$, and $3$ and $\bar{\psi}=\psi^{\dag}\gamma^0$.  There is
an implied sum on $i$ in the above action from $i=1$ to $3$.  The fields, $\psi$, are
four-component Grassmann spinors, and $N$ is the number of valleys.  In all, this
model results in $2N$ Dirac cones.  The factor of $2$ comes from spin in a regular
Dirac semimetal, or from the presence of a pair of Weyl nodes for each of the $N$
flavors in a Weyl semimetal.

Here and throughout this work, we will employ a quasi-relativistic notation, i.e., we
define $k=(k_0,\vec{k})$, $d^4k=dk_0\,d^3\vec{k}$, $\slashed{k}=k_0\gamma^0+v_F\vec{k}\cdot\vec{\gamma}$,
and $k^2=k_0^2+v_F^2|\vec{k}|^2$.  The Feynman rules associated with this action are
the following:
\begin{alignat}{2}
&\raisebox{-0.cm}{\scalebox{0.6}{\epsfig{file=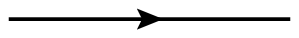}}} &\qquad & G_0(k)=\frac{i}{\slashed{k}}, \\
&\raisebox{-0.cm}{\scalebox{0.6}{\epsfig{file=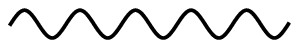}}} &\qquad & D_0(k)=\frac{g^2}{|\vec{k}|^2}=\frac{4\pi\alpha v_F}{|\vec{k}|^2}, \\
&\raisebox{-0.6cm}{\scalebox{0.6}{\epsfig{file=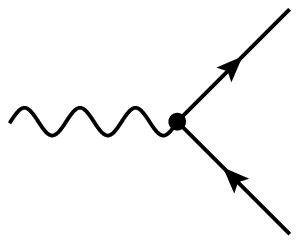}}} &\qquad & i \gamma^0.
\end{alignat}
Here, the straight lines are electron propagators, while the wavy lines are the Coulomb propagators.

\section{Polarization} \label{Sec:Polarization}
\begin{figure}
\includegraphics[width=0.5\columnwidth]{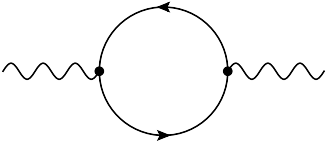}
\caption{Leading-order polarization bubble, $\Pi_B(q)$.}
\label{Fig:Polarization_FE}
\end{figure}
We now turn our attention to the calculation of the polarization.  We will first
quote the known first-order result, and then embark on a detailed calculation of
the second-order result.  A general feature of these contributions at all orders
is that they are ultraviolet divergent, and thus we introduce a momentum cutoff
$\Lambda$ into our theory.
\subsection{First order}
The first-order, noninteracting, contribution to the polarization, shown in Fig.\ \ref{Fig:Polarization_FE},
is well known\cite{AbrikosovJETP1971}, so we simply quote the result here for the
sake of completeness and leave the derivation to Appendix \ref{App:Polarization_FE}.  If one computes the polarization using the given action in
Eq.~\eqref{Eq:Action}, then this contribution is
\begin{align}
&\Pi_B(q) \cr
&=-\frac{|\vec{q}|^2}{24\pi^2v_F}N\ln\left\{\frac{[z^2+(2\lambda+1)^2][z^2+(2\lambda-1)^2]}{(z^2+1)^2}\right\} \cr
&+\frac{|\vec{q}|^2}{8\pi^2v_F}N\int_{-1}^{1}dy\,y(1-\frac{1}{3}y^2)\frac{2\lambda+y}{z^2+(2\lambda+y)^2}, \label{Eq:Polarization_FE}
\end{align}
where $z=\frac{q_0}{v_F|\vec{q}|}$ and $\lambda=\frac{\Lambda}{|\vec{q}|}$.  The
integral in this expression can be done analytically, but the result is rather
complicated and not particularly illuminating.  We see that $\Pi_B(q)$ is logarithmically
divergent as $\Lambda\to\infty$, the entire divergence coming from the first term.
The logarithmic and finite terms, the latter of which we will need later, in $\Pi_B(q)$
are
\begin{equation}
\Pi_B(q)\approx-\frac{|\vec{q}|^2}{g^2}\frac{2\alpha}{3\pi}N\left [\ln\left (\frac{\Lambda}{|\vec{q}|}\right )-\frac{1}{2}\ln\left (\frac{z^2+1}{4}\right )\right ]. \label{Eq:Polarization_FE_Div}
\end{equation}
Note that this expression has an additional factor of $2N$ compared to that found
in Ref.\ \onlinecite{AbrikosovJETP1971}.  This is because said reference considers
a single Dirac cone, while we consider a system with $2N$ flavors of Dirac cones.

\subsection{Second order}
\begin{figure}
\subfigure[]{\includegraphics[width=0.46\columnwidth]{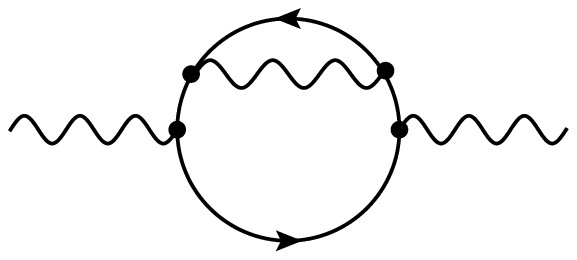}}\qquad
\subfigure[]{\includegraphics[width=0.46\columnwidth]{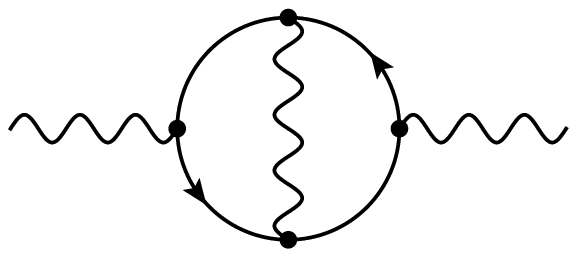}}
\caption{Self-energy (left) and vertex (right) corrections to the polarization, $\Pi_{SE}(q)$ and $\Pi_V(q)$,
respectively.}
\label{Fig:Polarization_SE}
\end{figure}
\begin{figure}
\includegraphics[width=0.49\columnwidth]{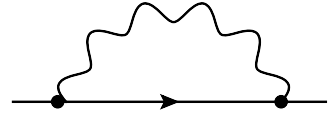}
\caption{One-loop electron self-energy, $\Sigma_1(q)$.}
\label{Fig:SelfEnergy_FE}
\end{figure}
We now consider the second-order correction.  Two diagrams, both depicted in
Fig.\ \ref{Fig:Polarization_SE}, contribute to this correction.  Note that the first
diagram contains the first-order electron self-energy $\Sigma_1(q)$, shown in
Fig.\ \ref{Fig:SelfEnergy_FE}, as a subdiagram.  We will simply quote the result
here, leaving the derivation to App.\ \ref{App:ElectronSE_FE}:
\begin{equation}
\Sigma_1(q)=i\frac{2\alpha}{3\pi} \left[\frac{4}{3}+\ln \left(\frac{\Lambda}{|\vec{q}|}\right )\right] v_F\vec{q}\cdot\vec{\gamma} \label{Eq:ElectronSE_FE}
\end{equation}

\subsubsection{``Self-energy'' correction}
We will first determine the ``self-energy'' term, which we denote by $\Pi_{SE}(q)$.
The integral for this correction is
\begin{equation}
\Pi_{SE}(q)=2N\int\frac{d^4k}{(2\pi)^4}\,\mbox{Tr}[\gamma^0G_0(k+q)\gamma^0G_0(k)\Sigma_1(k)G_0(k)],
\end{equation}
which includes a factor of $2$ due to the symmetry of the diagram.  We now substitute
in the bare Green's functions and self-energy and evaluate the trace.  The trace that
we must perform is
\begin{eqnarray}
&&\mbox{Tr}(\gamma^0\gamma^\mu\gamma^0\gamma^\nu\gamma^i\gamma^\rho) \cr
&&=4(2\delta^{\mu 0}-1)(\delta^{\mu\nu}\delta^{i\rho}-\delta^{\mu i}\delta^{\nu\rho}+\delta^{\mu\rho}\delta^{\nu i}).
\end{eqnarray}
Using this formula, we may rewrite the integral as
\begin{eqnarray}
\Pi_{SE}(q)&=&\frac{16\alpha}{3\pi}v_F N\int\frac{d^3\vec{k}}{(2\pi)^3}\,(I_1-I_2+I_3) \cr
&\times&\left [\frac{4}{3}+\ln\left (\frac{\Lambda}{|\vec{k}|}\right )\right ],
\end{eqnarray}
where $I_1$, $I_2$, and $I_3$ are integrals over $k_0$, given by
\begin{eqnarray}
I_1&=&2v_F\int_{-\infty}^{\infty}\frac{dk_0}{2\pi}\,\frac{k_0(k_0+q_0)|\vec{k}|^2}{(k_0^2+v_F^2|\vec{k}|^2)^2[(k_0+q_0)^2+v_F^2|\vec{k}+\vec{q}|^2]} \cr
&=&-\frac{|\vec{k}|[q_0^2-v_F^2(|\vec{k}|+|\vec{k}+\vec{q}|)^2]}{2[q_0^2+v_F^2(|\vec{k}|+|\vec{k}+\vec{q}|)^2]^2}, \\
I_2&=&2v_F^3\int_{-\infty}^{\infty}\frac{dk_0}{2\pi}\,\frac{\vec{k}\cdot(\vec{k}+\vec{q})|\vec{k}|^2}{(k_0^2+v_F^2|\vec{k}|^2)^2[(k_0+q_0)^2+v_F^2|\vec{k}+\vec{q}|^2]} \cr
&=&\left\{\frac{1}{2|\vec{k}|[q_0^2+v_F^2(|\vec{k}|+|\vec{k}+\vec{q}|)^2]}\right. \cr
&&+\frac{1}{|\vec{k}+\vec{q}|[q_0^2+v_F^2(|\vec{k}|+|\vec{k}+\vec{q}|)^2]} \cr
&&-\left.\frac{q_0^2}{|\vec{k}+\vec{q}|[q_0^2+v_F^2(|\vec{k}|+|\vec{k}+\vec{q}|)^2]^2}\right\}\vec{k}\cdot(\vec{k}+\vec{q}), \\
I_3&=&v_F\int_{-\infty}^{\infty}\frac{dk_0}{2\pi}\,\frac{\vec{k}\cdot(\vec{k}+\vec{q})}{(k_0^2+v_F^2|\vec{k}|^2)[(k_0+q_0)^2+v_F^2|\vec{k}+\vec{q}|^2]} \cr
&=&\frac{|\vec{k}|+|\vec{k}+\vec{q}|}{2|\vec{k}||\vec{k}+\vec{q}|[q_0^2+v_F^2(|\vec{k}|+|\vec{k}+\vec{q}|)^2]}\vec{k}\cdot(\vec{k}+\vec{q}).
\end{eqnarray}
If we now substitute all of these into $\Pi_{SE}(q)$, we obtain
\begin{eqnarray}
\Pi_{SE}(q)&=&-\frac{8\alpha}{3\pi}v_F N\int\frac{d^3\vec{k}}{(2\pi)^3}\,\frac{|\vec{k}|[q_0^2-v_F^2(|\vec{k}|+|\vec{k}+\vec{q}|)^2]}{[q_0^2+v_F^2(|\vec{k}|+|\vec{k}+\vec{q}|)^2]^2} \cr
&\times&\left [1-\frac{\vec{k}\cdot(\vec{k}+\vec{q})}{|\vec{k}||\vec{k}+\vec{q}|}\right ]\left [\frac{4}{3}+\ln\left (\frac{\Lambda}{|\vec{k}|}\right )\right ].
\end{eqnarray}
We now switch to a prolate spheroidal coordinate system, with the $z$ axis being the long
axis of the spheroids.  The system is defined as follows:
\begin{eqnarray}
k_{\perp,x}&=&\frac{1}{2}|\vec{q}|\sinh{\mu}\sin{\nu}\cos{\theta}, \\
k_{\perp,y}&=&\frac{1}{2}|\vec{q}|\sinh{\mu}\sin{\nu}\sin{\theta}, \\
k_{\|}&=&\frac{1}{2}|\vec{q}|(\cosh{\mu}\cos{\nu}-1), \label{Eq:ProSphCoord}
\end{eqnarray}
where $0\leq\mu<\infty$, $0\leq\nu<\pi$, and $0\leq\theta<2\pi$.  The Jacobian of this transformation
is
\begin{equation}
J=\frac{1}{8}|\vec{q}|^3\sinh{\mu}\sin{\nu}(\cosh^2{\mu}-\cos^2{\nu}).
\end{equation}
Some useful identities are
\begin{align}
|\vec{k}|&=\frac{1}{2}|\vec{q}|(\cosh{\mu}-\cos{\nu}), \\
|\vec{k}+\vec{q}|&=\frac{1}{2}|\vec{q}|(\cosh{\mu}+\cos{\nu}),
\end{align}
and
\begin{equation}
\vec{k}\cdot(\vec{k}+\vec{q})=\frac{1}{4}|\vec{q}|^2(\cosh^2{\mu}+\cos^2{\nu}-2).
\end{equation}
Using these identities, we may also write the Jacobian as $J=\tfrac{1}{2}|\vec{q}|\sinh{\mu}\sin{\nu}|\vec{k}||\vec{k}+\vec{q}|$.
Making this coordinate change, $\Pi_{SE}(q)$ becomes, after performing the (trivial) integral over $\theta$,
\begin{align}
&\Pi_{SE}(q)=-\frac{2}{\pi^2}\frac{|\vec{q}|^2}{g^2}\alpha^2 N\int_{0}^{\infty}d\mu\,\int_{0}^{\pi}d\nu\,\sinh{\mu}\sin{\nu} \nonumber \\
&\times\frac{(\cosh{\mu}-\cos{\nu})(z^2-\cosh^2{\mu})}{(z^2+\cosh^2{\mu})^2} \nonumber \\
&\times\sin^2{\nu}\left [\frac{2}{9}+\frac{1}{6}\ln\left (\frac{2\lambda}{\cosh{\mu}-\cos{\nu}}\right )\right ].
\end{align}
We now make another substitution, namely $x=\cosh{\mu}$ and $y=\cos{\nu}$.  Doing this, we obtain
\begin{align}
&\Pi_{SE}(q)=-\frac{2}{\pi^2}\frac{|\vec{q}|^2}{g^2}\alpha^2 N\int_{1}^{\infty}dx\,\int_{-1}^{1}dy \cr
&\times\frac{(x-y)(z^2-x^2)}{(z^2+x^2)^2}(1-y^2)\left [\frac{2}{9}+\frac{1}{6}\ln\left (\frac{2\lambda}{x-y}\right )\right ]. \quad
\end{align}
We note now that this integral is divergent, and thus we must impose a momentum cutoff $|\vec{k}|\leq\Lambda$.  In terms
of $x$ and $y$, this translates to imposing a cutoff of $2\lambda+y$ on the $x$
integral:
\begin{eqnarray}
\Pi_{SE}(q)&=&-\frac{2}{\pi^2}\frac{|\vec{q}|^2}{g^2}\alpha^2 N\int_{-1}^{1}dy\,\int_{1}^{2\lambda+y}dx \cr
&\times&\frac{(x-y)(z^2-x^2)}{(z^2+x^2)^2} \cr
&\times&(1-y^2)\left [\frac{2}{9}+\frac{1}{6}\ln\left (\frac{2\lambda}{x-y}\right )\right ].
\end{eqnarray}
We may simplify this expression further by interchanging the order of the $x$ and $y$ integrals,
\begin{eqnarray}
\int_{-1}^1 dy\,\int_1^{2\lambda+y}dx&\to&\int_1^{2\lambda-1}dx\,\int_{-1}^1 dy \cr
&+&\int_{2\lambda-1}^{2\lambda+1}dx\,\int_{x-2\lambda}^1 dy, \nonumber \\
\end{eqnarray}
and evaluating the $y$ integral.  Doing these, we find that
\begin{eqnarray}
\Pi_{SE}(q)&=&-\frac{2}{\pi^2}\frac{|\vec{q}|^2}{g^2}\alpha^2 N\left [\int_{1}^{2\lambda-1}dx\,\frac{z^2-x^2}{(z^2+x^2)^2}f_1(x)\right. \cr
&+&\left.\int_{2\lambda-1}^{2\lambda+1}dx\,\frac{z^2-x^2}{(z^2+x^2)^2}f_2(x)\right ],
\end{eqnarray}
where
\begin{eqnarray}
f_1(x)&=&\frac{1}{108}\left [\frac{3}{2}(x^4-6x^2-3)\ln\left (\frac{x+1}{x-1}\right )\right. \cr
&+&\left.x(49-3x^2)+24x\ln\left (\frac{2\lambda}{\sqrt{x^2-1}}\right )\right ], \\
f_2(x)&=&\frac{1}{864}\left\{-57\left (\frac{2\Lambda}{|\vec{q}|}\right )^4+160\left (\frac{2\Lambda}{|\vec{q}|}\right )^3x\right. \cr
&-&132\left (\frac{2\Lambda}{|\vec{q}|}\right )^2(x^2-1)+(x-1)^3(29x+75) \cr
&+&\left.12(x-1)^3(x+3)\ln\left (\frac{2\lambda}{x-1}\right )\right\}.
\end{eqnarray}
We may now extract the leading divergence of $\Pi_{SE}(q)$.  To do this, we take the derivative of the
integrals above with respect to $\lambda$, evaluate the result, and then expand in powers of $\lambda$.
A $\ln^2$ term would correspond to $\frac{\ln{\lambda}}{\lambda}$ with a coefficient twice that
of the $\ln^2$ term, while a simple $\ln$ term would correspond to $\frac{1}{\lambda}$.  The first
integral is the sole contributor to the divergence of $\Pi_{SE}(q)$; we find that it possesses both a
$\ln^2$ and a $\ln$ term.  We thus find that $\Pi_{SE}(q)$ is given by
\begin{eqnarray}
\Pi_{SE}(q)&\approx&\frac{|\vec{q}|^2}{g^2}\frac{2\alpha^2}{9\pi^2} N\ln^2\left (\frac{\Lambda}{|\vec{q}|}\right ) \nonumber \\
&+&\frac{|\vec{q}|^2}{g^2}\frac{2\alpha^2}{27\pi^2} N\left [\frac{2(z^2+4)}{z^2+1}\right. \nonumber \\
&+&\left.3\ln\left (\frac{4}{z^2+1}\right )\right ]\ln\left (\frac{\Lambda}{|\vec{q}|}\right ) \nonumber \\
&-&\frac{2}{\pi^2}\frac{|\vec{q}|^2}{g^2}\alpha^2Nf(z)+O\left (\frac{1}{\lambda}\right ). \label{Eq:PiSE}
\end{eqnarray}

\begin{figure}
\includegraphics[width=\columnwidth]{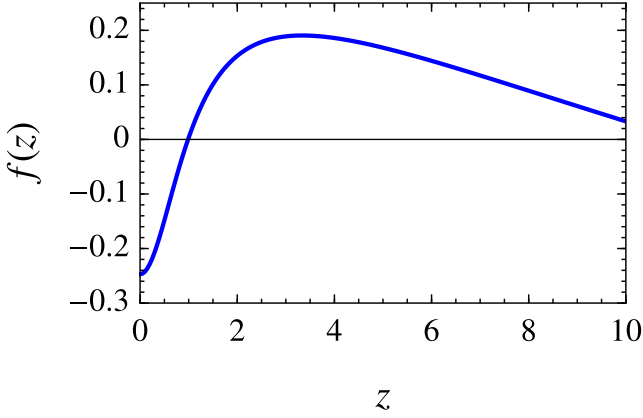}
\caption{Plot of the function $f(z)$ appearing in Eq.\ \eqref{Eq:PiSE}, giving the dependence of the finite
term in $\Pi_{SE}(q)$ as a function of $z=q_0/v_F|\vec{q}|$.}
\label{Fig:ConstTerm_PiSE}
\end{figure}
The finite term, $f(z)$, could, in principle, be found analytically, but the expression giving it is
extremely complicated, and thus we determine it numerically.  This term comes entirely from the
integral involving $f_1(x)$, as that involving $f_2(x)$ goes to zero as $\lambda\to\infty$.  We present
a plot of $f(z)$ in Fig.\ \ref{Fig:ConstTerm_PiSE}.

\subsubsection{Vertex correction to polarization}
We now turn our attention to the vertex contribution, $\Pi_V(q)$.  This contribution is given by
\begin{eqnarray}
\Pi_V(q)&=&-4\pi\alpha v_FN\int\frac{d^4k}{(2\pi)^4}\,\int\frac{d^4p}{(2\pi)^4} \cr
&\times&\mbox{Tr}\left [\gamma^0\frac{\slashed{k}}{k^2}\gamma^0\frac{\slashed{k}+\slashed{q}}{(k+q)^2}\gamma^0\frac{\slashed{p}+\slashed{q}}{(p+q)^2}\gamma^0\frac{\slashed{p}}{p^2}\right ]\frac{1}{|\vec{k}-\vec{p}|^2}. \nonumber \\
\end{eqnarray}
We now evaluate the trace and perform the integrals on $k_0$ and $p_0$.  The trace that we need
to evaluate is
\begin{eqnarray}
&&\mbox{Tr}(\gamma^0\gamma^\mu\gamma^0\gamma^\nu\gamma^0\gamma^\rho\gamma^0\gamma^\sigma) \cr
&&=4\delta^{\mu 0}\delta^{\rho 0}(2\delta^{\nu 0}\delta^{\sigma 0}-\delta^{\nu\sigma})-4\delta^{\mu 0}\delta^{\rho\neq 0}(\delta^{\nu 0}\delta^{\rho\sigma}+\delta^{\sigma 0}\delta^{\nu\rho}) \cr
&&-4\delta^{\mu\neq 0}\delta^{\rho 0}(\delta^{\mu\nu}\delta^{\sigma 0}+\delta^{\mu\sigma}\delta^{\nu 0}) \cr
&&+4\delta^{\mu\neq 0}\delta^{\rho\neq 0}(\delta^{\mu\nu}\delta^{\rho\sigma}-\delta^{\mu\rho}\delta^{\nu\sigma}+\delta^{\mu\sigma}\delta^{\nu\rho}).
\end{eqnarray}
Using this, we find that the integral splits into nine terms:
\begin{eqnarray}
J_1&=&-32\pi\alpha v_FN\int\frac{d^4k}{(2\pi)^4}\,\int\frac{d^4p}{(2\pi)^4}\,\frac{1}{|\vec{k}-\vec{p}|^2} \cr
&\times&\frac{k_0p_0(p_0+q_0)(k_0+q_0)}{k^2p^2(p+q)^2(k+q)^2}, \\
J_2&=&16\pi\alpha v_FN\int\frac{d^4k}{(2\pi)^4}\,\int\frac{d^4p}{(2\pi)^4}\,\frac{1}{|\vec{k}-\vec{p}|^2} \cr
&\times&\frac{k_0(p_0+q_0)p\cdot(k+q)}{k^2p^2(p+q)^2(k+q)^2}, \\
J_3&=&16\pi\alpha v_F^3N\int\frac{d^4k}{(2\pi)^4}\,\int\frac{d^4p}{(2\pi)^4}\,\frac{1}{|\vec{k}-\vec{p}|^2} \cr
&\times&\frac{k_0(k_0+q_0)\vec{p}\cdot(\vec{p}+\vec{q})}{k^2p^2(p+q)^2(k+q)^2}, \\
J_4&=&16\pi\alpha v_F^3N\int\frac{d^4k}{(2\pi)^4}\,\int\frac{d^4p}{(2\pi)^4}\,\frac{1}{|\vec{k}-\vec{p}|^2} \cr
&\times&\frac{k_0p_0(\vec{p}+\vec{q})\cdot(\vec{k}+\vec{q})}{k^2p^2(p+q)^2(k+q)^2}, \\
J_5&=&16\pi\alpha v_F^3N\int\frac{d^4k}{(2\pi)^4}\,\int\frac{d^4p}{(2\pi)^4}\,\frac{1}{|\vec{k}-\vec{p}|^2} \cr
&\times&\frac{p_0(p_0+q_0)\vec{k}\cdot(\vec{k}+\vec{q})}{k^2p^2(p+q)^2(k+q)^2}, \\
J_6&=&16\pi\alpha v_F^3N\int\frac{d^4k}{(2\pi)^4}\,\int\frac{d^4p}{(2\pi)^4}\,\frac{1}{|\vec{k}-\vec{p}|^2} \cr
&\times&\frac{(k_0+q_0)(p_0+q_0)\vec{k}\cdot\vec{p}}{k^2p^2(p+q)^2(k+q)^2}, \\
J_7&=&-16\pi\alpha v_F^5N\int\frac{d^4k}{(2\pi)^4}\,\int\frac{d^4p}{(2\pi)^4}\,\frac{1}{|\vec{k}-\vec{p}|^2} \cr
&\times&\frac{\vec{k}\cdot(\vec{k}+\vec{q})\vec{p}\cdot(\vec{p}+\vec{q})}{k^2p^2(p+q)^2(k+q)^2}, \\
J_8&=&16\pi\alpha v_F^3N\int\frac{d^4k}{(2\pi)^4}\,\int\frac{d^4p}{(2\pi)^4}\,\frac{1}{|\vec{k}-\vec{p}|^2} \cr
&\times&\frac{\vec{k}\cdot(\vec{p}+\vec{q})(k+q)\cdot p}{k^2p^2(p+q)^2(k+q)^2,} \\
J_9&=&-16\pi\alpha v_F^5N\int\frac{d^4k}{(2\pi)^4}\,\int\frac{d^4p}{(2\pi)^4}\,\frac{1}{|\vec{k}-\vec{p}|^2} \cr
&\times&\frac{\vec{k}\cdot\vec{p}(\vec{k}+\vec{q})\cdot(\vec{p}+\vec{q})}{k^2p^2(p+q)^2(k+q)^2}.
\end{eqnarray}
Evaluating the frequency integrals is a straightforward, if tedious, exercise; once we have done so
and simplified the result, we obtain
\begin{eqnarray}
\Pi_V(q)&=&-4\pi\alpha v_FN\int\frac{d^3\vec{k}}{(2\pi)^3}\,\int\frac{d^3\vec{p}}{(2\pi)^3}\,\frac{1}{|\vec{k}-\vec{p}|^2} \cr
&\times&\frac{1}{[q_0^2+v_F^2(|\vec{k}|+|\vec{k}+\vec{q}|)^2][q_0^2+v_F^2(|\vec{p}|+|\vec{p}+\vec{q}|)^2]} \cr
&\times&(-q_0^2Q_1+v_F^2Q_2),
\end{eqnarray}
where $Q_1$ and $Q_2$ are given by
\begin{eqnarray}
Q_1&=&\frac{\vec{k}\cdot\vec{p}}{|\vec{k}||\vec{p}|}-\frac{(\vec{k}+\vec{q})\cdot\vec{p}}{|\vec{k}+\vec{q}||\vec{p}|}-\frac{\vec{k}\cdot(\vec{p}+\vec{q})}{|\vec{k}||\vec{p}+\vec{q}|}+\frac{(\vec{k}+\vec{q})\cdot(\vec{p}+\vec{q})}{|\vec{k}+\vec{q}||\vec{p}+\vec{q}|}, \nonumber \\ \\
Q_2&=&\frac{(|\vec{k}|+|\vec{k}+\vec{q}|)(|\vec{p}|+|\vec{p}+\vec{q}|)}{|\vec{k}||\vec{k}+\vec{q}||\vec{p}||\vec{p}+\vec{q}|}[\vec{k}\cdot\vec{p}|\vec{q}|^2-(\vec{k}\cdot\vec{q})(\vec{p}\cdot\vec{q}) \cr
&+&(|\vec{k}|^2+\vec{k}\cdot\vec{q}-|\vec{k}||\vec{k}+\vec{q}|)(|\vec{p}|^2+\vec{p}\cdot\vec{q}-|\vec{p}||\vec{p}+\vec{q}|)]. \nonumber \\
\end{eqnarray}
To perform this integral, we introduce two sets of prolate spheroidal coordinates, one for $\vec{k}$
($\mu$, $\nu$, $\theta$) and one for $\vec{p}$ ($\mu'$, $\nu'$, $\theta'$).  The quantities $Q_1$ and
$Q_2$ become
\begin{eqnarray}
Q_1&=&\frac{4}{(\cosh^2{\mu}-\cos^2{\nu})(\cosh^2{\mu'}-\cos^2{\nu'})} \cr
&\times&[\cosh{\mu}\sin^2{\nu}\cosh{\mu'}\sin^2{\nu'} \cr
&+&\sinh{\mu}\sin{\nu}\cos{\nu}\sinh{\mu'}\sin{\nu'}\cos{\nu'}\cos(\theta-\theta')], \\
Q_2&=&\frac{4|\vec{q}|^2\cosh{\mu}\cosh{\mu'}}{(\cosh^2{\mu}-\cos^2{\nu})(\cosh^2{\mu'}-\cos^2{\nu'})} \cr
&\times&[\sin^2{\nu}\sin^2{\nu'}+\sinh{\mu}\sin{\nu}\sinh{\mu'}\sin{\nu'}\cos(\theta-\theta')]. \nonumber \\
\end{eqnarray}
Furthermore,
\begin{eqnarray}
|\vec{k}-\vec{p}|^2&=&\frac{1}{4}|\vec{q}|^2[\cosh^2{\mu}+\cos^2{\nu}+\cosh^2{\mu'}+\cos^2{\nu'} \cr
&&-2-2\cosh{\mu}\cos{\nu}\cosh{\mu'}\cos{\nu'} \cr
&&-2\sinh{\mu}\sin{\nu}\sinh{\mu'}\sin{\nu'}\cos(\theta-\theta')] \cr
&=&|\vec{q}|^2f(\mu,\nu,\theta;\mu',\nu',\theta').
\end{eqnarray}
We thus obtain
\begin{eqnarray}
\Pi_V(q)&=&\frac{|\vec{q}|^2}{g^2}\frac{\alpha^2}{256\pi^4}N\int_0^{\infty}d\mu\,\int_0^\pi d\nu\,\int_0^{2\pi}d\theta \cr
&\times&\int_0^{\infty}d\mu'\,\int_0^\pi d\nu'\,\int_0^{2\pi}d\theta'\,\frac{\sinh{\mu}\sin{\nu}\sinh{\mu'}\sin{\nu'}}{f(\mu,\nu,\theta;\mu',\nu',\theta')} \cr
&\times&\frac{1}{(z^2+\cosh^2{\mu})(z^2+\cosh^2{\mu'})}\left (z^2Q'_1-\frac{Q'_2}{|\vec{q}|^2}\right ), \nonumber \\
\end{eqnarray}
where $Q'_i=(\cosh^2{\mu}-\cos^2{\nu})(\cosh^2{\mu'}-\cos^2{\nu'})Q_i$.

Note that the integrand only depends on $\theta$ and $\theta'$ through their difference, $\theta-\theta'$,
and that it is a periodic function of both.  As a result, we may ``shift away'' one of these variables, say,
$\theta'$, thus making the integral over that variable trivial.  The other integral, in this case over $\theta$,
can then be done with the aid of the formulas,
\begin{eqnarray}
\int_0^{2\pi}d\theta\,\frac{1}{a+b\cos{\theta}}&=&\frac{2\pi}{\sqrt{a^2-b^2}}, \label{Eq:IntIdents1} \\
\int_0^{2\pi}d\theta\,\frac{\cos{\theta}}{a+b\cos{\theta}}&=&\frac{2\pi}{b}\left (1-\frac{a}{\sqrt{a^2-b^2}}\right ). \label{Eq:IntIdents2}
\end{eqnarray}
Doing these integrals, we find that $\Pi_V(q)$ splits into two parts, $\Pi_V(q)=\Pi_{V,E}(q)+\Pi_{V,H}(q)$.
These two parts are as follows.  $\Pi_{V,E}(q)$ is the ``easy'' part, which we can do analytically, while
$\Pi_{V,H}(q)$ is the ``hard'' part, which cannot be determined completely analytically.

The ``easy'' part comes from the $\frac{2\pi}{b}$ in Eq.\ \eqref{Eq:IntIdents2}, and is given by,
after changing variables to $x=\cosh{\mu}$ and $y=\cos{\nu}$,
\begin{equation}
\Pi_{V,E}(q)=\frac{|\vec{q}|^2}{g^2}\frac{\alpha^2}{8\pi^2}N(I_1^2-z^2I_2^2),
\end{equation}
where
\begin{eqnarray}
I_1&=&\int_1^{\infty}dx\,\int_{-1}^1 dy\,\frac{x}{z^2+x^2}, \\
I_2&=&\int_1^{\infty}dx\,\int_{-1}^1 dy\,\frac{y}{z^2+x^2}.
\end{eqnarray}
Note that $I_1$ is logarithmically divergent; we must therefore impose a cutoff, as before.  These integrals
may be done exactly; the results are
\begin{eqnarray}
I_1&=&\ln\left [\frac{z^2+(2\lambda-1)^2}{z^2+1}\right ]+(\lambda+\frac{1}{2})\ln\left [\frac{z^2+(2\lambda+1)^2}{z^2+(2\lambda-1)^2}\right ] \cr
&-&2+z\left [\tan^{-1}\left (\frac{2\lambda+1}{z}\right )-\tan^{-1}\left (\frac{2\lambda-1}{z}\right )\right ], \\
I_2&=&\lambda\ln\left [\frac{z^2+(2\lambda+1)^2}{z^2+(2\lambda-1)^2}\right ]-1+\frac{z^2+1-4\lambda^2}{2z} \cr
&\times&\left [\tan^{-1}\left (\frac{2\lambda+1}{z}\right )-\tan^{-1}\left (\frac{2\lambda-1}{z}\right )\right ]. \nonumber \\
\end{eqnarray}
The divergence of $\Pi_{V,E}(q)$ comes from $I_1$.  To determine the full divergence, we need both the $\ln$
and finite terms of $I_1$:
\begin{equation}
I_1\approx2\ln\left (\frac{\Lambda}{|\vec{q}|}\right )+\ln\left (\frac{4}{1+z^2}\right ).
\end{equation}
We can also determine the finite term; in fact, $I_2$ goes to zero as $\lambda\to\infty$, so the full finite
contribution is determined by $I_1$.  Therefore, the logarithmically divergent and finite terms coming from this
part of $\Pi_V(q)$ are
\begin{eqnarray}
\Pi_{V,E}(q)&\approx&\frac{|\vec{q}|^2}{g^2}\frac{\alpha^2}{2\pi^2}N\ln^2\left (\frac{\Lambda}{|\vec{q}|}\right ) \cr
&+&\frac{|\vec{q}|^2}{g^2}\frac{\alpha^2}{2\pi^2}N\ln\left (\frac{4}{1+z^2}\right )\ln\left (\frac{\Lambda}{|\vec{q}|}\right ) \cr
&+&\frac{|\vec{q}|^2}{g^2}\frac{\alpha^2}{8\pi^2}N\ln^2\left (\frac{4}{1+z^2}\right ).
\end{eqnarray}

The remaining terms in the integrals in Eqs. \eqref{Eq:IntIdents1} and \eqref{Eq:IntIdents2} give us $\Pi_{V,H}(q)$.
These are
\begin{eqnarray}
\Pi_{V,H}(q)&=&\frac{|\vec{q}|^2}{g^2}\frac{\alpha^2}{4\pi^2}N\int_0^{\infty}d\mu\,\int_0^{\pi}d\nu \cr
&\times&\int_0^{\infty}d\mu'\,\int_0^{\pi}d\nu'\,\frac{\sinh{\mu}\sin{\nu}\sinh{\mu'}\sin{\nu'}}{\sqrt{F(\mu,\nu;\mu',\nu')}} \cr
&\times&\frac{R_1-z^2R_2}{(z^2+\cosh^2{\mu})(z^2+\cosh^2{\mu'})},
\end{eqnarray}
where
\begin{eqnarray}
R_1&=&\cosh{\mu}\cosh{\mu'}(-\sin^2{\nu}\sin^2{\nu'}-\tfrac{1}{2}\cosh^2{\mu} \cr
&-&\tfrac{1}{2}\cos^2{\nu}-\tfrac{1}{2}\cosh^2{\mu'}-\tfrac{1}{2}\cos^2{\nu'}+1 \cr
&+&\cosh{\mu}\cos{\nu}\cosh{\mu'}\cos{\nu'}), \\
R_2&=&-\cosh{\mu}\sin^2{\nu}\cosh{\mu'}\sin^2{\nu'} \cr
&-&\tfrac{1}{2}\cos{\nu}\cos{\nu'}(\cosh^2{\mu}+\cos^2{\nu}+\cosh^2{\mu'} \cr
&+&\cos^2{\nu'}-2-2\cosh{\mu}\cos{\nu}\cosh{\mu'}\cos{\nu'}),
\end{eqnarray}
and
\begin{eqnarray}
F(\mu,\nu;\mu',\nu')&=&[\cosh(\mu+\mu')-\cos(\nu+\nu')] \cr
&\times&[\cosh(\mu+\mu')-\cos(\nu-\nu')] \cr
&\times&[\cosh(\mu-\mu')-\cos(\nu+\nu')] \cr
&\times&[\cosh(\mu-\mu')-\cos(\nu-\nu')].
\end{eqnarray}
This integral also has a logarithmic divergence.  In fact, it too contributes to the $\ln^2$ term
of $\Pi_V(q)$.  Fortunately, we can determine this term analytically.  We begin by introducing a
pair of $x$ and $y$ coordinates corresponding to each pair of $\mu$ and $\nu$ coordinates and by imposing
a cutoff, as before.  Doing this and simplifying the resulting expression, we obtain
\begin{eqnarray}
\Pi_{V,H}(q)&=&\frac{|\vec{q}|^2}{g^2}\frac{\alpha^2}{4\pi^2}N\int_{-1}^{1}dy\,\int_1^{2\lambda+y}dx \cr
&\times&\int_{-1}^{1}dy'\,\int_1^{2\lambda+y'}dx'\,\frac{1}{\sqrt{\bar{F}(x,y;x',y')}} \cr
&\times&\frac{\bar{R}_1-z^2\bar{R}_2}{(z^2+x^2)(z^2+x'^2)},
\end{eqnarray}
where $\lambda=\frac{\Lambda}{|\vec{q}|}$, $z=\frac{q_0}{v_F|\vec{q}|}$, and $\bar{F}$, $\bar{R}_1$, and
$\bar{R}_2$ are $F$, $R_1$, and $R_2$, respectively, rewritten in terms of the $x$ and $y$ coordinates.
We now take the derivative of this integral with respect to $\lambda$.  The $\ln^2{\lambda}$ term
of the integral will correspond to a $\frac{\ln{\lambda}}{\lambda}$ term in the derivative, with a coefficient
twice as large as that of the $\ln^2$ term.  The result that we obtain is
\begin{eqnarray}
\frac{\partial}{\partial\lambda}\Pi_{V,H}(q)&=&\frac{|\vec{q}|^2}{g^2}\frac{\alpha^2}{\pi^2}N\int_{-1}^{1}dy\,\int_1^{2\lambda+y}dx \cr
&\times&\int_{-1}^{1}dy'\,\frac{1}{\sqrt{\bar{F}(x,y;2\lambda+y',y')}} \cr
&\times&\frac{\bar{R}_1(x'=2\lambda+y')-z^2\bar{R}_2(x'=2\lambda+y')}{(z^2+x^2)[z^2+(2\lambda+y')^2]}. \nonumber \label{Eq:DPiVHDL} \\
\end{eqnarray}
We now expand the integrand in powers of $\frac{1}{\lambda}$.  The leading term is, after integrating over $y'$,
with respect to which the leading term is constant,
\begin{eqnarray}
\frac{\partial}{\partial\lambda}\Pi_{V,H}(q)&\approx&-\frac{|\vec{q}|^2}{g^2}\frac{\alpha^2}{2\pi^2}N\frac{1}{\lambda}\int_{-1}^{1}dy \cr
&\times&\int_1^{2\lambda+y}dx\,\frac{x}{x^2+z^2}.
\end{eqnarray}
This is the only term in the expansion that will contribute to the $\ln^2{\lambda}$ term of $\Pi_{V,H}(q)$.
It also contributes to the subleading $\ln{\lambda}$ term, and is in fact the only contribution with a coefficient
that is a function of $z$; all other terms only contribute $\ln{\lambda}$-divergent terms with constant coefficients.
The total constant contributed by these, which we will call $C$, must be determined numerically.  If we now evaluate
the integrals, we obtain
\begin{eqnarray}
\frac{\partial}{\partial\lambda}\Pi_{V,H}(q)&\approx&-\frac{|\vec{q}|^2}{g^2}\frac{\alpha^2}{2\pi^2}N\left\{\frac{2\ln{\lambda}}{\lambda}\right. \cr
&+&\left.\left [\ln\left (\frac{4}{1+z^2}\right )+C\right ]\frac{1}{\lambda}\right\}.
\end{eqnarray}
Therefore, the leading divergence of $\Pi_{V,H}(q)$ is
\begin{eqnarray}
\Pi_{V,H}(q)&\approx&-\frac{|\vec{q}|^2}{g^2}\frac{\alpha^2}{2\pi^2}N\left\{\ln^2\left (\frac{\Lambda}{|\vec{q}|}\right )\right. \cr
&+&\left.\left [\ln\left (\frac{4}{1+z^2}\right )+C\right ]\ln\left (\frac{\Lambda}{|\vec{q}|}\right )\right\}. \nonumber \\
\end{eqnarray}
The $\ln^2$ term exactly cancels that from the ``easy'' part, so that the whole diagram only has a simple logarithmic
divergence.  The full form of $\Pi_V(q)$ is thus
\begin{eqnarray}
\Pi_V(q)&=&\frac{|\vec{q}|^2}{g^2}\frac{\alpha^2}{4\pi^2}N\left [-2C\ln\left (\frac{\Lambda}{|\vec{q}|}\right )+g(z)\right ] \cr
&+&O\left (\frac{1}{\lambda}\right ), \label{Eq:PiVFull}
\end{eqnarray}
where $g(z)$ is a function that we will determine shortly.

\begin{figure}
\includegraphics[width=\columnwidth]{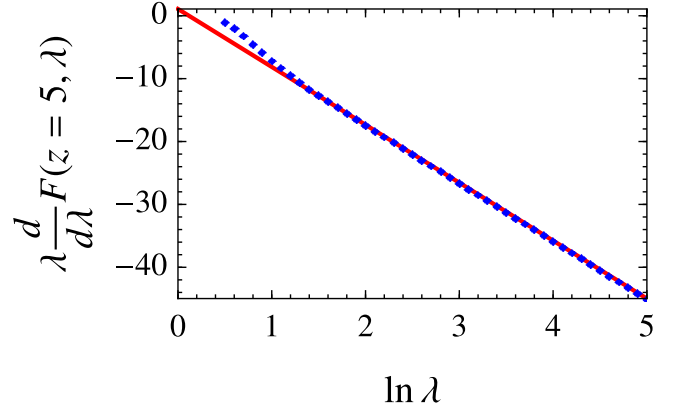}
\caption{Plot of $\lambda\frac{d}{d\lambda}F(z,\lambda)$ for $z=5$, where $\lambda=\Lambda/|\vec{q}|$ and $z=q_0/v_F|\vec{q}|$.  At large $\lambda$, this curve becomes approximately linear;
the slope of the line is exactly $-4$ and is twice the coefficient of the $\ln^2{\lambda}$ term, while the constant term gives us
the coefficient of the $\ln{\lambda}$ term.  The red line is a linear fit to the points at large $\lambda$.}
\label{Fig:dPiVdl}
\end{figure}

\begin{figure}
\includegraphics[width=\columnwidth]{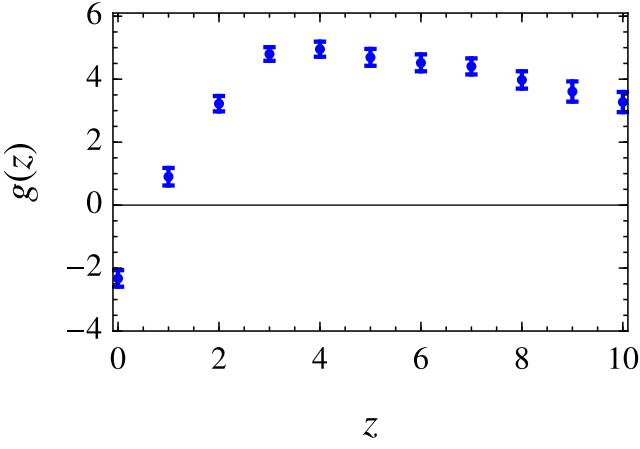}
\caption{Plot of the function $g(z)$ appearing in Eq.\ \eqref{Eq:PiVFull}, giving the dependence of the finite part
of $\Pi_V(q)$ as a function of $z=q_0/v_F|\vec{q}|$.}
\label{Fig:MD_Integral_Finite_Part}
\end{figure}
To obtain $C$, we numerically evaluate $\lambda\frac{\partial}{\partial\lambda}\Pi_{V,H}(q)$ for a large value of
$\lambda$ and $z=0$ using Eq.\ \eqref{Eq:DPiVHDL} and subtract off the $\ln{\lambda}$ term, which we know exactly.  In
the limit of large $\lambda$, $\lambda\frac{\partial}{\partial\lambda}\Pi_{V,H}(q)$ is a linear function of $\ln{\lambda}$,
the linear term giving the coefficient of the $\ln^2$ term and the constant term giving the coefficient of the $\ln$ term.
We note that $\Pi_{V,H}(q)$ and all two-loop contributions to the polarization that we determine have the form, $\frac{|\vec{q}|^2}{g^2}\frac{\alpha^2}{4\pi^2}NF(z,\lambda)$,
so that we only need to vary $z$ and $\lambda$.  To perform these numerical calculations, we made use of the \textsc{vegas} algorithm as implemented in the \textsc{cuba} package.  We obtained results with errors of $0.1\%$ or less; see Fig.\ \ref{Fig:dPiVdl} for
an example, in which we plot $\lambda\frac{d}{d\lambda}F(z,\lambda)$ for $z=5$.  We find that
\begin{align}
C\approx 1.33318 . \label{eq:C}
\end{align}
To obtain $g(z)$, we evaluate $\Pi_{V,H}(q)$ numerically for a large value of $\lambda$ and subtract off the logarithmically-divergent
terms; our result is plotted in Fig.\ \ref{Fig:MD_Integral_Finite_Part}.

\section{Second-order self-energy} \label{Sec:SelfEnergy_SE}
We now determine the electron self-energy to two loops.  There are three possible diagrams that contribute to this
order, and we discuss each in turn.

\subsection{Two-loop rainbow correction to self-energy}\label{sec:twolooprainbow}

\begin{figure}
\includegraphics[width=0.5\columnwidth]{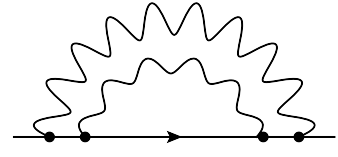}
\caption[Self-energy]{Two-loop rainbow correction to the electron self-energy, $\Sigma_{2b}(q)$.}
\label{fig:selfenergy2b}
\end{figure}

The first potential two-loop correction to the electron self-energy is shown in Fig.~\ref{fig:selfenergy2b}.
This diagram evaluates to
\begin{eqnarray}
\Sigma_{2b}(q)&=&-4\pi\alpha v_F\int\frac{d^4k}{(2\pi)^4}\frac{1}{|\vec{q}-\vec{k}|^2}\gamma^0G_0(k)\Sigma_1(k)G_0(k)\gamma^0\nonumber\\
&\propto&\int\frac{d^4k}{(2\pi)^4}\frac{1}{|\vec{q}-\vec{k}|^2}
\left[\frac{4}{3}+\ln \left(\frac{\Lambda}{|\vec{k}|}\right )\right]
\gamma^0\frac{\slashed k}{k^2}\vec{k}\cdot\vec\gamma\frac{\slashed k}{k^2}\gamma^0.\nonumber\\&&
\end{eqnarray}
Straightforward algebra reveals that the integral over $k_0$ vanishes identically:
\begin{eqnarray}
&&\int\frac{dk_0}{2\pi}\frac{1}{k^4}\gamma^0(k_0\gamma^0+v_F\vec{k}\cdot\vec\gamma)\vec{k}\cdot\vec\gamma(k_0\gamma^0+v_F\vec{k}\cdot\vec\gamma)\gamma^0\nonumber\\&&
=\vec{k}\cdot\vec\gamma\int\frac{dk_0}{2\pi}\frac{k_0^2-v_F^2|\vec{k}|^2}{(k_0^2+v_F^2|\vec{k}|^2)^2}=0.
\end{eqnarray}
Therefore, the full contribution vanishes identically:
\begin{equation}
\Sigma_{2b}(q)=0.\label{selfenergy2b}
\end{equation}

\subsection{Two-loop vertex correction to self-energy}
\begin{figure}
\includegraphics[width=0.5\columnwidth]{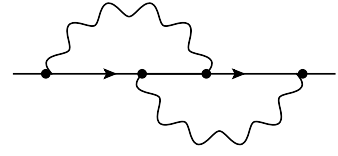}
\caption[Self-energy]{Two-loop vertex correction to the electron self-energy, $\Sigma_{2a}(q)$.}
\label{fig:selfenergy2a}
\end{figure}

{\it Reduction to a quadruple integral.}
The second two-loop correction to the electron self-energy is shown in Fig.~\ref{fig:selfenergy2a}.
This diagram has the value
\begin{eqnarray}
\Sigma_{2a}(q)&{=}&{-}16\pi^2 i\alpha^2 v_F^2{\int}\frac{d^4k}{(2\pi)^4}{\int}\frac{d^4p}{(2\pi)^4}\frac{1}{|\vec{q}{-}\vec{k}|^2}\frac{1}{|\vec{q}{-}\vec{p}|^2}\nonumber\\&&\qquad\qquad\times\gamma^0\frac{\slashed k}{k^2}
\gamma^0\frac{\slashed k{+}\slashed p{-}\slashed q}{(k{+}p{-}q)^2}\gamma^0\frac{\slashed p}{p^2}\gamma^0.
\end{eqnarray}
The product of gamma matrices in the integrand can be expanded as
\begin{eqnarray}
&&\gamma^0\gamma^\mu\gamma^0\gamma^\nu\gamma^0\gamma^\rho\gamma^0=\delta^{\mu0}\delta^{\nu0}\delta^{\rho0}\gamma^0{-}\delta^{\mu0}\delta^{\nu0}(1{-}\delta^{\rho0})\gamma^\rho \nonumber\\&&{-}\delta^{\mu0}(1{-}\delta^{\nu0})\delta^{\rho0}\gamma^\nu {-}\delta^{\mu0}(1{-}\delta^{\nu0})(1{-}\delta^{\rho0})\gamma^\nu\gamma^\rho\gamma^0 \nonumber\\&&{-}(1{-}\delta^{\mu0})\delta^{\nu0}\delta^{\rho0}\gamma^\mu {-}(1{-}\delta^{\mu0})\delta^{\nu0}(1{-}\delta^{\rho0})\gamma^\mu\gamma^\rho\gamma^0\nonumber\\&&{-}(1{-}\delta^{\mu0})(1{-}\delta^{\nu0})\delta^{\rho0}\gamma^\mu\gamma^\nu\gamma^0 \nonumber\\&&{+}(1{-}\delta^{\mu0})(1{-}\delta^{\nu0})(1{-}\delta^{\rho0})\gamma^\mu\gamma^\nu\gamma^\rho,
\end{eqnarray}
which leads to
\begin{eqnarray}
&&\Sigma_{2a}(q){=}\nonumber\\&&={-}16\pi^2 i\alpha^2 v_F^2{\int}\frac{d^4k}{(2\pi)^4}{\int}\frac{d^4p}{(2\pi)^4}\frac{1}{|\vec{q}{-}\vec{k}|^2}\frac{1}{|\vec{q}{-}\vec{p}|^2}\frac{1}{k^2p^2(k{+}p{-}q)^2}\nonumber\\&&\times\bigg\{k_0p_0(k_0{+}p_0{-}q_0)\gamma^0 {-}v_Fk_0(k_0{+}p_0{-}q_0)\vec{p}{\cdot}\vec\gamma\nonumber\\&&{-}v_Fk_0p_0(\vec{k}{+}\vec{p}{-}\vec{q}){\cdot}\vec\gamma{-}v_F^2k_0\gamma^0(\vec{k}{+}\vec{p}{-}\vec{q}){\cdot}\vec\gamma \vec{p}{\cdot}\vec\gamma\nonumber\\&&{-}v_Fp_0(k_0{+}p_0{-}q_0)\vec{k}{\cdot}\vec\gamma  {-}v_F^2(k_0{+}p_0{-}q_0)\gamma^0 \vec{k}{\cdot}\vec\gamma \vec{p}{\cdot} \vec\gamma\nonumber\\&&{-}v_F^2p_0\gamma^0\vec{k}{\cdot}\vec\gamma(\vec{k}{+}\vec{p}{-}\vec{q}){\cdot}\vec\gamma{+}v_F^3\vec{k}{\cdot}\vec\gamma(\vec{k}{+}\vec{p}{-}\vec{q}){\cdot}\vec\gamma \vec{p}{\cdot}\vec\gamma\bigg\}.\nonumber\\&&
\end{eqnarray}
As usual, we first perform the integrals over the energies $k_0$ and $p_0$:
\begin{eqnarray}
B_1&\equiv&{\int}\frac{dk_0}{2\pi}{\int}\frac{dp_0}{2\pi}\frac{k_0p_0(k_0{+}p_0{-}q_0)}{M(k,p)} \nonumber\\&{=}&\frac{1}{4}\frac{q_0}{q_0^2{+}v_F^2(|\vec{k}|{+}|\vec{p}|{+}|\vec{k}{+}\vec{p}{-}\vec{q}|)^2},\nonumber\\
\end{eqnarray}
\begin{eqnarray}
B_2&\equiv&{\int}\frac{dk_0}{2\pi}{\int}\frac{dp_0}{2\pi}\frac{k_0(k_0{+}p_0{-}q_0)}{M(k,p)} \nonumber\\&{=}&\frac{1}{4}\frac{|\vec{k}|{+}|\vec{p}|{+}|\vec{k}{+}\vec{p}{-}\vec{q}|}{|\vec{p}|[q_0^2{+}v_F^2(|\vec{k}|{+}|\vec{p}|{+}|\vec{k}{+}\vec{p}{-}\vec{q}|)^2]},\nonumber\\
\end{eqnarray}
\begin{eqnarray}
B_3&\equiv&{\int}\frac{dk_0}{2\pi}{\int}\frac{dp_0}{2\pi}\frac{k_0p_0}{M(k,p)} \nonumber\\&{=}&{-}\frac{1}{4}\frac{|\vec{k}|{+}|\vec{p}|{+}|\vec{k}{+}\vec{p}{-}\vec{q}|}{|\vec{k}{+}\vec{p}{-}\vec{q}|[q_0^2{+}v_F^2(|\vec{k}|{+}|\vec{p}|{+}|\vec{k}{+}\vec{p}{-}\vec{q}|)^2]},\nonumber\\
\end{eqnarray}
\begin{eqnarray}
B_4&\equiv&{\int}\frac{dk_0}{2\pi}{\int}\frac{dp_0}{2\pi}\frac{k_0}{M(k,p)} \nonumber\\&{=}&\frac{1}{4v_F^2}\frac{q_0}{|\vec{p}||\vec{k}{+}\vec{p}{-}\vec{q}|[q_0^2{+}v_F^2(|\vec{k}|{+}|\vec{p}|{+}|\vec{k}{+}\vec{p}{-}\vec{q}|)^2]},\nonumber\\
\end{eqnarray}
\begin{eqnarray}
B_5&\equiv&{\int}\frac{dk_0}{2\pi}{\int}\frac{dp_0}{2\pi}\frac{p_0(k_0{+}p_0{-}q_0)}{M(k,p)} \nonumber\\&{=}&\frac{1}{4}\frac{|\vec{k}|{+}|\vec{p}|{+}|\vec{k}{+}\vec{p}{-}\vec{q}|}{|\vec{k}|[q_0^2{+}v_F^2(|\vec{k}|{+}|\vec{p}|{+}|\vec{k}{+}\vec{p}{-}\vec{q}|)^2]},\nonumber\\
\end{eqnarray}
\begin{eqnarray}
B_6&\equiv&{\int}\frac{dk_0}{2\pi}{\int}\frac{dp_0}{2\pi}\frac{k_0{+}p_0{-}q_0}{M(k,p)} \nonumber\\&{=}&{-}\frac{1}{4v_F^2}\frac{q_0}{|\vec{k}||\vec{p}|[q_0^2{+}v_F^2(|\vec{k}|{+}|\vec{p}|{+}|\vec{k}{+}\vec{p}{-}\vec{q}|)^2]},\nonumber\\
\end{eqnarray}
\begin{eqnarray}
B_7&\equiv&{\int}\frac{dk_0}{2\pi}{\int}\frac{dp_0}{2\pi}\frac{p_0}{M(k,p)} \nonumber\\&{=}&\frac{1}{4v_F^2}\frac{q_0}{|\vec{k}||\vec{k}{+}\vec{p}{-}\vec{q}|[q_0^2{+}v_F^2(|\vec{k}|{+}|\vec{p}|{+}|\vec{k}{+}\vec{p}{-}\vec{q}|)^2]},\nonumber\\
\end{eqnarray}
\begin{eqnarray}
B_8&\equiv&{\int}\frac{dk_0}{2\pi}{\int}\frac{dp_0}{2\pi}\frac{1}{M(k,p)} \nonumber\\&{=}&\frac{1}{4v_F^2}\frac{|\vec{k}|{+}|\vec{p}|{+}|\vec{k}{+}\vec{p}{-}\vec{q}|}{|\vec{k}||\vec{p}||\vec{k}{+}\vec{p}{-}\vec{q}|[q_0^2{+}v_F^2(|\vec{k}|{+}|\vec{p}|{+}|\vec{k}{+}\vec{p}{-}\vec{q}|)^2]},\nonumber\\&&
\end{eqnarray}
with
\begin{eqnarray}
&&M(k,p)\equiv\nonumber\\&&(k_0^2{+}v_F^2|\vec{k}|^2)(p_0^2{+}v_F^2|\vec{p}|^2)[(k_0{+}p_0{-}q_0)^2{+}v_F^2|\vec{k}{+}\vec{p}{-}\vec{q}|^2].\nonumber\\&&
\end{eqnarray}
We then have
\begin{eqnarray}
&&\Sigma_{2a}(q){=}{-}16\pi^2 i\alpha^2 v_F^2{\int}\frac{d^3k}{(2\pi)^3}{\int}\frac{d^3p}{(2\pi)^3}\frac{1}{|\vec{q}{-}\vec{k}|^2}\frac{1}{|\vec{q}{-}\vec{p}|^2}\nonumber\\&&\times\bigg\{B_1\gamma^0{-}B_2v_F\vec{p}{\cdot}\vec\gamma {-}B_3v_F(\vec{k}{+}\vec{p}{-}\vec{q}){\cdot}\vec\gamma\nonumber\\&&{-}B_4v_F^2\gamma^0(\vec{k}{+}\vec{p}{-}\vec{q}){\cdot}\vec\gamma\vec{p}{\cdot}\vec\gamma{-}B_5v_F\vec{k}{\cdot}\vec\gamma {-}B_6v_F^2\gamma^0\vec{k}{\cdot}\vec\gamma\vec{p}{\cdot}\vec\gamma\nonumber\\&&{-}B_7v_F^2\gamma^0\vec{k}{\cdot}\vec\gamma(\vec{k}{+}\vec{p}{-}\vec{q}){\cdot}\vec\gamma {+}B_8v_F^3\vec{k}{\cdot}\vec\gamma(\vec{k}{+}\vec{p}{-}\vec{q}){\cdot}\vec\gamma\vec{p}{\cdot}\vec\gamma\bigg\}.\nonumber\\&&
\end{eqnarray}
The $\gamma^0$ component of this is
\begin{eqnarray}
&&{1\over4}\hbox{Tr}[\gamma^0\Sigma_{2a}(q)]{=}\nonumber\\&&={-}4\pi^2 i\alpha^2 v_F^2{\int}\frac{d^3k}{(2\pi)^3}{\int}\frac{d^3p}{(2\pi)^3}\frac{1}{|\vec{q}{-}\vec{k}|^2}\frac{1}{|\vec{q}{-}\vec{p}|^2}\big\{4B_1 \nonumber\\&&{-}4B_4v_F^2\vec{p}{\cdot}(\vec{k}{+}\vec{p}{-}\vec{q}){-}4B_6v_F^2\vec{k}{\cdot}\vec{p}{-}4B_7v_F^2\vec{k}{\cdot}(\vec{k}{+}\vec{p}{-}\vec{q})\big\}\nonumber\\
&&{=}{-}4\pi^2 i\alpha^2 v_F^2 q_0{\int}\frac{d^3k}{(2\pi)^3}{\int}\frac{d^3p}{(2\pi)^3}\frac{1}{|\vec{q}{-}\vec{k}|^2}\frac{1}{|\vec{q}{-}\vec{p}|^2}\nonumber\\&&\times\frac{1}{q_0^2{+}v_F^2(|\vec{k}|{+}|\vec{p}|{+}|\vec{k}{+}\vec{p}{-}\vec{q}|)^2} \bigg\{1{-}\nonumber\\&& \frac{\vec{p}}{|\vec{p}|}{\cdot}\frac{\vec{k}{+}\vec{p}{-}\vec{q}}{|\vec{k}{+}\vec{p}{-}\vec{q}|}{+}\frac{\vec{k}}{|\vec{k}|}{\cdot}\frac{\vec{p}}{|\vec{p}|} {-}\frac{\vec{k}}{|\vec{k}|}{\cdot}\frac{\vec{k}{+}\vec{p}{-}\vec{q}}{|\vec{k}{+}\vec{p}{-}\vec{q}|}\bigg\},\label{temporal2loop}
\end{eqnarray}
while the spatial components are
\begin{eqnarray}
&&{1\over4}\hbox{Tr}[\gamma^i\Sigma_{2a}(q)]{=}\nonumber\\&&={-}4\pi^2 i\alpha^2 v_F^2{\int}\frac{d^3k}{(2\pi)^3}{\int}\frac{d^3p}{(2\pi)^3}\frac{1}{|\vec{q}{-}\vec{k}|^2}\frac{1}{|\vec{q}{-}\vec{p}|^2}\Big\{{-}4B_2v_Fp_i\nonumber\\&&{-}4B_3v_F(k_i{+}p_i{-}q_i) {-}4B_5v_Fk_i{+}4B_8v_F^3\Big[k_i\vec{p}{\cdot}(\vec{k}{+}\vec{p}{-}\vec{q})\nonumber\\&&{-}(k_i{+}p_i{-}q_i)\vec{k}{\cdot}\vec{p}{+}p_i\vec{k}{\cdot}(\vec{k}{+}\vec{p}{-}\vec{q})\Big]\Big\}\nonumber\\
&&{=}{-}4\pi^2 i\alpha^2 v_F^2{\int}\frac{d^3k}{(2\pi)^3}{\int}\frac{d^3p}{(2\pi)^3}\frac{1}{|\vec{q}{-}\vec{k}|^2}\frac{1}{|\vec{q}{-}\vec{p}|^2} \nonumber\\&&\times\frac{|\vec{k}|{+}|\vec{p}|{+}|\vec{k}{+}\vec{p}{-}\vec{q}|}{q_0^2{+}v_F^2(|\vec{k}|{+}|\vec{p}|{+}|\vec{k}{+}\vec{p}{-}\vec{q}|)^2}\bigg\{{-}\frac{k_i}{|\vec{k}|}{-}\frac{p_i}{|\vec{p}|} \nonumber\\&&{+}\frac{k_i{+}p_i{-}q_i}{|\vec{k}{+}\vec{p}{-}\vec{q}|} {+}\frac{k_i(|\vec{p}|^2{-}\vec{p}{\cdot}\vec{q}){+}p_i(|\vec{k}|^2{-}\vec{k}{\cdot}\vec{q}){+}q_i\vec{k}{\cdot}\vec{p}}{|\vec{k}||\vec{p}||\vec{k}{+}\vec{p}{-}\vec{q}|}\bigg\}.\nonumber\\&&
\end{eqnarray}
If we choose the coordinates such that $\vec{q}{=}(0,0,|\vec{q}|)$, then it becomes apparent that the terms of the integrand which are proportional to $k_x,k_y$ or $p_x,p_y$ are odd functions of these variables, implying that these terms vanish upon integration. We may then make the replacement $k_i\to q_i\vec{k}{\cdot}\vec{q}/|\vec{q}|^2$, and similarly for $p_i$. The total integral is therefore proportional to $q_i$:
\begin{eqnarray}
&&{1\over4}\hbox{Tr}[\gamma^i\Sigma_{2a}(q)]{=}\frac{{-}4\pi^2 i\alpha^2 v_F^3q_i}{|\vec{q}|^2}{\int}\frac{d^3k}{(2\pi)^3}{\int}\frac{d^3p}{(2\pi)^3}\frac{1}{|\vec{q}{-}\vec{k}|^2}\frac{1}{|\vec{q}{-}\vec{p}|^2} \nonumber\\&&\times\frac{|\vec{k}|{+}|\vec{p}|{+}|\vec{k}{+}\vec{p}{-}\vec{q}|}{q_0^2{+}v_F^2(|\vec{k}|{+}|\vec{p}|{+}|\vec{k}{+}\vec{p}{-}\vec{q}|)^2}\bigg\{{-}\frac{\vec{k}{\cdot}\vec{q}}{|\vec{k}|}{-}\frac{\vec{p}{\cdot}\vec{q}}{|\vec{p}|}
{+}\frac{(\vec{k}{+}\vec{p}{-}\vec{q}){\cdot}\vec{q}}{|\vec{k}{+}\vec{p}{-}\vec{q}|}
\nonumber\\&&{+}\frac{|\vec{p}|^2\vec{k}{\cdot}\vec{q}{+}|\vec{k}|^2\vec{p}{\cdot}\vec{q}{+}|\vec{q}|^2\vec{k}{\cdot}\vec{p}{-}2(\vec{k}{\cdot}\vec{q})(\vec{p}{\cdot}\vec{q})}{|\vec{k}||\vec{p}||\vec{k}{+}\vec{p}{-}\vec{q}|}\bigg\}.\label{spatial2loop}\nonumber\\&&
\end{eqnarray}

{\it Extracting the divergence in the temporal part:-}
To extract the divergent logarithm term in the temporal part of the two{-}loop self{-}energy, Eq.~(\ref{temporal2loop}), we must examine the behavior of the quadruple integral in the region $|\vec{k}|,|\vec{p}|\gg|\vec{q}|$. In this regime, the integral reduces to
\begin{eqnarray}
\!\!\!&&{1\over4}\hbox{Tr}[\gamma^0\Sigma_{2a}(q)]={-}4\pi^2 i\alpha^2 v_F^2 q_0
\nonumber\\
\!\!\!&&\times{\int}\frac{d^3k}{(2\pi)^3}{\int}\frac{d^3p}{(2\pi)^3}\frac{1}{|\vec{k}|^2}\frac{1}{|\vec{p}|^2}\frac{1}{q_0^2{+}v_F^2(|\vec{k}|{+}|\vec{p}|{+}|\vec{k}{+}\vec{p}|)^2} \nonumber\\
\!\!\!&&\times\bigg\{1{-}\frac{\vec{p}{\cdot}(\vec{k}{+}\vec{p})}{|\vec{p}||\vec{k}{+}\vec{p}|} {-}\frac{\vec{k}{\cdot}(\vec{k}{+}\vec{p})}{|\vec{k}||\vec{k}{+}\vec{p}|}{+}\frac{\vec{k}{\cdot}\vec{p}}{|\vec{k}||\vec{p}|}\bigg\},
\end{eqnarray}
where we have discarded terms which become odd under the change of variable $\vec{k}\to{-}\vec{k}$, $\vec{p}\to{-}\vec{p}$ in the limit of large $|\vec{k}|, |\vec{p}|$. To perform the remaining integrals, we switch to prolate spheroidal coordinates:
\begin{eqnarray}
|\vec{p}|&=&\frac{|\vec{k}|}{2}\left(\cosh\mu-\cos\nu\right),\nonumber\\
|\vec{k}+\vec{p}|&=&\frac{|\vec{k}|}{2}\left(\cosh\mu+\cos\nu\right),\nonumber\\
\vec{k}\cdot\vec{p}&=&\frac{|\vec{k}|^2}{2}\left(\cosh\mu\cos\nu-1\right),\nonumber\\
\vec{p}\cdot(\vec{k}+\vec{p})&=&\frac{|\vec{k}|^2}{4}\left(\cosh^2\mu+\cos^2\nu-2\right),\nonumber\\
|\vec{k}|+|\vec{p}|+|\vec{k}+\vec{p}|&=&|\vec{k}|\left(\cosh\mu+1\right),\nonumber\\
d^3p&=&\frac{|\vec{k}|^3}{8}\sinh\mu\sin\nu(\cosh^2\mu-\cos^2\nu). \nonumber \\ \label{prolatespheroid}
\end{eqnarray}
Using spherical coordinates for $\vec{k}$ and performing the trivial integrations over $\theta$ and over the angular variables associated with $\vec{k}$, we find
\begin{eqnarray}
\!\!\!&&{1\over4}\hbox{Tr}[\gamma^0\Sigma_{2a}(q)]=
\nonumber\\\!\!\!&&=\frac{i\alpha^2 v_F^2q_0}{\pi^2}\int_0^{\Lambda}d|\vec{k}||\vec{k}|\int d\mu d\nu\frac{\sinh\mu\sin\nu}{q_0^2+v_F^2|\vec{k}|^2(\cosh\mu+1)^2} \cr
&&\times\frac{\sin^2\nu\sinh^2\tfrac{\mu}{2}}{(\cosh\mu-\cos\nu)^2}.
\end{eqnarray}
Next we perform the integration over $|\vec{k}|$:
\begin{eqnarray}
&&\int_0^\Lambda d|\vec{k}||\vec{k}|\frac{1}{q_0^2+v_F^2|\vec{k}|^2(1+\cosh\mu)^2} \cr
&&=\frac{1}{2(1+\cosh\mu)^2v_F^2}\ln\left(1+\frac{v_F^2\Lambda^2}{q_0^2}(1+\cosh\mu)^2\right), \nonumber \\
\end{eqnarray}
and define $x\equiv\cosh\mu$, $y\equiv\cos\nu$ to obtain:
\begin{eqnarray}
\!\!\!&&{1\over4}\hbox{Tr}[\gamma^0\Sigma_{2a}(q)]=
\nonumber\\\!\!\!&&=\frac{i\alpha^2 q_0}{4\pi^2}\int_1^\infty dx\frac{x-1}{(1+x)^2}\ln\left(1+\frac{v_F^2\Lambda^2}{q_0^2}(1+x)^2\right) \cr
&&\times\int_{-1}^1dy\frac{1-y^2}{(x-y)^2}\nonumber\\&&
=\frac{i\alpha^2 q_0}{\pi^2}\int_1^\infty dx\frac{x-1}{(1+x)^2}\ln\left(1+\frac{v_F^2\Lambda^2}{q_0^2}(1+x)^2\right) \cr
&&\times\left(x\hbox{arccoth}x-1\right)]\nonumber\\&&
=\frac{i\alpha^2 q_0}{2\pi^2}\left(10-\pi^2\right)\ln\frac{v_F\Lambda}{q_0}+\hbox{finite}\nonumber\\&&
=\frac{iq_0}{2\pi^2}\left(10-\pi^2\right)\alpha^2\ln\frac{\Lambda}{|\vec{q}|}+\hbox{finite}.
\end{eqnarray}

{\it Extracting the divergence in the spatial part.}
To extract the divergent logarithm term in the spatial part of the two{-}loop self{-}energy [Eq.~(\ref{spatial2loop})], we must examine the behavior of the quadruple integral in the region $|\vec{k}|,|\vec{p}|\gg|\vec{q}|$. It helps to first redefine $\vec{k}\to\vec{k}{+}\vec{q}$:
\begin{eqnarray}
&&\!\!\!\!\!{1\over4}\hbox{Tr}[\gamma^i\Sigma_{2a}(q)]{=}\frac{{-}4\pi^2 i\alpha^2 v_F^3 q_i}{|\vec{q}|^2}{\int}\frac{d^3k}{(2\pi)^3}{\int}\frac{d^3p}{(2\pi)^3}\frac{1}{|\vec{k}|^2}\frac{1}{|\vec{q}{-}\vec{p}|^2} \nonumber\\&&\!\!\!\!\!\times\frac{|\vec{k}{+}\vec{q}|{+}|\vec{p}|{+}|\vec{k}{+}\vec{p}|}{q_0^2{+}v_F^2(|\vec{k}{+}\vec{q}|{+}|\vec{p}|{+}|\vec{k}{+}\vec{p}|)^2}\bigg\{{-}\frac{(\vec{k}{+}\vec{q}){\cdot}\vec{q}}{|\vec{k}{+}\vec{q}|} {-}\frac{\vec{p}{\cdot}\vec{q}}{|\vec{p}|}{+}\frac{(\vec{k}{+}\vec{p}){\cdot}\vec{q}}{|\vec{k}{+}\vec{p}|}
\nonumber\\&&\!\!\!\!\!{+}\frac{|\vec{p}|^2(\vec{k}{+}\vec{q}){\cdot}\vec{q}{+}|\vec{k}{+}\vec{q}|^2\vec{p}{\cdot}\vec{q}{+}|\vec{q}|^2\vec{k}{\cdot}\vec{p}{-}2(\vec{k}{\cdot}\vec{q})(\vec{p}{\cdot}\vec{q}) {-}|\vec{q}|^2\vec{p}{\cdot}\vec{q}}{|\vec{k}{+}\vec{q}||\vec{p}||\vec{k}{+}\vec{p}|}\bigg\}.\nonumber\\&&
\end{eqnarray}
We then make the following expansions in the large momentum limit:
\begin{eqnarray}
&&|\vec{k}{+}\vec{q}|{\approx}|\vec{k}|{+}\frac{\vec{k}{\cdot}\vec{q}}{|\vec{k}|},\; \frac{1}{|\vec{k}{+}\vec{q}|}{\approx}\frac{1}{|\vec{k}|}{-}\frac{\vec{k}{\cdot}\vec{q}}{|\vec{k}|^3},\; \frac{1}{|\vec{p}{-}\vec{q}|}{\approx}\frac{1}{|\vec{p}|}{+}\frac{\vec{p}{\cdot}\vec{q}}{|\vec{p}|^3},\nonumber\\
&&\frac{|\vec{k}{+}\vec{q}|{+}|\vec{p}|{+}|\vec{k}{+}\vec{p}|}{q_0^2{+}v_F^2(|\vec{k}{+}\vec{q}|{+}|\vec{p}|{+}|\vec{k}{+}\vec{p}|)^2}{\approx} \frac{1}{q_0^2{+}v_F^2(|\vec{k}|{+}|\vec{p}|{+}|\vec{k}{+}\vec{p}|)^2}\nonumber\\&&\times\bigg[ |\vec{k}|{+}|\vec{p}|{+}|\vec{k}{+}\vec{p}|{+} \frac{q_0^2{-}v_F^2(|\vec{k}|{+}|\vec{p}|{+}|\vec{k}{+}\vec{p}|)^2}{q_0^2{+}v_F^2(|\vec{k}|{+}|\vec{p}|{+}|\vec{k}{+}\vec{p}|)^2}\frac{\vec{k}{\cdot}\vec{q}}{|\vec{k}|}\bigg] \nonumber\\&&\qquad\qquad\equiv Q(\vec{k},\vec{p}){+}R(\vec{k},\vec{p})\frac{\vec{k}{\cdot}\vec{q}}{|\vec{k}|},
\end{eqnarray}
where the last equality defines the functions $Q(\vec{k},\vec{p})$ and $R(\vec{k},\vec{p})$. We then have
\begin{eqnarray}
&&{1\over4}\hbox{Tr}[\gamma^i\Sigma_{2a}(q)]{=}\frac{{-}4\pi^2 i\alpha^2 v_F^3 q_i}{|\vec{q}|^2}{\int}\frac{d^3k}{(2\pi)^3}{\int}\frac{d^3p}{(2\pi)^3}\frac{1}{|\vec{k}|^2|\vec{p}|^2} \nonumber\\&&\times\left(1{+}2\frac{\vec{p}{\cdot}\vec{q}}{|\vec{p}|^2}\right)\left[Q(\vec{k},\vec{p}){+}R(\vec{k},\vec{p})\frac{\vec{k}{\cdot}\vec{q}}{|\vec{k}|}\right]\bigg\{ {-}\frac{\vec{p}{\cdot}\vec{q}}{|\vec{p}|} {+}\frac{(\vec{k}{+}\vec{p}){\cdot}\vec{q}}{|\vec{k}{+}\vec{p}|}
\nonumber\\&&{-}\frac{(\vec{k}{+}\vec{q}){\cdot}\vec{q}}{|\vec{k}|}\left(1{-}\frac{\vec{k}{\cdot}\vec{q}}{|\vec{k}|^2}\right)
{+}\frac{|\vec{p}|(\vec{k}{+}\vec{q}){\cdot}\vec{q}}{|\vec{k}||\vec{k}{+}\vec{p}|}\left(1{-}\frac{\vec{k}{\cdot}\vec{q}}{|\vec{k}|^2}\right)
\nonumber\\&&{+}\frac{|\vec{k}|\vec{p}{\cdot}\vec{q}}{|\vec{p}||\vec{k}{+}\vec{p}|}\left(1{+}\frac{\vec{k}{\cdot}\vec{q}}{|\vec{k}|^2}\right)
{+}\frac{|\vec{q}|^2\vec{k}{\cdot}\vec{p}}{|\vec{k}||\vec{p}||\vec{k}{+}\vec{p}|}\left(1{-}\frac{\vec{k}{\cdot}\vec{q}}{|\vec{k}|^2}\right) \nonumber\\&&{-}2\frac{(\vec{k}{\cdot}\vec{q})(\vec{p}{\cdot}\vec{q})}{|\vec{k}||\vec{p}||\vec{k}{+}\vec{p}|}\left(1{-}\frac{\vec{k}{\cdot}\vec{q}}{|\vec{k}|^2}\right) {-}\frac{|\vec{q}|^2\vec{p}{\cdot}\vec{q}}{|\vec{k}||\vec{p}||\vec{k}{+}\vec{p}|}\left(1{-}\frac{\vec{k}{\cdot}\vec{q}}{|\vec{k}|^2}\right)\bigg\}.\label{spatial2loop2}\nonumber\\&&
\end{eqnarray}
The terms that scale as the inverse sixth power in the momenta $\vec{k},\vec{p}$ give rise to a logarithmic divergence. These are the terms we are interested in. There are also terms in Eq.~(\ref{spatial2loop2}) which scale as the inverse fifth power and so would seem to produce a linear divergence. However, these terms vanish identically as can be seen by performing a coordinate transformation $\vec{k}\to{-}\vec{k}$, $\vec{p}\to{-}\vec{p}$. We isolate each of the terms which contribute to the logarithmic divergence in the following series of integrals:
\begin{equation}
\Xi_1{=}{\int}\frac{d^3k}{(2\pi)^3}{\int}\frac{d^3p}{(2\pi)^3}\frac{Q(\vec{k},\vec{p})}{|\vec{k}|^2|\vec{p}|^2}\left(\frac{(\vec{k}{\cdot}\vec{q})^2}{|\vec{k}|^3}{-}\frac{|\vec{q}|^2}{|\vec{k}|}\right),
\end{equation}
\begin{equation}
\Xi_2{=}{-}{\int}\frac{d^3k}{(2\pi)^3}{\int}\frac{d^3p}{(2\pi)^3}\frac{Q(\vec{k},\vec{p})}{|\vec{k}|^2|\vec{p}|^2}\frac{|\vec{p}|}{|\vec{k}{+}\vec{p}|}\left(\frac{(\vec{k}{\cdot}\vec{q})^2}{|\vec{k}|^3}{-}\frac{|\vec{q}|^2}{|\vec{k}|}\right),
\end{equation}
\begin{equation}
\Xi_3{=}{\int}\frac{d^3k}{(2\pi)^3}{\int}\frac{d^3p}{(2\pi)^3}\frac{Q(\vec{k},\vec{p})}{|\vec{k}|^2|\vec{p}|^2}\frac{(\vec{k}{\cdot}\vec{q})(\vec{p}{\cdot}\vec{q})}{|\vec{k}||\vec{p}||\vec{k}{+}\vec{p}|},
\end{equation}
\begin{equation}
\Xi_4{=}{\int}\frac{d^3k}{(2\pi)^3}{\int}\frac{d^3p}{(2\pi)^3}\frac{Q(\vec{k},\vec{p})}{|\vec{k}|^2|\vec{p}|^2}\frac{|\vec{q}|^2(\vec{k}{\cdot}\vec{p})}{|\vec{k}||\vec{p}||\vec{k}{+}\vec{p}|},
\end{equation}
\begin{equation}
\Xi_5{=}{-}2{\int}\frac{d^3k}{(2\pi)^3}{\int}\frac{d^3p}{(2\pi)^3}\frac{Q(\vec{k},\vec{p})}{|\vec{k}|^2|\vec{p}|^2}\frac{(\vec{k}{\cdot}\vec{q})(\vec{p}{\cdot}\vec{q})}{|\vec{k}||\vec{p}||\vec{k}{+}\vec{p}|}{=}{-}2\Xi_3,
\end{equation}
\begin{equation}
\Xi_6{=}{-}2{\int}\frac{d^3k}{(2\pi)^3}{\int}\frac{d^3p}{(2\pi)^3}\frac{Q(\vec{k},\vec{p})}{|\vec{k}|^2|\vec{p}|^2}\frac{(\vec{k}{\cdot}\vec{q})(\vec{p}{\cdot}\vec{q})}{|\vec{k}||\vec{p}|^2},
\end{equation}
\begin{equation}
\Xi_7{=}{-}2{\int}\frac{d^3k}{(2\pi)^3}{\int}\frac{d^3p}{(2\pi)^3}\frac{Q(\vec{k},\vec{p})}{|\vec{k}|^2|\vec{p}|^2}\frac{(\vec{p}{\cdot}\vec{q})^2}{|\vec{p}|^3},
\end{equation}
\begin{equation}
\Xi_8{=}2{\int}\frac{d^3k}{(2\pi)^3}{\int}\frac{d^3p}{(2\pi)^3}\frac{Q(\vec{k},\vec{p})}{|\vec{k}|^2|\vec{p}|^2}\frac{(\vec{p}{\cdot}\vec{q})(\vec{k}{+}\vec{p}){\cdot}\vec{q}}{|\vec{p}|^2|\vec{k}{+}\vec{p}|},
\end{equation}
\begin{equation}
\Xi_9{=}2{\int}\frac{d^3k}{(2\pi)^3}{\int}\frac{d^3p}{(2\pi)^3}\frac{Q(\vec{k},\vec{p})}{|\vec{k}|^2|\vec{p}|^2}\frac{(\vec{k}{\cdot}\vec{q})(\vec{p}{\cdot}\vec{q})}{|\vec{k}||\vec{p}||\vec{k}{+}\vec{p}|}{=}2\Xi_3,
\end{equation}
\begin{equation}
\Xi_{10}{=}2{\int}\frac{d^3k}{(2\pi)^3}{\int}\frac{d^3p}{(2\pi)^3}\frac{Q(\vec{k},\vec{p})}{|\vec{k}|^2|\vec{p}|^2}\frac{|\vec{k}|(\vec{p}{\cdot}\vec{q})^2}{|\vec{p}|^3|\vec{k}{+}\vec{p}|},
\end{equation}
\begin{equation}
\Xi_{11}{=}{-}{\int}\frac{d^3k}{(2\pi)^3}{\int}\frac{d^3p}{(2\pi)^3}\frac{R(\vec{k},\vec{p})}{|\vec{k}|^2|\vec{p}|^2}\frac{(\vec{k}{\cdot}\vec{q})^2}{|\vec{k}|^2},
\end{equation}
\begin{equation}
\Xi_{12}{=}{-}{\int}\frac{d^3k}{(2\pi)^3}{\int}\frac{d^3p}{(2\pi)^3}\frac{R(\vec{k},\vec{p})}{|\vec{k}|^2|\vec{p}|^2}\frac{(\vec{k}{\cdot}\vec{q})(\vec{p}{\cdot}\vec{q})}{|\vec{k}||\vec{p}|},
\end{equation}
\begin{equation}
\Xi_{13}{=}{\int}\frac{d^3k}{(2\pi)^3}{\int}\frac{d^3p}{(2\pi)^3}\frac{R(\vec{k},\vec{p})}{|\vec{k}|^2|\vec{p}|^2}\frac{(\vec{k}{\cdot}\vec{q})(\vec{k}{+}\vec{p}){\cdot}\vec{q}}{|\vec{k}||\vec{k}{+}\vec{p}|},
\end{equation}
\begin{equation}
\Xi_{14}{=}{\int}\frac{d^3k}{(2\pi)^3}{\int}\frac{d^3p}{(2\pi)^3}\frac{R(\vec{k},\vec{p})}{|\vec{k}|^2|\vec{p}|^2}\frac{|\vec{p}|(\vec{k}{\cdot}\vec{q})^2}{|\vec{k}|^2|\vec{k}{+}\vec{p}|},
\end{equation}
\begin{equation}
\Xi_{15}{=}{\int}\frac{d^3k}{(2\pi)^3}{\int}\frac{d^3p}{(2\pi)^3}\frac{R(\vec{k},\vec{p})}{|\vec{k}|^2|\vec{p}|^2}\frac{(\vec{k}{\cdot}\vec{q})(\vec{p}{\cdot}\vec{q})}{|\vec{p}||\vec{k}{+}\vec{p}|},
\end{equation}
To perform these integrals, we again make use of prolate spheroidal coordinates for $\vec{p}$ and ordinary spherical coordinates for $\vec{k}$, in which case we have
\begin{eqnarray}
Q(\vec{k},\vec{p})&{\to}&\frac{|\vec{k}|(\cosh\mu{+}1)}{q_0^2{+}v_F^2|\vec{k}|^2(\cosh\mu{+}1)^2}\equiv \widetilde Q(\mu,|\vec{k}|),\nonumber\\
R(\vec{k},\vec{p})&{\to}&\frac{q_0^2{-}v_F^2|\vec{k}|^2(\cosh\mu{+}1)^2}{\left[q_0^2{+}v_F^2|\vec{k}|^2(\cosh\mu{+}1)^2\right]^2}\equiv\widetilde R(\mu,|\vec{k}|).\nonumber\\&&
\end{eqnarray}
In addition to the relations given in Eq.~(\ref{prolatespheroid}), we also make use of the following:
\begin{eqnarray}
\vec{k}\cdot\vec{q}&=&|\vec{k}||\vec{q}|\cos\theta_k, \cr
\vec{p}\cdot\vec{q}&=&\frac{|\vec{k}|}{2}[q_x\sinh\mu\sin\nu\cos\theta+q_y\sinh\mu\sin\nu\sin\theta \cr
&+&q_z(\cosh\mu\cos\nu-1)].
\end{eqnarray}
Here, $\theta_k$ is the polar coordinate associated with $\vec{k}$. After we plug these expressions into the $\Xi_i$, we first perform the integrations over $\theta$ as well as over $\phi_k$, the azimuthal coordinate associated with $\vec{k}$. All the remaining four-dimensional integrals are then functions of $q_x$ and $q_y$ only in the combination $q_x^2+q_y^2$, which we may rewrite as $|\vec{q}|^2-q_z^2$. We then set $q_z=|\vec{q}|\cos\theta_k$ and perform all the $\theta_k$ integrations, which are easily done. Replacing $\cosh\mu\to x$, $\cos\nu\to y$, we then find the following results:
\begin{eqnarray}
\sum_{i=1}^{10}\Xi_i&=&\frac{|\vec{q}|^2}{12\pi^4}\int_1^\infty dx\int_{-1}^1dy\int_0^\Lambda d|\vec{k}| \cr
&\times&\frac{|\vec{k}|(1+x)(x+y-2)(1-y^2)}{[q_0^2+v_F^2|\vec{k}|^2(1+x)^2](x-y)^3},\nonumber\\
\sum_{i=11}^{15}\Xi_i&=&\frac{|\vec{q}|^2}{24\pi^4}\int_1^\infty dx\int_{-1}^1dy\int_0^\Lambda d|\vec{k}| \cr
&\times&\frac{|\vec{k}|[q_0^2-v_F^2|\vec{k}|^2(1+x)^2](x-1)(1-y^2)}{[q_0^2+v_F^2|\vec{k}|^2(1+x)^2]^2(x-y)^2}. \nonumber \\
\end{eqnarray}
Each 3D integral is logarithmically divergent, and the coefficient of the divergence can be computed straightforwardly by first performing the $|\vec{k}|$ integrals exactly and keeping only the term proportional to $\ln(v_F\Lambda/q_0)$. The remaining integrals on $x$ and $y$ can also be done exactly. The results are
\begin{eqnarray}
\sum_{i=1}^{10}\Xi_i&=&-\frac{(10-\pi^2)|\vec{q}|^2}{12\pi^4v_F^2}\ln\left (\frac{\Lambda}{|\vec{q}|}\right )+\hbox{finite},\nonumber\\
\sum_{i=11}^{15}\Xi_i&=&-\frac{(10-\pi^2)|\vec{q}|^2}{24\pi^4v_F^2}\ln\left (\frac{\Lambda}{|\vec{q}|}\right )+\hbox{finite},
\end{eqnarray}
and thus
\begin{equation}
\sum_{i=1}^{15}\Xi_i=-\frac{(10-\pi^2)|\vec{q}|^2}{8\pi^4v_F^2}\ln\left (\frac{\Lambda}{|\vec{q}|}\right )+\hbox{finite},
\end{equation}
\begin{eqnarray}
{1\over4}\hbox{Tr}[\gamma^i\Sigma_{2a}(q)]&=&\frac{4\pi^2 i\alpha^2 v_F^3 q_i}{|\vec{q}|^2}\frac{(10-\pi^2)|\vec{q}|^2}{8\pi^4v_F^2}\ln\left (\frac{\Lambda}{|\vec{q}|}\right ) \cr
&=&\frac{i(10-\pi^2)}{2\pi^2}\alpha^2v_Fq_i\ln\left (\frac{\Lambda}{|\vec{q}|}\right ).
\end{eqnarray}
Combining this with what we obtained for the temporal part, we then have for the full diagram
\begin{equation}
\Sigma_{2a}(q)=i\frac{10-\pi^2}{2\pi^2}\left(q_0\gamma^0+v_F\vec{q}\cdot\vec{\gamma}\right)\alpha^2\ln\left (\frac{\Lambda}{|\vec{q}|}\right ).
\end{equation}

\subsection{Two{-}loop bubble correction to self{-}energy}

\begin{figure}
\includegraphics[width=0.5\columnwidth]{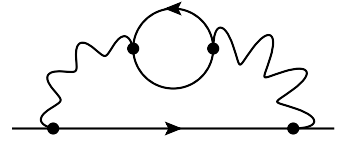}
\caption[Bubble]{Two-loop bubble correction to the electron self-energy, $\Sigma_{2c}(q)$.}
\label{fig:selfenergy2c}
\end{figure}

The two-loop bubble correction to the self-energy is shown in Fig.\ \ref{fig:selfenergy2c}. This diagram evaluates to
\begin{equation}
\Sigma_{2c}(q)={-}{\int}\frac{d^4k}{(2\pi)^4}\gamma^0G_0(q{-}k)\gamma^0\Pi_B(k)V_k^2,
\end{equation}
where $\Pi_B(q)$ is the result for the one{-}loop bubble diagram given in Eq.\ \eqref{Eq:Polarization_FE_Div}. To compute this integral, we will devise a streamlined approach to computing self-energy diagrams that contain vacuum polarization subdiagrams. In other words, consider the more general situation where $\Pi_B(k)$ is replaced by an arbitrary polarization function $\Pi(k)$. Dimensional analysis tells us that $\Pi(k)$ must have the following general form:
\begin{equation}
\Pi(k)=\frac{|\vec{k}|^2}{v_F}\left[F(z)\ln\left (\frac{\Lambda}{|\vec{k}|}\right )+G(z)\right],
\end{equation}
where $z=k_0/(v_F|\vec{k}|)$ as usual, and $F(z)$ and $G(z)$ are arbitrary functions. Using this form, we can rewrite our expression for the corresponding self-energy $\Sigma(q)$ as
\begin{eqnarray}
\Sigma(q)&=&-8\pi i\alpha^2 v_F^2\int_{-\infty}^\infty dz\int\frac{d^3k}{(2\pi)^3}\frac{1}{|\vec{k}|}\gamma^0\frac{\slashed q-\slashed k}{(q-k)^2}\gamma^0 \cr
&\times&\left[F(z)\ln\left(\frac{\Lambda}{|\vec{k}|}\right)+G(z)\right].
\end{eqnarray}
We are only interested in the logarithmically divergent terms, so we expand the fermion propagator in the limit of large $\vec{k}$:
\begin{eqnarray}
\gamma^0\frac{\slashed q-\slashed k}{(q-k)^2}\gamma^0&\to&\frac{1}{v_F^2|\vec{k}|^2}\left[\frac{1-z^2}{(1+z^2)^2}q_0\gamma^0\right. \cr
&-&\left.\frac{z^2+1/3}{(1+z^2)^2}v_F\vec{q}\cdot\vec{\gamma}\right].
\end{eqnarray}
We then have
\begin{eqnarray}
\Sigma(q)&=&-\frac{4i\alpha^2}{\pi}\int_{-\infty}^\infty dz\left[\frac{1-z^2}{(1+z^2)^2}q_0\gamma^0\right. \cr
&-&\left.\frac{z^2+1/3}{(1+z^2)^2}v_F\vec{q}\cdot\vec{\gamma}\right]\int_{|\vec{q}|}^\Lambda d|\vec{k}| \frac{1}{|\vec{k}|}\left[F(z)\ln\left (\frac{\Lambda}{|\vec{k}|}\right )\right. \cr
&+&\left.G(z)\right].
\end{eqnarray}
The integral over $|\vec{k}|$ is easily performed, allowing us to express the self-energy as
\begin{eqnarray}
\Sigma(q)&=&-\frac{4i\alpha^2}{\pi}\left\{\left[A_2q_0\gamma^0-B_2v_F\vec{q}\cdot\vec{\gamma}\right]\ln^2\left (\frac{\Lambda}{|\vec{q}|}\right )\right. \cr
&+&\left.\left[A_1q_0\gamma^0-B_1v_F\vec{q}\cdot\vec{\gamma}\right]\ln\left (\frac{\Lambda}{|\vec{q}|}\right )\right\},
\end{eqnarray}
where
\begin{eqnarray}
A_2&=&\frac{1}{2}\int_{-\infty}^\infty dzF(z)\frac{1-z^2}{(1+z^2)^2}, \cr
B_2&=&\frac{1}{2}\int_{-\infty}^\infty dzF(z)\frac{z^2+1/3}{(1+z^2)^2},\nonumber\\
A_1&=&\int_{-\infty}^\infty dzG(z)\frac{1-z^2}{(1+z^2)^2}, \cr
B_1&=&\int_{-\infty}^\infty dzG(z)\frac{z^2+1/3}{(1+z^2)^2}.
\end{eqnarray}

We are now ready to evaluate $\Sigma_{2c}(q)$. From the expression for $\Pi_B(q)$ given in Eq.\ \eqref{Eq:Polarization_FE_Div}, we have
\begin{equation}
F(z)=-\frac{N}{6\pi^2}, \qquad G(z)=\frac{N}{12\pi^2}\ln\left (\frac{1+z^2}{4}\right ).
\end{equation}
Plugging these functions into the above expressions for $A_1$, etc. we find
\begin{equation}
A_2=0, \ B_2=-\frac{N}{18\pi}, \ A_1=-\frac{N}{12\pi}, \ B_1=\frac{N}{36\pi}.
\end{equation}
These results then lead to the final expression for the self-energy correction:
\begin{eqnarray}
\Sigma_{2c}(q)&=&-\frac{2N}{9\pi^2}i\alpha^2v_F\vec{q}\cdot\vec{\gamma}\ln^2\left (\frac{\Lambda}{|\vec{q}|}\right ) \cr
&+&\frac{iN}{9\pi^2}\left[3q_0\gamma^0+v_F\vec{q}\cdot\vec{\gamma}\right]\alpha^2\ln\left (\frac{\Lambda}{|\vec{q}|}\right ).
\end{eqnarray}

Combining $\Sigma_{2a}$ and $\Sigma_{2c}$, we then obtain the full second-order electron self-energy:
\begin{eqnarray}
\Sigma_2(q)&=&-\frac{2N}{9\pi^2}i\alpha^2v_F\vec{q}\cdot\vec{\gamma}\ln^2\left (\frac{\Lambda}{|\vec{q}|}\right ) \cr
&+&i\left[\left(\frac{15+N}{3\pi^2}-\frac{1}{2}\right)q_0\gamma^0+\left(\frac{45+N}{9\pi^2}-\frac{1}{2}\right)v_F\vec{q}\cdot\vec{\gamma}\right] \cr
&\times&\alpha^2\ln\left (\frac{\Lambda}{|\vec{q}|}\right ).
\end{eqnarray}

\section{Renormalization} \label{Sec:Renormalization}
We will now demonstrate the renormalizability of our theory and derive the renormalization
group (RG) equation for the interaction strength, Fermi velocity, and quasiparticle residue to second order in $\alpha$.  We will also derive identities that
serve as valuable checks for our previous results.

All of the results derived in the previous section depend on an ultraviolet momentum cutoff scale $\Lambda$.  In a complete theory (which is unknown) that takes into account the full band structure and interaction without any approximation, the cutoff dependence would be absent since the energy dispersion will not be precisely linear all the way to large momenta in such an exact theory, and therefore no explicit cutoff will be necessary. The cutoff indicates the energy/momentum regime where our effective low-energy theory defined by our action in Eq.\ \eqref{Eq:Action} that describes the excitations at the Dirac cones is no longer valid. This high-momentum regime can be, for example, identified with the Brillouin zone boundary, the position of a van Hove singularity in the band structure, or simply the inverse lattice constant.  We emphasize, however, that even such an unknown exact band theory, if it includes quantum interaction effects correctly, would manifest the ultraviolet renormalization effects in the sense that as the theory is used to calculate lower and lower momentum or energy properties of the system, it would manifest a logarithmic divergence as the Dirac point is approached, but the effective cutoff momentum entering such an exact theory would itself be also scale-dependent and not like the fixed cutoff $\Lambda$ appearing in our formalism, which is based on a model linear Dirac band dispersion. In our effective theory framework, the cutoff separates the space of low-energy modes with known dispersion and interaction from the {\it a priori} unknown high-energy modes, and when restricting internal loop momenta to $k < \Lambda$, we essentially discard the latter. In a complete renormalized theory, these contributions must be taken into account separately. For observables measured at small scales $p \ll \Lambda$, the high-energy contributions arise from modes fluctuating over time and distance scales $\sim 1/\Lambda$, i.e., they appear as essentially local corrections. A simple power-counting argument shows that only the self-energy, the vertex, and the polarization have a strong dependence on the high-energy cutoff, and we account for the missing high-energy contributions by adding local terms to the action. These terms are known as counterterms. The coefficients of the counterterms are {\it a priori} unknown but should be chosen in such a way that our computation gives the experimentally observable renormalized parameters, which only depend on the cutoff in subleading order ${\cal O}(p/\Lambda)$. As we discuss in the following, this procedure allows us to remove any strong cutoff dependence from the theory, yet introduces a dependence of the renormalized parameters on a low-energy renormalization scale. The divergence structure of the theory also determines the running of the renormalized parameters.

This section is structured as follows: First, in Sec.~\ref{sec:renpt}, we introduce the renormalized Hamiltonian and discuss the relation between bare and renormalized parameters. This is followed by a detailed and abstract discussion of counter-term renormalization for 3D Dirac materials in Sec.~\ref{sec:identities}. In particular, we derive various identities that link the coefficients of logarithmic divergences in self-energy, vertex, and polarization to the beta functions for coupling, Fermi velocity, and field strength. In Sec.~\ref{sec:rganalysis}, we use the results of the first part of this paper to derive the beta functions explicitly. The two-loop RG equations have some additional corrections stemming from the insertion of lower-order counterterms in the one-loop diagrams, and we include a derivation of these diagrams in this section. Based on the results for the beta function, we comment on the critical interaction strength $\alpha_c$ at which the RG flow has a fixed point, possibly, but not necessarily, indicating a breakdown of the perturbative expansion.

\subsection{Renormalized perturbation theory}\label{sec:renpt}

We first formulate a renormalized theory, in which we introduce counterterms that remove logarithmic
divergences as a function of the momentum cutoff $\Lambda$ from all results obtained from the
theory.  To do so, we relabel all fields and constants in the original theory with a superscript
$B$ to denote their ``bare'' values, and then introduce renormalized fields and constants.  In other
words, we rewrite the action as
\begin{eqnarray}
S&=&-\sum_{a=1}^{N}\int dt\,d^3\vec{R}\,(\bar{\psi}_a^B\gamma^0\partial_0\psi_a^B+v_F\bar{\psi}_a^B\gamma^i\partial_i\psi_a^B \cr
&+&\varphi^B\bar{\psi}_a^B\gamma^0\psi_a^B)+\frac{1}{2(g^B)^2}\int dt\,d^3\vec{R}\,(\partial_i \varphi^B)^2. \label{Eq:BareAction}
\end{eqnarray}
We first rescale the electron field, $\psi=Z_\psi^{-1/2}\psi^B$, and the Coulomb field, $\varphi = Z_\varphi^{-1/2} \varphi^B$.
If we perform this field rescaling and rearrange terms, the action becomes
\begin{equation}
S_B=S_0+S_{\rm ct},
\end{equation}
where
\begin{eqnarray}
S_0 &=& - \sum_{a=1}^N \int dt\,d^3\vec{R}\,(\bar{\psi}_a \gamma^0 \partial_t \psi_a + v_F \bar{\psi}_a \gamma^i \partial_i \psi_a \cr
 &+& \varphi\bar{\psi}_a \gamma^0 \psi_a) + \frac{1}{2g^2} \int dt\,d^3\vec{R}\,(\partial_i\varphi)^2, \\
S_{\rm ct} &=& \sum_{a=1}^N \int dt\,d^3\vec{R}\,(\delta_0 \bar{\psi}_a \gamma^0 \partial_t \psi_a + \delta_1 v_F  \bar{\psi}_a \gamma^i \partial_i \psi_a \cr
&+& \delta_v\varphi \bar{\psi}_a \gamma^0 \psi_a) - \frac{\delta_p}{2g^2} \int dt\,d^3\vec{R}\,(\partial_i\varphi)^2.
\end{eqnarray}
The first part of the action, $S_0$, is just the ``bare'' action, but with a {\it renormalized} Fermi velocity
$v_F$ and charge $g$ that are defined at a momentum scale $\mu$, which we will call the renormalization scale.
The second part, $S_{\rm ct}$, gives us the counterterms mentioned above. As discussed, they have a precise meaning in that they account for the high-energy physics that is not captured by the low-energy theory~\eqref{Eq:Action}. The $\delta$ coefficients in this
part of the action are given by
\begin{align}
\delta_0 &= 1 - Z_\psi, \label{Eq:Delta0} \\
\delta_1 &= 1 - \frac{v_F^B}{v_F} Z_\psi, \label{Eq:Delta1} \\
\delta_v &= 1 - Z_\psi Z_\varphi^{1/2}, \label{Eq:Deltav} \\
\delta_p &= 1 - \frac{g^2}{g_B^2} Z_\varphi. \label{Eq:Deltap}
\end{align}
The Feynman rules associated with $S_0$ are the same as those given earlier for the ``bare'' action.
Those for the counterterms are as follows.
\begin{alignat}{2}
&\raisebox{-0.05cm}{\scalebox{0.6}{\epsfig{file=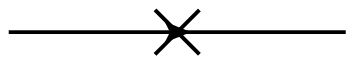}}} &\qquad &  - i (\delta_0 k_0 \gamma^0 + v_F \delta_1 \vec{k} \cdot \vec{\gamma}) \label{Eq:SE1CT} \\
&\raisebox{-0.1cm}{\scalebox{0.6}{\epsfig{file=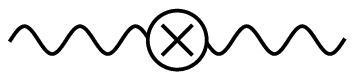}}} &\qquad & |\vec{k}|^2 \frac{\delta_p}{g^2} \label{Eq:P1CT} \\
&\raisebox{-0.6cm}{\scalebox{0.6}{\epsfig{file=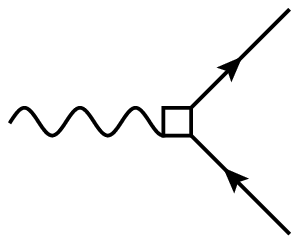}}} &\qquad & - i \gamma^0 \delta_v \label{Eq:V1CT}
\end{alignat}

\subsection{Identities}\label{sec:identities}
As we have seen, when computing correlation functions defined by the theory given by Eq.\ \eqref{Eq:BareAction},
\begin{align}
&G^B_{n,m}[\{x_i\}, \{y_j\}; \alpha^B, v_F^B, \Lambda] \cr
&= \langle {\cal T} \psi_a^B(x_1) \ldots \psi_a^B(x_n) \varphi^B(y_1) \ldots \varphi^B(y_m)\rangle , \label{Eq:Correlator2}
\end{align}
we will encounter divergences as a function of the momentum cutoff $\Lambda$ in our theory. The Green's functions
acquire an explicit cutoff dependence as indicated on the left-hand side.  Our goal in defining the renormalized
theory above is to eliminate this cutoff dependence, trading it for a dependence on the scale $\mu$.  The renormalized
correlators determined from this theory,
\begin{align}
&G_{n,m}(x_1, \ldots, x_n; y_1, \ldots, y_m; \alpha, v_F, \mu) \cr
&= Z_\psi^{-n/2}G^B_{n,m}[x_1, \ldots, x_n; y_1, \ldots, y_m; \alpha^B, v_F^B, \Lambda], \nonumber \\ \label{Eq:Relbareren}
\end{align}
where
\begin{align}
&G_{n,m}(\{x_i\}, \{y_i\};\alpha, v_F, \mu) \cr
&= \langle {\cal T} \psi_a(x_1) \ldots \psi_a(x_n) \varphi(y_1) \ldots \varphi(y_m)\rangle, \label{Eq:Correlator1}
\end{align}
thus do not depend on the cutoff.  The renormalized Green's functions are instead functions of the renormalization scale
$\mu$, the renormalized Fermi velocity $v_F$, and the renormalized charge $g^2$.  As we will show in the following, these renormalized
quantities are running quantities that depend on the scale $\mu$, i.e., $g^2=g^2(\mu)$ and $v_F=v_F(\mu)$, with an
evolution that is governed by renormalization group equations. The cutoff dependence is removed by allowing the bare
parameters (the bare Fermi velocity $v_F^B$ and the bare residue $Z_\psi$) to depend on the cutoff so as to cancel
the divergences order by order in perturbation theory:
\begin{alignat}{2}
Z_\psi &= 1 - F_\psi(\alpha, \Lambda, \mu), \label{Eq:Zexp} \\
Z_\varphi &= 1- F_\varphi(\alpha, \Lambda, \mu), \label{Eq:E2exp} \\
v_F^B &= v_F \left [1 - F_{v_F}(\alpha, \Lambda, \mu)\right ], \label{Eq:Vexp} \\
(g^B)^2 &= g^2 \left [1 - F_{g}(\alpha, \Lambda, \mu)\right ], \label{Eq:Eexp}
\end{alignat}
where
\begin{alignat}{2}
F_\psi(\alpha, \Lambda, \mu) &= \sum_{n\geq 1} z_{\psi,n}(\alpha) \ln^n\left (\frac{\Lambda}{\mu}\right ), \\
F_\varphi(\alpha, \Lambda, \mu) &= \sum_{n\geq 1} z_{\varphi,n}(\alpha) \ln^n\left (\frac{\Lambda}{\mu}\right ), \\
F_{v_F}(\alpha, \Lambda, \mu) &= \sum_{n\geq 1} v_n(\alpha) \ln^n\left (\frac{\Lambda}{\mu}\right ), \\
F_{g}(\alpha, \Lambda, \mu) &= \sum_{n\geq 1} g_n(\alpha) \ln^n\left (\frac{\Lambda}{\mu}\right ).
\end{alignat}
The dependence on powers of $\ln\left (\frac{\Lambda}{\mu}\right )$ of these functions is required to cancel the divergences
in $\Lambda$ coming from the bare action.  The bare fine-structure constant $\alpha_B$ can similarly be written as
\begin{alignat}{2}
\alpha_B &= \alpha[1 - F_\alpha(\alpha, \Lambda, \mu)],\label{Eq:alphaBfromalpha}
\end{alignat}
where
\begin{alignat}{2}
F_\alpha(\alpha, \Lambda, \mu) = \sum_{n\geq 1} f_n(\alpha) \ln^n\left (\frac{\Lambda}{\mu}\right ).
\end{alignat}
The functions $f_n$ can be expressed as a linear combination of the various $v_n$'s and $e_n$'s, i.e.,
\begin{align}
f_1(\alpha) &= g_1(\alpha) - v_1(\alpha) , \\
f_2(\alpha) &= g_2(\alpha) - v_2(\alpha) + g_1(\alpha) v_1(\alpha) - v_1(\alpha)^2 , \\
f_3(\alpha) &= g_3(\alpha) - v_3(\alpha) + g_2(\alpha) v_1(\alpha) + g_1(\alpha) v_1(\alpha)^2 \cr
&+ g_1(\alpha) v_2(\alpha) - 2 v_1(\alpha) v_2(\alpha) - v_1(\alpha)^3 ,
\end{align}
and so on.

Let us now consider the bare $n$-particle correlator defined in Eq.\ \eqref{Eq:Correlator2}.  Note that \textit{by the definition
of} $S_B$ in Eq.\ \eqref{Eq:BareAction} it does not depend on $\mu$, i.e.,
\begin{align}
\mu \frac{d}{d\mu} G_{n,m}^B[\{x_i\}, \{y_i\}; \alpha^B, v_F^B, \Lambda] &= 0 . \label{Eq:Running}
\end{align}
We can use this fact to derive the Callan-Symanzik equation for the $n$-particle correlator, Eq.\ \eqref{Eq:Correlator1}, by
combining Eq.\ \eqref{Eq:Running} with Eq.\ \eqref{Eq:Relbareren}.  This gives
\begin{align}
\mu \frac{d}{d\mu}\left.\left [Z_\psi^{n/2}G_{n,m}(\{x_i\}, \{y_i\}; \alpha, v_F, \mu)\right ]\right |_{g^B, v_F^B} &= 0 .
\end{align}
The Callan-Symanzik equation then follows by using the chain rule:
\begin{align}
&\left (\mu \frac{\partial}{\partial \mu} + \beta_\alpha \frac{\partial}{\partial\alpha} + v_F \gamma_{v_F} \frac{\partial}{\partial v_F} + n \gamma_\psi\right ) \cr
&\times G_{n,m}(\{x_i\}, \{y_i\}; g^2, v_F, \mu) = 0,
\end{align}
where we define
\begin{align}
\beta_\alpha &= \mu \frac{d\alpha}{d\mu} \biggr|_{\alpha_B}, \\
v_F \gamma_{v_F} &= \left.\mu\frac{dv_F}{d\mu}\right|_{\alpha_B}, \\
\gamma_\psi\sqrt{Z_\psi} &= \left.\mu\frac{d}{d\mu}\sqrt{Z_\psi}\right|_{\alpha_B}, \\
\gamma_\varphi \sqrt{Z_\varphi} &= \mu \frac{d}{d\mu} \sqrt{Z_\varphi} \biggr|_{\alpha_B}.
\end{align}
Note that, \textit{by construction}, the $n$-particle Green's function $G_{n,m}(\{x_i\}; g, v_F, \mu)$ is finite, for we
determine the coefficients in Eqs.\ \eqref{Eq:Zexp}, \eqref{Eq:Vexp}, and \eqref{Eq:Eexp} precisely to remove the divergent
terms.  This implies that $\beta_g$, $\gamma_{v_F}$, and $\gamma_\psi$ must be free of any divergences in $\Lambda$.  In
addition, it turns out that $\beta_\alpha$, $\gamma_{v_F}$, $\gamma_\psi$, and $\gamma_\varphi$ \textit{are independent of $v_F$}.

Let us now relate $\beta_\alpha$, $\beta_{v_F}$, $\gamma_\psi$, and $\gamma_\varphi$ to the coefficients of the divergent pieces
in Eqs.\ \eqref{Eq:Zexp} and\ \eqref{Eq:Vexp}. First, let us consider the coupling $\alpha$.  Since the bare coupling $\alpha^B$
cannot depend on $\mu$,
\begin{equation}
\left.\mu \frac{d}{d\mu}\alpha^B \right|_{\alpha^B} = 0.
\end{equation}
If we now substitute in Eq.\ \eqref{Eq:alphaBfromalpha}, we get
\begin{equation}
\left.\mu\frac{d}{d\mu}\{\alpha[1 - F_\alpha(\alpha, \Lambda, \mu)]\}\right|_{\alpha^B}=0,
\end{equation}
or
\begin{align}
&\beta_\alpha[1-F_\alpha(\alpha,\Lambda,\mu)] \cr
&-\left.\mu \alpha\frac{d}{d\mu}[1-F_\alpha(\alpha,\Lambda,\mu)]\right|_{g^B,v_F^B}=0.
\end{align}
We can rewrite this as
\begin{align}
\beta_\alpha &= F_\alpha(\alpha,\Lambda,\mu)\beta_\alpha \cr
&-\alpha\left (\beta_\alpha\frac{\partial}{\partial\alpha}+v_F\gamma_{v_F}\frac{\partial}{\partial v_F}+\mu\frac{\partial}{\partial\mu}\right )F_\alpha(\alpha,\Lambda,\mu). \nonumber \\
\end{align}
Since $\beta_\alpha$ must be finite, it is completely determined by $f_1(\alpha)$:
\begin{equation}
\beta_\alpha=-\alpha f_1(\alpha).
\end{equation}
Higher-order terms obey the recursion relation,
\begin{align}
(n+1)f_{n+1}(\alpha)&=\beta_\alpha\left [f_n(\alpha)+\alpha\frac{\partial f_n}{\partial\alpha}\right ] \cr
&+\alpha v_F\gamma_{v_F}\frac{\partial f_n}{\partial v_F}. \label{Eq:AlphaRecRel}
\end{align}
We can derive similar relations for the Fermi velocity anomalous dimension $\gamma_{v_F}$.
Requiring that the bare Fermi velocity be constant in $\mu$ yields
\begin{align}
\gamma_{v_F} &= \gamma_{v_F} F_{v_F}(\alpha, \Lambda,\mu) + \left [\beta_\alpha\frac{\partial}{\partial\alpha}\right. \cr
 &+ \left. v_F\gamma_{v_F} \frac{\partial}{\partial v_F}+ \mu \frac{\partial}{\partial \mu}\right ] F_{v_F}(\alpha, \Lambda,\mu) ,
\end{align}
from which, by a similar method as for $g$ above, we obtain the relations,
\begin{align}
\gamma_{v_F} &= - v_1(\alpha) \label{eq:vFan}
\end{align}
and
\begin{align}
(n+1)v_{n+1}(\alpha) &= \left (\gamma_{v_F} + \beta_\alpha\frac{\partial}{\partial\alpha} + v_F\gamma_{v_F}\frac{\partial}{\partial v_F}\right ) \cr
&\times v_n(\alpha) . \label{Eq:vFRecRel}
\end{align}
We may also derive such a relation for the field anomalous dimensions $\gamma_\psi(\alpha)$ and $\gamma_\varphi(\alpha)$.
The relations for the renormalization of $\psi$ are
\begin{align}
\gamma_\psi(\alpha) &= \frac{1}{2} z_{\psi,1}(\alpha) \label{eq:Zpsian}
\end{align}
and
\begin{align}
(n+1) z_{\psi,n+1}(\alpha) &= \biggl(\beta_\alpha(\alpha) \frac{\partial}{\partial \alpha} - 2 \gamma_\psi(\alpha)\biggr) z_{\psi,n}(\alpha),
\end{align}
and those for $\varphi$ are
\begin{align}
\gamma_\varphi(\alpha) &= \frac{1}{2} z_{1,\varphi}(\alpha) \label{eq:Zphian}
\end{align}
and
\begin{align}
(n+1) z_{\varphi,n+1}(\alpha) &= \biggl(\beta_\alpha(\alpha) \frac{\partial}{\partial \alpha} - 2 \gamma_\varphi(\alpha)\biggr) z_{\varphi,n}(\alpha) . \label{eq:chargerecursion}
\end{align}

\subsection{RG analysis}\label{sec:rganalysis}

We now pursue the RG analysis for our system, deriving the counterterms and RG equations
and performing some important checks on our results.

\subsubsection{One-loop RG analysis}

The divergent part of the first-order self-energy [Eq.\ \eqref{Eq:ElectronSE_FE}] is
\begin{align}
\Sigma_1(q) &= i \frac{2 \alpha}{3 \pi} \ln\left (\frac{\Lambda}{|\vec{q}|}\right ) v_F \vec{q} \cdot \vec{\gamma} + {\rm finite}.
\end{align}
This divergence can be canceled out by the self energy-like counterterm \eqref{Eq:SE1CT}.
This, in fact, yields the values of $\delta_0$ and $\delta_1$ to ${\cal O}(\alpha)$; we find
that $\delta_0={\cal O}(\alpha^2)$, and thus there is no field strength renormalization $Z_\psi$
to this order, and that
\begin{align}
\delta_1 &= \frac{2 \alpha}{3 \pi} \ln\left (\frac{\Lambda}{\mu}\right ) + {\cal O}(\alpha^2).
\end{align}
The leading-order polarization, as we see from Eq.\ \eqref{Eq:Polarization_FE_Div}, is divergent as well.
This divergence may be canceled by the Coulomb propagatorlike counterterm, Eq.\ \eqref{Eq:P1CT}.
This defines the renormalization $\delta_p$ to order ${\cal O}(\alpha)$:
\begin{align}
\delta_p &= \frac{2 \alpha N}{3 \pi} \ln\left (\frac{\Lambda}{\mu}\right ) + {\cal O}(\alpha^2).
\end{align}

Finally, the vertex correction at $\vec{q}=0$ is given by:
\begin{align}
\raisebox{-0.9cm}{\scalebox{0.6}{\epsfig{file=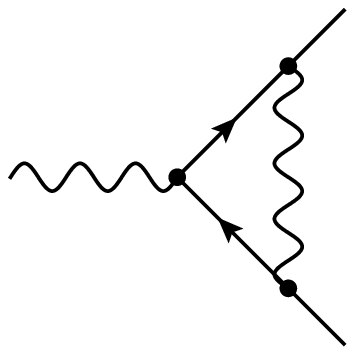}}} \ &\sim \int \frac{d^4k}{(2\pi)^4} \, \frac{g^2}{|\vec{k}|^2} i \gamma^0 \frac{i}{\slashed{k}} i \gamma^0 \frac{i}{\slashed{k}} i \gamma^0 = 0 . \label{eq:vertex}
\end{align}
Any divergence here would be canceled by the vertex counterterm, Eq.\ \eqref{Eq:V1CT}.  We see that there
is no divergence, and thus $\delta_v={\cal O}(\alpha^2)$.

\subsubsection{Two-loop RG analysis}

The divergent parts of the two-loop contributions to polarization, given by Eqs.\ \eqref{Eq:PiSE} and
\eqref{Eq:PiVFull}, are
\begin{eqnarray}
\Pi_{SE}(q)&\approx&\frac{|\vec{q}|^2}{g^2}\frac{2\alpha^2}{9\pi^2}N\left \{\ln^2\left (\frac{\Lambda}{|\vec{q}|}\right )+\frac{1}{3}\left [\frac{2(z^2+4)}{z^2+1}\right.\right. \cr
&+&\left.\left.3\ln\left (\frac{4}{z^2+1}\right )\right ]\ln\left (\frac{\Lambda}{|\vec{q}|}\right )\right \}
\end{eqnarray}
and
\begin{eqnarray}
\Pi_V(q)&\approx&-\frac{|\vec{q}|^2}{g^2}\frac{\alpha^2}{2\pi^2}NC\ln\left (\frac{\Lambda}{|\vec{q}|}\right ).
\end{eqnarray}
\begin{figure}
\includegraphics[width=0.5\columnwidth]{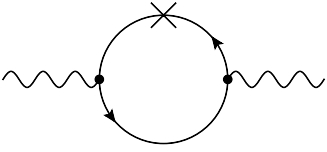}
\caption{Diagram corresponding to the counterterm, $\Pi_{SE,CT}(q)$, for the two-loop self-energy
polarization diagram.}
\label{Fig:P2CTb}
\end{figure}
The diagram corresponding to the counterterm that cancels out the divergence in $\Pi_{SE}(q)$ is shown in
Fig.\ \ref{Fig:P2CTb}, and we will denote it by $\Pi_{SE,CT}(q)$.  We find that it is given by
\begin{align}
&\Pi_{SE,CT}(q) \cr
&= 2i\delta_1 N \int \frac{d^4k}{(2\pi)^4} \, {\rm tr} \, \biggl[i \gamma^0 \frac{i}{\slashed{k} + \slashed{q}} i \gamma^0 \frac{i}{\slashed{k}} v_F \vec{k} \cdot \vec{\gamma} \frac{i}{\slashed{k}}\biggr] \cr
&= - \frac{|\vec{q}|^2}{g^2} \delta_1 \frac{2\alpha}{3\pi}N\left [\ln \frac{\Lambda}{|\vec{q}|}-\frac{z^2}{z^2+1}-\frac{1}{2}\ln\left (\frac{z^2+1}{4}\right )\right ] \cr
&= - \frac{|\vec{q}|^2}{g^2} \frac{4\alpha^2}{9\pi^2}N\ln \frac{\Lambda}{\mu}\left [\ln \frac{\Lambda}{|\vec{q}|}-\frac{z^2}{z^2+1}-\frac{1}{2}\ln\left (\frac{z^2+1}{4}\right )\right ]. \nonumber \\
\end{align}
Since the vertex renormalization is zero to leading order (cf. Eq.\ \eqref{eq:vertex}), there is no contribution from
\begin{align}
\raisebox{-0.7cm}{\scalebox{0.6}{\epsfig{file=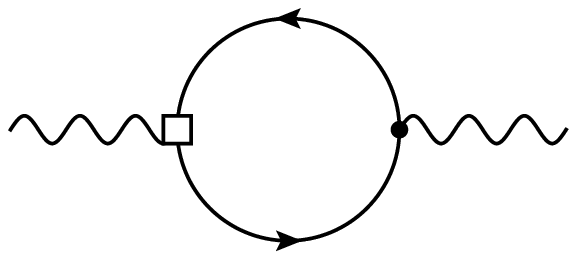}}} \ &= {\cal O}(\alpha^3) .
\end{align}
If we now add $\Pi_{SE}$ and $\Pi_{SE,CT}$, we find that it partially cancels the divergence from $\Pi_{SE}$;
all that is left behind are $\ln^2$ and $\ln$ terms with constant coefficients:
\begin{align}
&\Pi_{SE}(q)+\Pi_{SE,CT}(q) \cr
&=\frac{|\vec{q}|^2}{g^2}\frac{2\alpha^2}{9\pi^2}N\left \{\ln^2\left (\frac{\mu}{|\vec{q}|}\right )+\frac{1}{3}\left [\frac{2(z^2+4)}{z^2+1}\right.\right. \cr
&+\left.\left. 3\ln\left (\frac{4}{z^2+1}\right )\right ]\ln\left (\frac{\mu}{|\vec{q}|}\right )\right \} \cr
&+\frac{|\vec{q}|^2}{g^2}\frac{2\alpha^2}{9\pi^2}N\left [-\ln^2\left (\frac{\Lambda}{\mu}\right )+\frac{8}{3}\ln\left (\frac{\Lambda}{\mu}\right )\right ]
\end{align}
This remaining divergence plus that coming from $\Pi_V$ may then be canceled by adding the appropriate terms
to $\delta_p$.  Upon doing so, $\delta_p$ becomes
\begin{equation}
\delta_p=\frac{2\alpha}{3\pi}N\left [1+\left(\frac{3}{4}C-\frac{8}{9}\right)\frac{\alpha}{\pi}\right ]\ln\left (\frac{\Lambda}{\mu}\right )+\frac{2\alpha^2}{9\pi^2}N\ln^2\left (\frac{\Lambda}{\mu}\right ).
\end{equation}

\begin{figure}
\subfigure[]{\includegraphics[width=0.45\columnwidth]{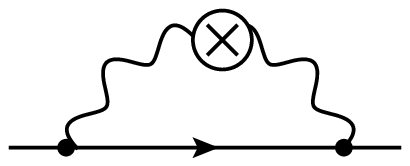}}\qquad
\subfigure[]{\includegraphics[width=0.45\columnwidth]{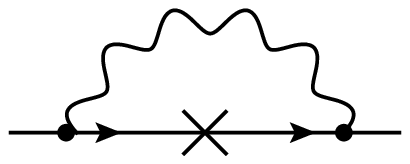}}
\subfigure[]{\includegraphics[width=0.45\columnwidth]{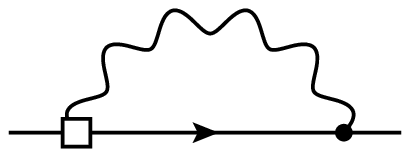}}
\caption{(a) Coulomb counterterm contribution to the second-order self-energy counterterm diagrams, $\Sigma_{2ct}(q)$.
This is the only non-vanishing contribution. (b) Self-energy counterterm contribution.  This diagram vanishes for the
same reason that the ``rainbow'' diagram is zero.  (c) Vertex counterterm contribution.  This diagram only contributes
at orders higher than $\alpha^2$.}
\label{Fig:SE2CT}
\end{figure}
We now turn our attention to the two-loop self-energy diagrams.  There is only one nonvanishing second-order
self-energy counterterm diagram, namely the one involving the counterterm Coulomb propagator, shown in Fig.\
\ref{Fig:SE2CT}. This diagram evaluates to
\begin{eqnarray}
\Sigma_{2ct}&=&-i\delta_p\int\frac{d^4k}{(2\pi)^4}\gamma^0\frac{\slashed k+\slashed q}{(k+q)^2}\gamma^0\frac{g^4}{|\vec{k}|^4}\frac{|\vec{k}|^2}{g^2} \cr
&=&\delta_p\Sigma_1(q)=i\frac{4N}{9\pi^2}\alpha^2v_F\vec{q}\cdot\vec{\gamma}\ln\frac{\Lambda}{|\vec{q}|}\ln\frac{\Lambda}{\mu}.
\end{eqnarray}
This has precisely the form needed to remove the $\ln^2$ term from $\Sigma_2(q)$ provided we add appropriate second-order terms to $\delta_1$:
\begin{eqnarray}
\delta_1&=&\frac{2\alpha}{3\pi}\ln\left (\frac{\Lambda}{\mu}\right )+\frac{2N}{9\pi^2}\alpha^2\ln^2\left (\frac{\Lambda}{\mu}\right ) \cr
&+&\left(\frac{45+N}{9\pi^2}-\frac{1}{2}\right)\alpha^2\ln\left (\frac{\Lambda}{\mu}\right )+{\cal O}(\alpha^3).
\end{eqnarray}
$\delta_0$ also receives contributions at second order:
\begin{equation}
\delta_0=\left(\frac{15+N}{3\pi^2}-\frac{1}{2}\right)\alpha^2\ln\left (\frac{\Lambda}{\mu}\right )+{\cal O}(\alpha^3).
\end{equation}

\subsubsection{Beta functions and RG equations}

We are now in a position to determine the RG equations describing the various couplings in our theory.  We do
this by determining the coefficients in the $F$ functions defined by Eqs.\ \eqref{Eq:Zexp}--\eqref{Eq:Eexp}.
As we saw above, we can then immediately read off the beta functions and anomalous dimensions from these and
thus obtain the associated RG equations.

Let us begin with $Z_\psi$.  From Eq.\ \eqref{Eq:Delta0}, we see that, to second order in $\alpha$, it is just
\begin{equation}
Z_\psi=1-\left(\frac{15+N}{3\pi^2}-\frac{1}{2}\right)\alpha^2\ln\left (\frac{\Lambda}{\mu}\right ).
\end{equation}
This immediately tells us $z_{1,\psi}(\alpha)$,
\begin{equation}
z_{1,\psi}(\alpha)=\left(\frac{15+N}{3\pi^2}-\frac{1}{2}\right)\alpha^2,
\end{equation}
and thus $\gamma_\psi$,
\begin{equation}
\gamma_\psi=\left(\frac{15+N}{6\pi^2}-\frac{1}{4}\right)\alpha^2.
\end{equation}
The RG equation for $Z_\psi$ is thus
\begin{equation}
\frac{d\ln{Z_\psi}}{d\ln{\mu}}=\left(\frac{15+N}{3\pi^2}-\frac{1}{2}\right)\alpha^2. \label{eq:rgZ}
\end{equation}

Next, we will look at $v_F$.  We first solve Eq.\ \eqref{Eq:Delta1} for $\frac{v_F^B}{v_F}$:
\begin{equation}
\frac{v_F^B}{v_F}=\frac{1-\delta_1}{Z_\psi}=\frac{1-\delta_1}{1-\delta_0}
\end{equation}
Using our results, we find that, to order $\alpha^2$, $\frac{v_F^B}{v_F}$ is
\begin{eqnarray}
\frac{v_F^B}{v_F}&=&1-\frac{2\alpha}{3\pi}\ln\left (\frac{\Lambda}{\mu}\right )+\frac{2N}{9\pi^2}\alpha^2\ln\left (\frac{\Lambda}{\mu}\right ) \cr
&-&\frac{2N}{9\pi^2}\alpha^2\ln^2\left (\frac{\Lambda}{\mu}\right ).
\end{eqnarray}
This immediately tells us $v_1(\alpha)$ to second order:
\begin{equation}
v_1(\alpha)=\frac{2\alpha}{3\pi}-\frac{2N}{9\pi^2}\alpha^2.
\end{equation}
The RG equation for $v_F$ is then
\begin{equation}
\frac{d\ln{v_F}}{d\ln{\mu}}=\gamma_{v_F}=-v_1(\alpha)=-\frac{2\alpha}{3\pi}+\frac{2N}{9\pi^2}\alpha^2. \label{eq:rgvF}
\end{equation}
It is interesting to note that the second-order correction to the RG flow of the velocity only depends on the self-energy correction coming from the RPA-type bubble diagram shown in Fig.\ \ref{fig:selfenergy2c}, while the contribution coming from Fig.\ \ref{fig:selfenergy2a} cancels out exactly.

\begin{figure}
\includegraphics[width=\columnwidth]{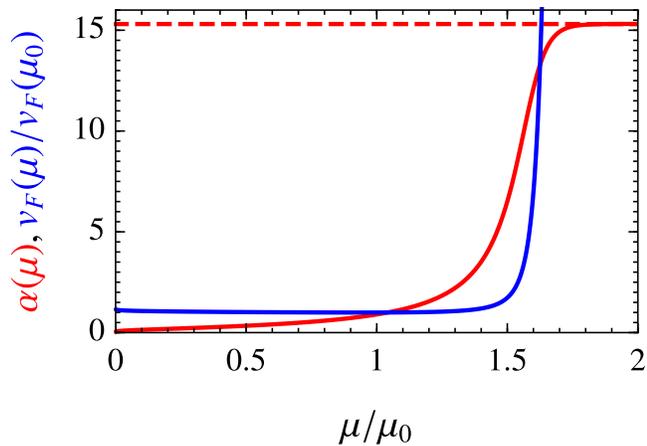}
\caption{Numerical solution of the RG equations for $\alpha$ (red) and $v_F$ (blue) for $\alpha(\mu_0)=0.9$ and $N=12$, which are typical values for
the pyrochlore iridates.  Since $\alpha(\mu_0)$ is far below $\alpha_c$, indicated by the red dashed line, we find that it increases for small $\mu$,
then saturates to $\alpha_c$.  $v_F$, on the other hand, decreases as $\mu$ increases up to about $\mu\approx 0.95\mu_0$, then starts increasing.  This
is better illustrated in Fig.\ \ref{Fig:RGEquNumInt_vFOnly} below.}
\label{Fig:RGEquNumInt}
\end{figure}

\begin{figure}
\includegraphics[width=\columnwidth]{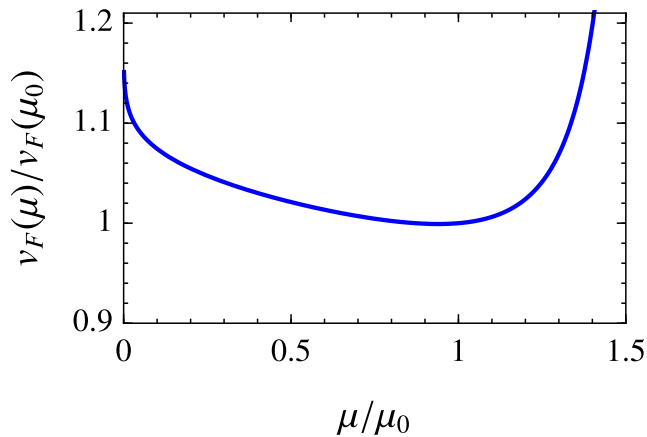}
\caption{Plot of $v_F$ alone, with the range for $\mu$ adjusted to better illustrate the non-monotonic behavior of $v_F$.}
\label{Fig:RGEquNumInt_vFOnly}
\end{figure}

Next, we look at $Z_\varphi$.  As is shown in App.\ \ref{App:SE_Vertex}, $1-\delta_v=Z_\psi$, and
thus, to second order in $\alpha$, $Z_\varphi=1$.  We therefore find that $Z_\varphi$ remains constant
for all values of $\mu$.  Finally, we consider $g^2$.  Solving Eq.\ \eqref{Eq:Deltap} for $\frac{(g^B)^2}{g^2}$
gives us
\begin{equation}
\frac{(g^B)^2}{g^2}=\frac{1}{1-\delta_p}.
\end{equation}
Expanding to second order in $\alpha$, we get
\begin{eqnarray}
\frac{(g^B)^2}{g^2}&=&1+\frac{2\alpha}{3\pi}N\left [1+\left(\frac{3}{4}C-\frac{8}{9}\right)\frac{\alpha}{\pi}\right ]\ln\left (\frac{\Lambda}{\mu}\right ) \cr
&+&\frac{2\alpha^2}{9\pi^2}N(2N+1)\ln^2\left (\frac{\Lambda}{\mu}\right ).
\end{eqnarray}
From this, we can now read off $g_1(\alpha)$:
\begin{equation}
g_1(\alpha)=-\frac{2\alpha}{3\pi}N\left [1+\left(\frac{3}{4}C-\frac{8}{9}\right)\frac{\alpha}{\pi}\right ].
\end{equation}
We can now determine $f_1(\alpha)$ and thus the RG equation for $\alpha$.  The former is
\begin{equation}
f_1(\alpha)=-\frac{2}{3}(N+1)\frac{\alpha}{\pi}-\left(\frac{1}{2}C-\frac{22}{27}\right)N\left (\frac{\alpha}{\pi}\right )^2, \label{Eq:f1OalphaSq}
\end{equation}
and the latter is
\begin{equation}
\frac{d\alpha}{d\ln{\mu}}=-\alpha f_1(\alpha)=\frac{2(N+1)}{3\pi}\alpha^2+\frac{27C-44}{54\pi^2}N\alpha^3. \label{eq:rgalpha}
\end{equation}
These equations agree exactly with those that we would obtain directly from the Callan-Symanzik equation;
we show this in Appendix \ref{App:CSAnalysis}.  If we calculate the value of the coefficient of the $\alpha^3$
term, then we find that it is negative:
\begin{equation}
\frac{27C-44}{54\pi^2}=-0.0150184
\end{equation}
Therefore, we find that there are two fixed points for this equation, similarly to the case of graphene\cite{BarnesPRB2014},
$\alpha=0$ and
\begin{equation}
\alpha_c=\frac{36\pi}{44-27C}\left (1+\frac{1}{N}\right )=14.1298\left (1+\frac{1}{N}\right ).
\end{equation}
If we start with a value below the latter critical value, $\alpha<\alpha_c$, at a large scale $\mu$, then $\alpha$ will go to zero as we decrease $\mu$.  On the other hand, if we start with $\alpha>\alpha_c$, then we instead find that $\alpha$ diverges to infinity as $\mu$ decreases, indicating a runaway to infinite coupling. Similar behavior was also found in graphene\cite{BarnesPRB2014}, although in that case the corresponding critical coupling is much smaller: $\alpha_c=0.78$. In the present context of 3D Dirac materials, however, an additional interesting feature, not present in 2D graphene, arises. Notice that the second-order correction to the velocity RG equation, Eq.\ \eqref{eq:rgvF} also differs in sign from the leading-order term, implying the existence of a second special value of $\alpha$, this time at $\alpha_*=3\pi/N$. If we start with $\alpha<\alpha_*$, then the velocity grows monotonically as $\mu$ decreases, while for $\alpha>\alpha_*$ the velocity is initially suppressed as $\mu$ decreases. Moreover, since $\alpha_*<\alpha_c$ for all values of $N$, there exists a window of couplings, $\alpha_*<\alpha<\alpha_c$, in which both the coupling and velocity become smaller with decreasing $\mu$. This behavior persists until $\alpha$ falls below $\alpha_*$, at which point the velocity reverses course and grows as the low-energy, non-interacting fixed point is approached. Thus, we find that for sufficiently large interaction strengths, the velocity exhibits non-monotonic behavior as the weak-coupling fixed point is approached at the Dirac point.  We illustrate this behavior in Figs.\ \ref{Fig:RGEquNumInt} and  \ref{Fig:RGEquNumInt_vFOnly}, in which we show a numerical solution of the RG equations for $\alpha$ and $v_F$ for $\alpha(\mu_0)=0.9$ and $N=12$, typical values for the pyrochlore iridates.  This nonmonotonic running of the velocity for $\alpha_* < \alpha < \alpha_c$ is in total contrast to graphene velocity renormalization, which is always monotonic for all coupling constants.

One noteworthy feature of the RG flow for the Fermi velocity that creates a major contrast to graphene is the fact that, unlike in graphene, where the Fermi velocity diverges at low energy\cite{EliasNatPhys2011,YuPNAS2013,BarnesPRB2014} (at least for a strictly non-relativistic, instantaneous interaction; in the real system with relativistic interactions, this growth will cease when the Fermi velocity reaches the speed of light) the Fermi velocity remains finite in the limit that the energy scale goes to zero (i.e., at the Dirac point).  This can be traced back to the fact that, in graphene, charge does not renormalize, as shown in Table \ref{Tab:Graphenevs3DDirac}.  As a result, since graphene becomes weakly interacting at low-energy scales, the Fermi velocity \textit{must} diverge.  This is not the case for 3D Dirac materials.  Because charge \textit{can} renormalize in this case, there is another way for the system to become weakly interacting at low energy, and that is for the charge to go to zero.  This is exactly what happens here.  As a result, the Fermi velocity remains finite, even at low energies; it simply renormalizes to a different value. Thus, the nature of ultraviolet renormalization in 3D and 2D Dirac materials is qualitatively different.

As in the case of graphene\cite{BarnesPRB2014}, the appearance of $\alpha_c$ and $\alpha_*$ in 3D Dirac systems may be a manifestation of the breakdown of perturbation theory (due to the asymptotic nature of the perturbative expansion in $\alpha$) rather than real quantum phase transitions, particularly since these points precisely correspond to the second-order corrections becoming comparable in magnitude to the leading-order results. Following Dyson's original argument for QED\cite{DysonPR1952}, we can obtain an estimate for the order of the expansion beyond which the perturbative results can no longer be trusted\cite{BarnesPRB2014,KolomeiskyArXiv2014}:
\begin{equation}
n\approx\frac{3.09797}{\alpha^{3/2}}.\label{eq:breakdown}
\end{equation}
Here, $n$ is the expected perturbative order up to which the loop expansion in the effective fine structure constant is asymptotic, with the perturbation series diverging beyond the $n$th order.  As an example, let us consider Cd${}_3$As${}_2$: the experimental values\cite{LiuNatMater2014,NeupaneNatCommun2014,BorisenkoPRL2014} of the Fermi velocity are typically in the range $10^5-10^6$m/s, while the dielectric constant has been measured to be $\kappa\sim20-40$,\cite{ZivitzPRB1974,JayGerinSSC1977} suggesting coupling values in the range $\alpha\sim0.1-1$. Equation \eqref{eq:breakdown} then suggests that our second-order perturbative analysis is likely to be valid for coupling values toward the lower end of this range, but as $\alpha$ approaches unity, the results become questionable. If we interpret $\alpha_*$ as a signature of the breakdown of the perturbative expansion, then we would further conclude that the validity of perturbation theory also depends on $N$ since $\alpha_*$ is inversely proportional to $N$. For a material like Cd${}_3$As${}_2$ for which $N=1$, we find $\alpha_*=3\pi\approx9.4$, which is more or less consistent with Eq.\ \eqref{eq:breakdown}. However, for the pyrochlore iridates\cite{WanPRB2011} with $N=12$, we instead obtain $\alpha_*=\pi/4\approx0.8$, which may call into question the applicability of perturbation theory for these materials if their effective interaction strength $\alpha$ satisfies $\alpha\gtrsim\alpha_*$. The fact that perturbation theory would seem to be more reliable in the regime of smaller $N$ is interesting given that this is precisely complementary to the regime in which RPA or a large-$N$ expansion is valid\cite{HofmannPRL2014,HofmannArXiv2014}. It would appear that perturbation theory and RPA/large-$N$ are in this sense complementary to one another.

In contrast to QED, where the bare coupling is fixed to be $\sim \tfrac{1}{137}$ so that the perturbative expansion is asymptotic to a very high order, the 3D Dirac materials enable, in principle, an investigation of both the running of the coupling for a specific material and the analysis of the asymptotic nature of the underlying field theory by varying the Dirac material so as to modify the bare coupling from a value as small as $0.1$ to as large as $1$.

\subsubsection{Consistency check}

We will now apply the recursion relations for the counterterms, Eqs.\ \eqref{Eq:AlphaRecRel} and \eqref{Eq:vFRecRel},
to our results to ensure that they are satisfied, thus providing a compelling check on our calculations.  To be exact,
we can show that our results for $f_1(\alpha)$ and $v_1(\alpha)$, at order $\alpha$, along with the beta functions
and anomalous dimensions determined in the previous section, give the same values of $f_2(\alpha)$ and $v_2(\alpha)$
to order $\alpha^2$ that we determined directly from our two-loop calculations.

Let us start with the relation for $\alpha$, Eq.\ \eqref{Eq:AlphaRecRel}.  We already found $f_1(\alpha)$ in the
previous section, so we only need to determine $f_2(\alpha)$.  Using the previously-stated formula for this function,
we find that, to order $\alpha^2$,
\begin{eqnarray}
f_2(\alpha)&=&g_2(\alpha)-v_2(\alpha)+g_1(\alpha)v_1(\alpha)-v_1(\alpha)^2 \cr
&=&-\frac{4\alpha^2}{9\pi^2}(N+1)^2.
\end{eqnarray}
Now let us check that we obtain this result from our recursion relation.  If we substitute $f_1(\alpha)$ and $\beta_\alpha$
into Eq.\ \eqref{Eq:AlphaRecRel} and only keep the lowest-order terms, we obtain
\begin{eqnarray}
2\alpha f_2(\alpha)&=&-\frac{2(N+1)}{3\pi}\alpha^2\left (1+\alpha\frac{\partial}{\partial\alpha}\right )\frac{2(N+1)}{3\pi}\alpha \cr
&=&-\frac{8(N+1)^2}{9\pi^2}\alpha^3.
\end{eqnarray}
This gives us the value of $f_2(\alpha)$ stated above.

We may similarly apply the relation for $v_F$, Eq.\ \eqref{Eq:vFRecRel}.  We may immediately read off $v_2(\alpha)$
from our previous results:
\begin{equation}
v_2(\alpha)=\frac{2N}{9\pi^2}\alpha^2
\end{equation}
Again, we wish to show that we obtain this same result from the recursion relation.  Substituting $v_1(\alpha)$, $\gamma_{v_F}$,
and $\beta_\alpha$ into Eq.\ \eqref{Eq:vFRecRel} and keeping only the lowest-order terms, we find that
\begin{eqnarray}
2v_2(\alpha)&=&\left (-\frac{2\alpha}{3\pi}+\frac{2(N+1)}{3\pi}\alpha^2\frac{\partial}{\partial\alpha}\right )\frac{2\alpha}{3\pi} \cr
&=&\frac{4N}{9\pi^2}\alpha^2.
\end{eqnarray}
We thus obtain the same value of $v_2(\alpha)$ stated above.

\section{Experimental detection}\label{Sec:Experiment}

In this section, we discuss how the running of the quasiparticle properties discussed in the previous sections can be detected in experiment. Such an experiment has to extract either the Fermi velocity or the fine-structure constant over a parameter range to check for scaling. Previous experiments on 2D Dirac materials had precise control over temperature, doping, and disorder, all of which are competing scales that cut off the intrinsic renormalization group flow at low energies, possibly masking the intrinsic Dirac point physics. For example, varying the doping by means of an external gate potential, the Fermi velocity renormalization induces an increase in the Fermi velocity as smaller densities (corresponding to smaller Fermi energies) are probed\cite{EliasNatPhys2011}.

It appears that at the present time, a comparable control over 3D Dirac material properties is still lacking. While initial experiments have reported control over the bulk detuning by varying the surface doping\cite{LiuScience2014}, this might not yet be sufficient to provide definite signatures of renormalization. In addition, typical samples can have strong disorder. However, a clear observation of scaling does not require the suppression of all relevant perturbations, but rather a hierarchy of scales where one parameter (for example, the temperature) can be tuned to large values compared to any other scale (for example, the doping). In this section, we discuss this largely overlooked alternative way of detecting renormalization effects by varying the temperature.

This section is structured as follows: First, we give an example for the integrated charge renormalization to leading order in $\alpha$ which can be obtained in closed analytical form, thereby introducing the Landau pole scale which signifies a divergence in the renormalized coupling. Second, we illustrate the effects of finite temperature on the Drude weight as obtained from kinetic theory as a generic example of an observable that can be detected in optical conductivity measurements. We show that the Drude weight of an extrinsic system (i.e., with finite doping) at finite temperature assumes the same value as the Drude weight of an intrinsic system, i.e., increasing the temperature probes the intrinsic system irrespective of initial detuning. The renormalization is then apparent in an additional temperature dependence of the quasiparticle parameters giving rise to a deviation from the intrinsic unrenormalized linear-in-temperature scaling. In the third subsection, we present a calculation of the plasmon behavior to leading order in RPA, which naturally incorporates the effect of finite doping and the renormalization of $\alpha$.  Thus, by studying plasmon properties experimentally as a function of doping density and/or temperature, it should be possible to verify the ultraviolet renormalization and the RG flow of 3D Dirac system many-body effects.

\subsection{Coupling constant renormalization and Landau pole}\label{sec:examplergintegrated}

To illustrate the coupling constant renormalization, we consider the renormalization group flow of the coupling to one loop. As derived in Eq.~\eqref{eq:rgalpha}, the renormalization group equation is
\begin{align}
\frac{d\alpha}{d\ln \mu} &= \frac{2 (N+1) \alpha^2}{3\pi}. \label{eq:rgoneloop}
\end{align}
If the beta function has negative sign, the coupling becomes weaker with increasing energy scale -- this is the situation encountered in quantum chromodynamics, where the perturbative high-energy limit is known as `asymptotic freedom'. In the present theory, however, the coupling increases with increasing energy scale, just like in quantum electrodynamics. Starting in a perturbative regime at small energy, the system becomes nonperturbative at higher energy. Solving the renormalization group equation~\eqref{eq:rgoneloop} with the boundary condition $\alpha(\mu_0) = \alpha_0$ (the measured value of the charge at a scale $\mu_0$), we obtain:
\begin{align}
\alpha(\mu) &= \frac{\alpha_0}{1 + \dfrac{2 (N+1) \alpha_0}{3 \pi} \ln \dfrac{\mu_0}{\mu}} . \label{Eq:RGOneLoopSoln}
\end{align}
It turns out that this solution has a divergence at a scale $\mu = \Lambda_L = \mu_0 e^{3 \pi/2 (N+1)\alpha_0}$ known as the Landau pole. We can express the renormalized coupling~\eqref{Eq:RGOneLoopSoln} in terms of the Landau pole:
\begin{align}
\alpha(\mu) &= \frac{3 \pi}{2 (N+1)} \ln^{-1} \dfrac{\Lambda_L}{\mu} . \label{eq:chargeren}
\end{align}
The Landau pole is thus the effective parameter that characterizes the interacting Dirac semimetal. At scales close to $\Lambda_L$, the coupling diverges and the theory becomes strongly interacting, even if we start with a weakly interacting theory at low energy. The theory makes no prediction for $\Lambda_L$; it is a material property that must be taken from experiment. Strictly speaking, as the renormalized coupling diverges, our first-order theory loses validity, and a more complete and fully nonperturbative calculation is required in order to establish if the divergence is cut off by nonperturbative effects or if the coupling truly diverges. Our theory, however, makes the unambiguous prediction that the charge should increase logarithmically at high energy or temperature (setting $\mu = T$). By analogy with quantum electrodynamics, one would expect that the pole is preempted by a phase transition, although this remains to be established for Dirac semimetals\cite{GockelerPRL1998}.

We mention that by definition our leading-order beta function [Eq.\ \eqref{eq:rgoneloop}] has no asymptotic freedom (i.e., there is a positive sign on the right hand side for all $\alpha$ values); in fact, we have the Landau pole at very high energy where the running coupling diverges.  We note, however (as discussed in Sec.\ \ref{Sec:Introduction} of our paper), that our calculated beta function to second order [c.f. Eq.\ \eqref{eq:rgalpha}] does have a negative sign if $\alpha>\alpha_c$ with the renormalized coupling running to infinity at lower-energy scales (i.e., behaving superficially similar to the QCD situation) provided we start in the strong-coupling regime of $\alpha>\alpha_c$.  In such a strong-coupling situation (which arises only when the second-order perturbative correction is comparable to the first-order term, and hence may very well be unreliable within a loop expansion theory), the Dirac system manifests stronger (weaker) coupling at low (high) energy, superficially mimicking QCD behavior.

\subsection{Optical conductivity and Drude weight}
\begin{figure}[t!]
\includegraphics[width=\columnwidth]{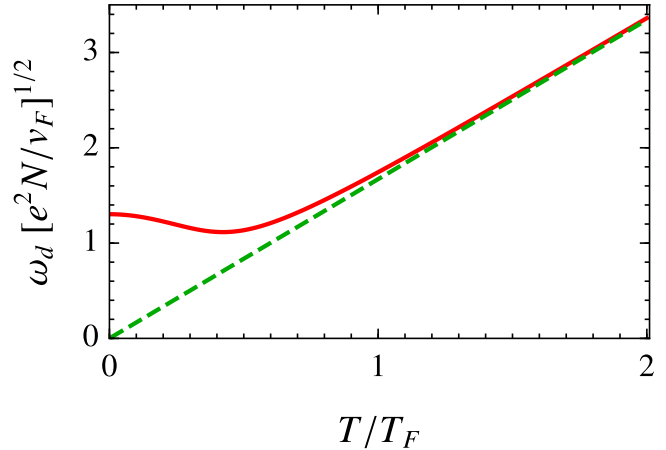}
\caption{Drude weight for extrinsic (red line) and intrinsic (green dashed line) Dirac materials. At high temperature, there is no difference between the intrinsic and the extrinsic Drude weight, thus manifestly showing that temperature can act as an appropriate running scale to study the intrinsic Dirac point physics.}
\label{fig:drude}
\end{figure}

In this section, we show the competing effects of temperature and doping on the Drude weight of a 3D Dirac material and demonstrate that even a moderate temperature of order $T \sim \varepsilon_F/2$, where $\epsilon_F$ is the doping-induced Fermi energy, is sufficient to probe the intrinsic response of the system without detuning. The Drude-Boltzmann form of the low-frequency conductivity takes the form
\begin{align}
\sigma(\omega) &= - \frac{\omega_d^2}{4\pi} \frac{1}{\omega - i/\tau} ,
\end{align}
where $\omega_d^2$ is the Drude weight and $1/\tau$ is a transport relaxation time induced by disorder or interactions. Within kinetic theory, the Drude weight is given by (a detailed derivation is given in Appendix \ref{sec:drude})
\begin{align}
\omega_d^2= 4 \pi e^2 \frac{2 v_F^2}{3} \frac{D_F}{\varepsilon_F^2} \int_0^\infty d\varepsilon \, \varepsilon^2 \, \biggl(- \frac{\partial f^0(\varepsilon)}{\partial \varepsilon} \biggr) ,\label{eq:dw}
\end{align}
where $D_F = N \varepsilon_F^2/[\pi^2 v_F^3]$ is the density of states and $f_0$ the Fermi-Dirac distribution. The red curve in Fig.~\ref{fig:drude} shows the extrinsic Drude weight as a function of temperature. The low-temperature Drude weight (which is dominated by intraband excitations) starts off at a constant value with a Sommerfeld correction that decreases its value at small but finite temperature. The Drude weight assumes a minimum and then increases at high temperature where it has a linear temperature dependence. For comparison, we include a plot of the intrinsic Drude weight as given by Eq.~\eqref{eq:dw} with the chemical potential set to zero. It is apparent from the figure that at high temperature both curves are equal and the intrinsic Drude weight is measured, even if the system started off at finite doping. The renormalization should induce a superlinear high-temperature dependence reflecting the intrinsic semimetallic many-body effects as discussed in the next section. 

The computation of the Drude weight within kinetic theory gives the same results as a diagrammatic calculation using the Kubo formula to the leading one-loop order\cite{AshbyPRB2014}. Note that in the intrinsic limit, higher-order diagrams are expected to introduce nonlocal corrections to the conductivity\cite{RosensteinPRB2014,RosensteinArXiv2015}. Here, we consider the extrinsic case, for which this effect does not appear.

\subsection{Plasmon dispersion}

\begin{figure}[t!]
\includegraphics[width=\columnwidth]{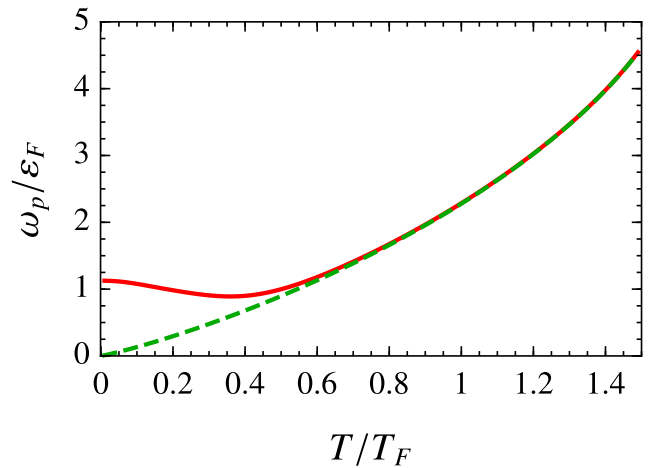}
\caption{Plasmon dispersion for extrinsic (red line) and intrinsic (green dashed line) systems with $\Lambda_L = 8 \varepsilon_F$.  The manifest superlinearity in the temperature-dependent plasmon dispersion of the doped system arises from the ultraviolet renormalization with temperature also acting as the cutoff for $T\ll T_F$, as expected.
}
\label{fig:plasmon}
\end{figure}

We now consider the plasmon dispersion which is given by the pole of the dielectric function. It is related to the Drude-Boltzmann form of the conductivity by
\begin{align}
\varepsilon(\omega) &= 1 - \frac{4\pi i}{\omega} \sigma(\omega) = 1 - \frac{\omega_d^2}{\omega^2} = 0 . \label{eq:plasmon}
\end{align}
A full solution of the plasmon dispersion to leading order in RPA was presented in Ref.~\onlinecite{HofmannArXiv2015}, and will not be repeated here. Instead, we expand on this work by considering the effect of charge renormalization. The defining expression for the dielectric function does not involve the bare coupling $\alpha$, Fermi velocity, and cutoff separately, but only combined in the form of the Landau pole without any additional divergent terms. The full solution of Eq.~\eqref{eq:plasmon} then gives the plasmon dispersion as a function of temperature, Fermi energy, and Landau pole. The plasmon dispersion is shown in Fig.~\ref{fig:plasmon} for both doped extrinsic (red line) and undoped intrinsic (green line) Dirac systems using a Landau pole scale of $\Lambda_L = 8 \varepsilon_F$. As discussed in the previous section, doping does not influence the plasmon at high temperature and the dispersion is equal for intrinsic and extrinsic configurations. The full calculation of the plasmon dispersion which includes the coupling constant renormalization predicts a superlinear high-temperature scaling of
\begin{align}
\omega_d \sim \frac{T}{\ln \Lambda_L/T} \sim \alpha(T) T .
\end{align}
This superlinear scaling is valid at high temperature and is cut off a low temperature due to finite doping, as seen in Fig.~\ref{fig:plasmon}. While simple dimensional analysis would only predict a linear temperature dependence of the intrinsic plasmon,\cite{HofmannArXiv2015} the correct result displays a superlinear scaling through the logarithmic temperature dependence due to the renormalization of the coupling as discussed in Sec.~\ref{sec:examplergintegrated}. Note that the result for the intrinsic plasmon agrees with the Drude weight~\eqref{eq:dw} if we assume a constant $\kappa$ and replace the bare value of charge and Fermi velocity by the corresponding renormalized expression~\eqref{eq:chargeren}.

An experimental observation of such a super-linearity in the measured temperature dependence of the plasmon dispersion would be strong evidence for the ultraviolet renormalization of interaction effects in Dirac materials.  We emphasize that since 3D Dirac materials cannot be doped in situ in a continuous manner using an external gate (as can be done for 2D graphene layers), temperature-dependent measurements (in the regime of temperature being larger than the Fermi energy associated with the doping density) are probably the simplest technique to characterize their intrinsic Dirac point properties.  Such temperature dependence will have a low-temperature cutoff (or saturation) due to the remnant doping in the sample, as shown in Fig.\ \ref{fig:plasmon}, but above this cutoff scale of the extrinsic Fermi energy, the system should reflect the RG scaling defined by its Dirac point properties.

\section{Conclusion} \label{Sec:Conclusion}

We have calculated the electron self-energy and polarizability for Dirac-Weyl materials up
to second order in the effective fine-structure constant $\alpha$ and for arbitrary number of pairs of Dirac cones $N$ (arising, for example, from valley degeneracy associated with band-structure effects). As expected, we find that log-squared divergences arise at second order as necessary for the renormalizability of the theory, and we have used these to perform many self-consistency checks of our results. After calculating the counterterms necessary to cure these divergences, we extracted from them the many-body renormalization effects that are evident in properties such as the quasiparticle residue, Fermi velocity, and coupling strength.  Our derivation of the RG flows of the quasiparticle properties reveal the existence of a critical point, $\alpha_c=14.13(1+1/N)$, above which the coupling flows to larger values at low energies. This behavior is very analogous to what occurs in graphene, which also exhibits such a fixed point, albeit at a much lower value of the coupling: $\alpha_c=0.78$\cite{BarnesPRB2014}. Following Dyson's original argument\cite{DysonPR1952,BarnesPRB2014}, we have argued that the appearance of $\alpha_c$ may be an artifact of the asymptotic nature of the perturbative series and not a real feature of the system, but only measurements can determine whether there is a strong-coupling fixed point in 3D Dirac systems.   It is conceivable that our perturbative expansion is correctly indicating the presence of a strong-coupling fixed point for the system, but the precise value of the critical coupling strength can only be correctly evaluated by non-perturbative methods.

Unlike graphene, however, Dirac-Weyl materials exhibit a second special value of the coupling, $\alpha_*=3\pi/N$, at which the flow of the velocity reverses direction. This implies that for couplings in the range $\alpha_*<\alpha<\alpha_c$, both the interaction strength and Fermi velocity initially decrease with the energy scale until the coupling reaches the value $\alpha_*$, at which point the velocity flow reverses and begins to climb to larger values as the energy scale is reduced further. At this point, it is unclear whether $\alpha_*$ should be viewed as giving rise to an experimentally observable non-monotonic behavior in the Fermi velocity, or if it should instead be seen as an indication that perturbation theory cannot be trusted for coupling strengths comparable to or larger than $\alpha_*$ (as is likely to be the case for $\alpha_c$). This is a very interesting avenue for further exploration given that $\alpha_*$ may be in the experimentally relevant regime for materials possessing a large Dirac cone multiplicity, such as the pyrochlore iridates.  We hope that future experimental work in 3D Dirac materials will look for possible signatures of nonmonotonicity in the measured Fermi velocity as a function of doping density (at low temperatures) or temperature (at low doping densities) to check for the predicted existence of $\alpha_*$ in the theory.

We also discuss the experimental detection of ultraviolet renormalization effects by computing the Drude weight and the plasmon dispersion up to leading order in the fine-structure constant and the random phase approximation, respectively. There are two main findings: first, by performing computations at finite doping and temperature, we establish that the intrinsic renormalization group results which we obtain in this work govern the behavior of the system at high temperature, i.e., the behavior of a Dirac material is the same as for an intrinsic system irrespective of initial detuning. This general result (that holds not only for the observables discussed in this paper) is of particular relevance for current experiments which are not able to precisely control doping and disorder. Second, we find that the high-temperature Drude weight including renormalization and the plasmon dispersion scale in a superlinear form as $\sim T/\ln\Lambda_L/T$. While the term linear in temperature can be predicted from dimensional analysis, the logarithmic scaling is a direct manifestation of renormalization. The strength of the renormalization effect is set by the Landau pole scale $\Lambda_L$, a scale which resembles similar findings in QED at one-loop order\cite{PeskinSchroederBook} and which provides a renormalization-flow invariant parameter that may be used to fit our results to experimental data. The superlinear temperature scaling should provide a direct signature of renormalization effects in current and future experiments on Dirac materials.

\acknowledgements
This work was supported by LPS-MPO-CMTC.

\appendix

\section{First-order polarization bubble} \label{App:Polarization_FE}
Here, we give the details of how to determine the first-order polarization
bubble $\Pi_B(q)$, shown in Fig.\ \ref{Fig:Polarization_FE}.  The expression
that we obtain is
\begin{eqnarray}
\Pi_B(q)&=&N\int\frac{d^4k}{(2\pi)^4}\,\mbox{Tr}[\gamma^0G_0(k+q)\gamma^0G_0(k)] \cr
&=&-N\int\frac{d^4k}{(2\pi)^4}\,\mbox{Tr}\left [\gamma^0\frac{\slashed{k}+\slashed{q}}{(k+q)^2}\gamma^0\frac{\slashed{k}}{k^2}\right ].
\end{eqnarray}
Evaluating the trace, we obtain
\begin{equation}
\Pi_B(q)=-4N\int\frac{d^4k}{(2\pi)^4}\,\frac{k_0(k_0+q_0)-v_F^2\vec{k}\cdot(\vec{k}+\vec{q})}{(k_0^2+v_F^2|\vec{k}|^2)[(k_0+q_0)^2+v_F^2|\vec{k}+\vec{q}|^2]}.
\end{equation}
Evaluating the integral over $k_0$, we get
\begin{eqnarray}
\Pi_B(q)&=&-2N\int\frac{d^3\vec{k}}{(2\pi)^3}\,\frac{|\vec{k}|+|\vec{k}+\vec{q}|}{q_0^2+v_F^2(|\vec{k}|+|\vec{k}+\vec{q}|)^2} \cr
&\times&\left [1-\frac{\vec{k}\cdot(\vec{k}+\vec{q})}{|\vec{k}||\vec{k}+\vec{q}|}\right ].
\end{eqnarray}
If we now switch to prolate spheroidal coordinates, as given in Eq.\ \eqref{Eq:ProSphCoord},
we obtain, after performing the (trivial) $\theta$ integration,
\begin{eqnarray}
\Pi_B(q)&=&-\frac{|\vec{q}|^2}{8\pi^2v_F}N\int_0^{\infty}d\mu\,\int_0^\pi d\nu\,\sinh{\mu}\sin{\nu} \cr
&\times&\frac{\cosh{\mu}}{z^2+\cosh^2{\mu}}(1-\cos^2{\nu}).
\end{eqnarray}
We now make another coordinate change, $x=\cosh{\mu}$ and $y=\cos{\nu}$.
Note, however, that the integral has a logarithmic divergence, and thus
we introduce a cutoff $|\bk|\leq\Lambda$ on the momentum.  Doing these,
we obtain
\begin{equation}
\Pi_B(q)=-\frac{|\vec{q}|^2}{8\pi^2v_F}N\int_{-1}^1 dy\,\int_1^{2\lambda+y} dx\,\frac{x}{z^2+x^2}(1-y^2).
\end{equation}
If we now evaluate the $x$ integral, we get
\begin{equation}
\Pi_B(q)=-\frac{|\vec{q}|^2}{8\pi^2v_F}N\int_{-1}^1 dy\,(1-y^2)\ln\left [\frac{z^2+(2\lambda+y)^2}{z^2+1}\right ].
\end{equation}
If we now perform an integration by parts, we obtain Eq.\ \eqref{Eq:Polarization_FE}
in the main text.

\section{First-order electron self-energy} \label{App:ElectronSE_FE}
The diagram corresponding to the first-order electron self-energy correction is
shown in Fig.\ \ref{Fig:SelfEnergy_FE}, and will be denoted by $\Sigma_1(q)$.  The
expression that we obtain for this diagram is
\begin{equation}
\Sigma_1(q)=-\int\frac{d^4k}{(2\pi)^4}\,\gamma^0G_0(k+q)\gamma^0D_0(-k),
\end{equation}
or, upon substituting in the expressions for the bare Green's function $G_0(k+q)$
and the bare Coulomb propagator,
\begin{equation}
\Sigma_1(q)=-ig^2\int\frac{d^4k}{(2\pi)^4}\,\gamma^0\frac{\slashed{k}+\slashed{q}}{(k+q)^2}\gamma^0\frac{1}{|\vec{k}|^2}.
\end{equation}
The integral on $k_0$ may easily be done; after evaluating it, we obtain
\begin{equation}
\Sigma_1(q)=-\frac{1}{2}ig^2\int\frac{d^3\vec{k}}{(2\pi)^3}\,\gamma^0\frac{(\vec{k}+\vec{q})\cdot\vec{\gamma}}{|\vec{k}|^2|\vec{k}+\vec{q}|}\gamma^0.
\end{equation}
Since $\gamma^0$ anticommutes with $\gamma^i$, we may rewrite this as
\begin{equation}
\Sigma_1(q)=\frac{1}{2}ig^2\int\frac{d^3\vec{k}}{(2\pi)^3}\,\frac{(\vec{k}+\vec{q})\cdot\vec{\gamma}}{|\vec{k}|^2|\vec{k}+\vec{q}|}.
\end{equation}
We now note that the integrand is symmetric under rotations of $\vec{k}$ about
the axis along $\vec{q}$.  Because of this, we may make the replacement,
\begin{equation}
(\vec{k}+\vec{q})\cdot\vec{\gamma}\to\frac{(\vec{k}+\vec{q})\cdot\vec{q}}{|\vec{q}|^2}\vec{q}\cdot\vec{\gamma},
\end{equation}
thus obtaining
\begin{equation}
\Sigma_1(q)=i\frac{g^2}{2|\vec{q}|^2}\int\frac{d^3\vec{k}}{(2\pi)^3}\,\frac{(\vec{k}+\vec{q})\cdot\vec{q}}{|\vec{k}|^2|\vec{k}+\vec{q}|}\vec{q}\cdot\vec{\gamma}.
\end{equation}
We now switch to spherical coordinates, with the $+z$ axis in the direction of
$\vec{q}$.  The integral now becomes
\begin{eqnarray}
\Sigma_1(q)&=&i\frac{g^2}{2|\vec{q}|^2}\frac{1}{(2\pi)^2}\int_0^{\Lambda}d|\vec{k}|\,\int_0^{\pi}d\theta\,\sin{\theta} \cr
&\times&\frac{|\vec{q}|^2+|\vec{k}||\vec{q}|\cos{\theta}}{\sqrt{|\vec{k}|^2+|\vec{q}|^2+2|\vec{k}||\vec{q}|\cos{\theta}}}\vec{q}\cdot\vec{\gamma}.
\end{eqnarray}
Here, we impose a cutoff $\Lambda$ on the $|\vec{k}|$ integral because the integral
is divergent.  Na\"ively, we would expect this integral to have a linear divergence,
but we will see that it is only logarithmic.  Let us make the substitutions, $|\vec{k}|=|\vec{q}|\kappa$
and $x=\cos{\theta}$.  The integral now becomes
\begin{equation}
\Sigma_1(q)=\frac{1}{8}i\left (\frac{g}{\pi}\right )^2\int_0^{\Lambda/|\vec{q}|}d\kappa\,\int_{-1}^{1}dx\,\frac{1+\kappa x}{\sqrt{1+\kappa^2+2\kappa x}}\vec{q}\cdot\vec{\gamma}.
\end{equation}
These integrals can be performed exactly; doing so, we obtain Eq.\ \eqref{Eq:ElectronSE_FE}
in the main text.

\section{Second-order vertex diagrams} \label{App:SE_Vertex}

In this appendix, we compute the two-loop corrections to the vertex and show that $Z_\varphi=1$
identically to second order.

\subsection{Two-loop self-energy correction to vertex function}
\begin{figure}
\includegraphics[width=0.5\columnwidth]{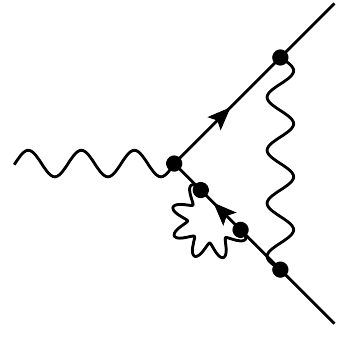}
\caption{Two-loop self-energy vertex diagram, ${\cal V}_{SE}(p,q)$.}
\label{Fig:V2SE}
\end{figure}
The first diagram that we compute is the two-loop self-energy diagram, Fig.\ \ref{Fig:V2SE}.
It is given by
\begin{equation}
{\cal V}_{SE}(p,q)=\int\frac{d^4k}{(2\pi)^4}i\gamma^0\frac{i}{\slashed k-\slashed p}i\gamma^0\frac{i}{\slashed k}\Sigma_1(k)\frac{i}{\slashed k}i\gamma^0\frac{g^2}{|\vec{k}-\vec{p}+\vec{q}|^2}.
\end{equation}
We isolate the logarithmic divergence by setting $p=0$:
\begin{equation}
{\cal V}_{SE}(0,q)\sim\int\frac{d^4k}{(2\pi)^4}i\gamma^0\frac{i}{\slashed k}i\gamma^0\frac{i}{\slashed k}\Sigma_1(k)\frac{i}{\slashed k}i\gamma^0\frac{g^2}{|\vec{k}+\vec{q}|^2}.
\end{equation}
Expanding the gamma matrix part, we find
\begin{eqnarray}
&&{\cal V}_{SE}(0,q)\sim-\frac{2g^2\alpha v_F}{3\pi}i\int\frac{d^4k}{(2\pi)^4}\frac{1}{k^6|\vec{k}+\vec{q}|^2} \cr
&&\times\left[\left(3v_Fk_0^2|\vec{k}|^2-v_F^3|\vec{k}|^4\right)\gamma^0\right. \cr
&&+\left.\left(k_0^3-3v_F^2k_0|\vec{k}|^2\right)\vec{k}\cdot\vec{\gamma}\right]
\left[\frac{4}{3} + \ln\frac{\Lambda}{|\vec{k}|}\right]
.
\end{eqnarray}
It turns out that the $k_0$ integration vanishes identically, and thus ${\cal V}_{SE}(p,q)$ is ultraviolet finite.

\subsection{Two-loop vacuum polarization bubble correction to vertex function}
\begin{figure}
\includegraphics[width=0.5\columnwidth]{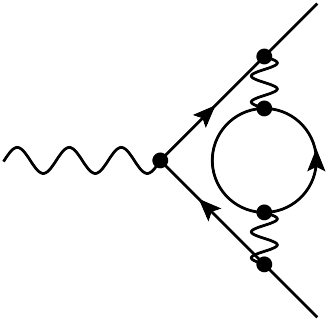}
\caption{Two-loop vacuum polarization bubble vertex diagram, ${\cal V}_B(p,q)$.}
\label{Fig:V2B}
\end{figure}
Next, we consider the vertex function diagram with the vacuum polarization bubble inserted in the internal Coulomb
line, Fig.\ \ref{Fig:V2B}, and we again set $p=0$ to extract the divergence:
\begin{equation}
{\cal V}_B(0,q)=\int\frac{d^4k}{(2\pi)^4}i\gamma^0\frac{i}{\slashed k}i\gamma^0\frac{i}{\slashed k}i\gamma^0\frac{g^4}{|\vec{k}+\vec{q}|^4}\Pi_B(k+q).
\end{equation}
Since the divergent part of $\Pi_B$ does not depend on frequency, the integration over $k_0$ for this term will be identical to what we had for the one-loop vertex function diagram [Eq.~(\ref{eq:vertex})], and it will thus vanish. Thus, the only potential divergence comes from the finite term in $\Pi_B$:
\begin{equation}
{\cal V}_B(0,q)\sim\frac{ig^4N}{12\pi^2v_F}\gamma^0\int\frac{d^4k}{(2\pi)^4}\frac{k_0^2-v_F^2|\vec{k}|^2}{k^4|\vec{k}|^2}\ln\frac{1+z^2}{4}.
\end{equation}
Rewriting $k_0$ in terms of $z$ and performing the trivial integration over $\vec{k}$, we find
\begin{equation}
{\cal V}_B(0,q)\sim\frac{ig^4N}{48\pi^5v_F^2}\gamma^0\int_{-\infty}^\infty dz\frac{z^2-1}{(1+z^2)^2}\ln\frac{1+z^2}{4}\ln\frac{\Lambda}{|\vec{q}|}.
\end{equation}
The integral over $z$ evaluates to $\pi$, leaving us with
\begin{equation}
{\cal V}_B(0,q)=i\frac{N}{3\pi^2}\alpha^2\gamma^0\ln\frac{\Lambda}{|\vec{q}|}+\hbox{finite}.
\end{equation}

\subsection{Two-loop parallel Coulomb line correction to vertex function}
\begin{figure}
\includegraphics[width=0.5\columnwidth]{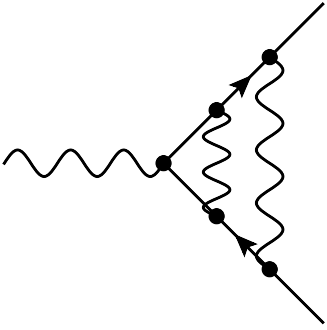}
\caption{Two-loop parallel Coulomb line vertex diagram, ${\cal V}_{PC}(q)$.}
\label{Fig:V2PC}
\end{figure}
We now consider the diagram with two parallel Coulomb lines, Fig.\ \ref{Fig:V2PC}.  The expression for
this diagram is
\begin{eqnarray}
&&{\cal V}_{PC}(p,q)=\int\frac{d^4k}{(2\pi)^4}\int\frac{d^4\ell}{(2\pi)^4}i\gamma^0\frac{i}{\slashed q+\slashed \ell}i\gamma^0\frac{i}{\slashed q+\slashed k+\slashed \ell} \cr
&&\times i\gamma^0\frac{i}{\slashed p+\slashed q+\slashed k+\slashed \ell}i\gamma^0\frac{i}{\slashed p+\slashed q+\slashed \ell}i\gamma^0\frac{g^2}{|\vec{k}|^2}\frac{g^2}{|\vec{\ell}|^2}.
\end{eqnarray}
We isolate the potential logarithmic divergence by setting $p=q=0$ in the integrand. If we then redefine $k\to k-\ell$, then it becomes straightforward to perform the integration over $k_0$:
\begin{equation}
\int dk_0\frac{\gamma^0\slashed k\gamma^0\slashed k}{k^4}=\int dk_0\frac{k_0^2-v_F^2|\vec{k}|^2-2k_0v_F\vec{k}\cdot\vec{\gamma}\gamma^0}{k^4}=0.
\end{equation}
We conclude that this diagram has no ultraviolet divergence.

\subsection{Two-loop crossed Coulomb line correction to vertex function}
\begin{figure}
\includegraphics[width=0.5\columnwidth]{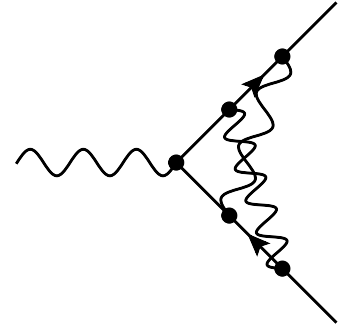}
\caption{Two-loop crossed Coulomb line vertex diagram, ${\cal V}_{CC}(q)$.}
\label{Fig:V2CC}
\end{figure}
Here, we consider the diagram with two crossed Coulomb lines, Fig.\ \ref{Fig:V2CC}.  The expression that
we obtain is
\begin{eqnarray}
&&{\cal V}_{CC}(p,q)=\int\frac{d^4k}{(2\pi)^4}\int\frac{d^4\ell}{(2\pi)^4}i\gamma^0\frac{i}{\slashed q+\slashed \ell}i\gamma^0\frac{i}{\slashed q+\slashed k+\slashed \ell} \cr
&&\times i\gamma^0\frac{i}{\slashed p+\slashed q+\slashed k+\slashed \ell}i\gamma^0\frac{i}{\slashed p+\slashed q+\slashed k}i\gamma^0\frac{g^2}{|\vec{k}|^2}\frac{g^2}{|\vec{\ell}|^2}.
\end{eqnarray}
We again set $p=q=0$, redefine $\ell\to-\ell$ and then $k\to k+\ell$ to obtain
\begin{equation}
{\cal V}_{CC}(p,q)\sim-ig^4\int\frac{d^4k}{(2\pi)^4}\int\frac{d^4\ell}{(2\pi)^4}\frac{\gamma^0\slashed \ell\gamma^0\slashed k\gamma^0\slashed k\gamma^0(\slashed k+\slashed\ell)\gamma^0}{\ell^2k^4(k+\ell)^2|\vec{k}+\vec{\ell}|^2|\vec{\ell}|^2}.
\end{equation}
Expanding out the gamma matrix product and throwing away terms which integrate to zero, we are left with
\begin{eqnarray}
&&{\cal V}_{CC}(p,q)\sim-ig^4\gamma^0\int\frac{d^4k}{(2\pi)^4}\int\frac{d^4\ell}{(2\pi)^4} \cr
&&\times\frac{1}{\ell^2k^4(k+\ell)^2|\vec{k}+\vec{\ell}|^2|\vec{\ell}|^2}\left\{(k_0^2-v_F^2|\vec{k}|^2)(\ell_0^2-v_F^2|\vec{\ell}|^2)\right. \cr
&&-\left.3k_0\ell_0v_F^2|\vec{k}|^2+k_0^3\ell_0 -\left[4k_0\ell_0+3k_0^2-v_F^2|\vec{k}|^2\right]v_F^2\vec{k}\cdot\vec{\ell}\right\}. \nonumber \\
\end{eqnarray}
Performing the integrations over $k_0$ and $\ell_0$, this becomes
\begin{eqnarray}
&&{\cal V}_{CC}(p,q)\sim-i\gamma^0\frac{g^4}{4v_F^2}\int\frac{d^3k}{(2\pi)^3}\int\frac{d^3\ell}{(2\pi)^3} \frac{1}{|\vec{\ell}|^2|\vec{k}+\vec{\ell}|^3} \cr
&&\times\frac{|\vec{k}|+|\vec{\ell}|-|\vec{k}+\vec{\ell}|}{(|\vec{k}|+|\vec{\ell}|+|\vec{k}+\vec{\ell}|)^2}\left(1+\frac{\vec{k}\cdot\vec{\ell}}{|\vec{k}||\vec{\ell}|}\right).
\end{eqnarray}
Next, we switch to prolate spherical coordinates in $\vec{k}$, yielding
\begin{eqnarray}
&&{\cal V}_{CC}(p,q)\sim-i\gamma^0\frac{g^4}{16\pi^2v_F^2}\int\frac{d^3\ell}{(2\pi)^3}\frac{1}{|\vec{\ell}|^3}\int d\mu d\nu \cr
&&\times\sinh\mu\sin\nu\frac{1-\cos\nu}{(1+\cosh\mu)^2} \cr
&&\times\frac{\cosh\mu\cos\nu+\cosh\mu-\cos\nu-1}{(\cosh\mu+\cos\nu)^2}.
\end{eqnarray}
Performing the integrations over $\vec{\ell}$ and $\nu$, we find
\begin{eqnarray}
&&{\cal V}_{CC}(p,q)\sim i\gamma^0\frac{g^4}{8\pi^4v_F^2}\ln\frac{\Lambda}{|\vec{q}|}\int d\mu\frac{\sinh\mu}{(1+\cosh\mu)^2} \cr
&&\times(\cosh\mu-1)(1+\cosh\mu\ln\tanh\frac{\mu}{2}).
\end{eqnarray}
The $\mu$-integral evaluates to $-\tfrac{10-\pi^2}{4}$, giving
\begin{equation}
{\cal V}_{CC}(p,q)=-i\gamma^0\frac{10-\pi^2}{2\pi^2}\alpha^2\ln\frac{\Lambda}{|\vec{q}|}+\hbox{finite}.
\end{equation}

\subsection{Two-loop vertex correction to vertex function}
\begin{figure}
\includegraphics[width=0.5\columnwidth]{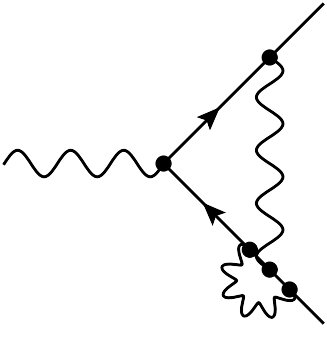}
\caption{Two-loop vertex correction vertex diagram, ${\cal V}_V(q)$.}
\label{Fig:V2V}
\end{figure}
Finally, we consider the two-loop vertex correction diagram, Fig.\ \ref{Fig:V2V}.  This diagram
gives us
\begin{eqnarray}
{\cal V}_{V}(p,q)&=&\int\frac{d^4k}{(2\pi)^4}\int\frac{d^4\ell}{(2\pi)^4}i\gamma^0\frac{i}{\slashed q+\slashed \ell}i\gamma^0\frac{i}{\slashed q+\slashed k+\slashed \ell} \cr
&\times&i\gamma^0\frac{i}{\slashed q+\slashed k}i\gamma^0\frac{i}{\slashed p+\slashed q+\slashed k}i\gamma^0\frac{g^2}{|\vec{k}|^2}\frac{g^2}{|\vec{\ell}|^2} \cr
&\sim&ig^4\gamma^0\int\frac{d^4k}{(2\pi)^4}\int\frac{d^4\ell}{(2\pi)^4}\frac{\slashed\ell\gamma^0(\slashed k+\slashed\ell)\gamma^0\slashed k\gamma^0\slashed k\gamma^0}{\ell^2(k+\ell)^2k^4|\vec{k}|^2|\vec{\ell}|^2}. \nonumber \\
\end{eqnarray}
We proceed by observing that
\begin{equation}
\gamma^0(\slashed k+\slashed\ell)\gamma^0\slashed k\gamma^0\slashed k\gamma^0 = \gamma^0\slashed k\gamma^0\slashed k\gamma^0(\slashed k+\slashed\ell)\gamma^0+\dots,
\end{equation}
where the $\ldots$ represent terms which vanish upon integration over $k$ and $\ell$. This means that the product of gamma matrices evaluates to precisely what he had in the previous subsection:
\begin{eqnarray}
&&{\cal V}_{V}(p,q)\sim ig^4\gamma^0\int\frac{d^4k}{(2\pi)^4}\int\frac{d^4\ell}{(2\pi)^4}\frac{1}{\ell^2k^4(k+\ell)^2|\vec{k}|^2|\vec{\ell}|^2} \cr
&&\times\left\{(k_0^2-v_F^2|\vec{k}|^2)(\ell_0^2-v_F^2|\vec{\ell}|^2)-3k_0\ell_0v_F^2|\vec{k}|^2\right. \cr
&&+\left.k_0^3\ell_0 -\left[4k_0\ell_0+3k_0^2-v_F^2|\vec{k}|^2\right]v_F^2\vec{k}\cdot\vec{\ell}\right\},
\end{eqnarray}
and the integrations over $k_0$ and $\ell_0$ are also identical:
\begin{eqnarray}
&&{\cal V}_{V}(p,q)\sim i\gamma^0\frac{g^4}{4v_F^2}\int\frac{d^3k}{(2\pi)^3}\int\frac{d^3\ell}{(2\pi)^3} \frac{1}{|\vec{k}|^2|\vec{\ell}|^2|\vec{k}+\vec{\ell}|} \cr
&&\times\frac{|\vec{k}|+|\vec{\ell}|-|\vec{k}+\vec{\ell}|}{(|\vec{k}|+|\vec{\ell}|+|\vec{k}+\vec{\ell}|)^2}\left(1+\frac{\vec{k}\cdot\vec{\ell}}{|\vec{k}||\vec{\ell}|}\right).
\end{eqnarray}
Again using prolate spherical coordinates, we find
\begin{eqnarray}
&&{\cal V}_{V}(p,q)\sim i\gamma^0\frac{g^4}{16\pi^2v_F^2}\int\frac{d^3\ell}{(2\pi)^3}\frac{1}{|\vec{\ell}|^3}\int d\mu d\nu \sinh\mu\sin\nu \cr
&&\frac{1-\cos\nu}{(1+\cosh\mu)^2} \frac{\cosh\mu\cos\nu+\cosh\mu-\cos\nu-1}{(\cosh\mu-\cos\nu)^2}.
\end{eqnarray}
This is nearly identical to the analogous expression for ${\cal V}_{CC}(p,q)$, except for the overall sign and the sign in the denominator of the final factor in the integrand. The latter comes from the fact that now we had $|\vec{k}|^2$ in the denominator instead of $|\vec{k}+\vec{\ell}|^2$ as before. Despite this sign difference in the denominator, the integral over $\nu$ evaluates to precisely the same value as before, so that the only difference compared with ${\cal V}_{CC}(p,q)$ is the overall sign:
\begin{equation}
{\cal V}_{V}(p,q)=i\gamma^0\frac{10-\pi^2}{2\pi^2}\alpha^2\ln\frac{\Lambda}{|\vec{q}|}+\hbox{finite}.
\end{equation}
It would then seem that the divergences of ${\cal V}_{CC}(p,q)$ and ${\cal V}_{V}(p,q)$ would cancel each other. However, ${\cal V}_{V}(p,q)$ receives an extra symmetry factor of 2.

\subsection{Two-loop vertex function}

Summing up the various contributions to the vertex function, we find the total result
\begin{eqnarray}
&&{\cal V}^{(2)}(p,q)={\cal V}_B(p,q)+{\cal V}_{CC}(p,q)+2{\cal V}_V(p,q)+\hbox{finite}\cr &&=i\gamma^0\left(\frac{N}{3\pi^2}+\frac{10-\pi^2}{2\pi^2}\right)\alpha^2\ln\frac{\Lambda}{|\vec{q}|}+\hbox{finite}.\label{twoloopvertex}
\end{eqnarray}
This implies that the relevant counterterm coefficient is
\begin{equation}
\delta_v=\left(\frac{N}{3\pi^2}+\frac{10-\pi^2}{2\pi^2}\right)\alpha^2\ln\frac{\Lambda}{\mu}+{\cal O}(\alpha^3)=\delta_0+{\cal O}(\alpha^3).
\end{equation}
As pointed out in the main text, this implies that $Z_\varphi=1$ to second order in $\alpha$. This confirms the requirement $\delta_v=\delta_0$ stemming from gauge invariance.

\section{Callan-Symanzik RG analysis} \label{App:CSAnalysis}
Here, we present an alternative means of deriving the RG equations for our system, namely
through direct application of the Callan-Symanzik equation.  We will find that our results
agree with those derived in the main text, providing a valuable consistency check.

\subsection{Photon two-point function}

The Callan-Symanzik equation for the photon two-point function expanded to third order in $\alpha$ reads
\begin{eqnarray}
&&\left(\mu\partial_\mu+\beta_\alpha\partial_\alpha+v_F\gamma_{v_F}\partial_{v_F}+2\gamma_\varphi\right) \cr
&&\times\left[V_q+V_q^2\Pi^{(0)}+V_q^2\Pi^{(1)}+V_q^3(\Pi^{(0)})^2\right]=0.\label{CSeqn}
\end{eqnarray}
Here, $V_q=4\pi\alpha v_F/|\vec{q}|^2$ is the Coulomb interaction, and $\Pi^{(0)}$ and $\Pi^{(1)}$ are the
leading- and next-to-leading-order vacuum polarization functions. Summing up the relevant diagrams and their
respective counterterms from Secs.~\ref{Sec:Polarization} and \ref{Sec:Renormalization} yields the following expressions for the renormalized vacuum polarization functions:
\begin{equation}
\Pi^{(0)}=-\frac{N|\vec{q}|^2}{6\pi^2v_F}\ln\left (\frac{\mu}{|\vec{q}|}\right )+\frac{N|\vec{q}|^2}{12\pi^2v_F}\ln\left (\frac{1+z^2}{4}\right ),
\end{equation}
\begin{eqnarray}
\Pi^{(1)}&=&\frac{N\alpha|\vec{q}|^2}{18\pi^3v_F}\ln^2\left (\frac{\mu}{|\vec{q}|}\right )-\frac{N\alpha|\vec{q}|^2}{18\pi^3v_F}\left[\ln\left (\frac{1+z^2}{4}\right )\right. \cr
&+&\left.\frac{2z^2}{1+z^2}\right]\ln\left (\frac{\mu}{|\vec{q}|}\right )-\frac{N\alpha|\vec{q}|^2}{8\pi^3v_F}\left(C-\frac{32}{27}\right)\ln\left (\frac{\mu}{|\vec{q}|}\right ). \nonumber \\
\end{eqnarray}
In $\Pi^{(1)}$, the term involving the constant $C$ comes from the two-loop vertex correction to the polarizability.
All other terms in $\Pi^{(1)}$ come from the self-energy correction to the polarizability.

We can use Eq.\ \eqref{CSeqn} to derive equations for $\beta_\alpha$, $\gamma_{v_F}$, and $\gamma_\varphi$ by
plugging in $\Pi^{(0)}$ and $\Pi^{(1)}$ and solving the resulting equation order by order in $\alpha$.  The
first-order equation we obtain is trivial:
\begin{equation}
\beta_\alpha^{(1)}\partial_\alpha\frac{4\pi\alpha v_F}{|\vec{q}|^2}=0,
\end{equation}
and simply tells us that the first-order beta function for $\alpha$ is zero: $\beta_\alpha^{(1)}=0$.

The second-order equation reads as
\begin{equation}
V_q^2\mu\partial_\mu\Pi^{(0)}+\beta_\alpha^{(2)}\partial_\alpha V_q+v_F\gamma_{v_F}^{(1)}\partial_{v_F}V_q+2\gamma_\varphi^{(1)}V_q=0.\label{CSeqnPi2}
\end{equation}
Plugging in the expressions for $V_q$ and $\Pi^{(0)}$, we obtain the following relation:
\begin{equation}
\beta_\alpha^{(2)}=\frac{2N}{3\pi}\alpha^2-\alpha\gamma_{v_F}^{(1)}-2\alpha\gamma_\varphi^{(1)}.\label{betaalpha2}
\end{equation}

Next, we consider the third-order equation which follows from Eq.\ \eqref{CSeqn}:
\begin{eqnarray}
&&V_q^2\mu\partial_\mu\Pi^{(1)}+V_q^3\mu\partial_\mu(\Pi^{(0)})^2+\beta_\alpha^{(3)}\partial_\alpha V_q \cr
&&+\beta_\alpha^{(2)}\partial_\alpha(V_q^2\Pi^{(0)})+v_F\gamma_{v_F}^{(2)}\partial_{v_F}V_q+v_F\gamma_{v_F}^{(1)}\partial_{v_F}(V_q^2\Pi^{(0)}) \cr
&&+2\gamma_\varphi^{(2)}V_q=0.\label{CSeqnPi3}
\end{eqnarray}
Plugging in $\Pi^{(0)}$, $\Pi^{(1)}$, $\beta_\alpha^{(2)}$, $\gamma_{v_F}^{(1)}$, we find that the $z$ dependence cancels out, and we are left with the following relation:
\begin{equation}
\beta_\alpha^{(3)}+\alpha\gamma_{v_F}^{(2)}+2\alpha\gamma_\varphi^{(2)}-\frac{\alpha^3}{2\pi^2}N\left(C-\frac{32}{27}\right)=0.\label{betaalpha3}
\end{equation}
The fact that the $z$ dependence cancels out of the Callan-Symanzik equation constitutes a nontrivial consistency check of our results for the vacuum polarizability.

In order to obtain explicit expressions for $\beta_\alpha^{(2)}$ and $\beta_\alpha^{(3)}$, we must first calculate the first- and second-order contributions to $\gamma_{v_F}$ and $\gamma_\varphi$ from the Callan-Symanzik equations for the electron two-point function and electron-photon three-point function.

\subsection{Electron two-point function}

We can obtain $\gamma_{v_F}$ as well as the scaling function for the electron field strength, $\gamma_\psi$, from the Callan-Symanzik equation for the electron two-point function expanded to second order:
\begin{eqnarray}
&&\left(\mu\partial_\mu+\beta_\alpha\partial_\alpha+v_F\gamma_{v_F}\partial_{v_F}+2\gamma_\psi\right)\left[\frac{i}{\slashed q}+\frac{i}{\slashed q}\Sigma_1\frac{i}{\slashed q}+\frac{i}{\slashed q}\Sigma_2\frac{i}{\slashed q}\right. \cr
&&+\left.\frac{i}{\slashed q}\Sigma_1\frac{i}{\slashed q}\Sigma_1\frac{i}{\slashed q}\right]=0, \label{CSeqnSigma}
\end{eqnarray}
This equation involves the renormalized first- and second-order electron self-energies, which we obtain by summing up the results from Secs.~\ref{Sec:SelfEnergy_SE} and \ref{Sec:Renormalization}:
\begin{equation}
\Sigma_1=\frac{2\alpha}{3\pi}iv_F\vec{q}\cdot\vec{\gamma}\ln\left (\frac{\mu}{|\vec{q}|}\right ).
\end{equation}
\begin{eqnarray}
\Sigma_2&=&-\frac{2N}{9\pi^2}i\alpha^2v_F\vec{q}\cdot\vec{\gamma}\ln^2\left (\frac{\mu}{|\vec{q}|}\right ) \cr
&+&i\left[\left(\frac{15+N}{3\pi^2}-\frac{1}{2}\right)q_0\gamma^0\right. \cr
&+&\left.\left(\frac{45+N}{9\pi^2}-\frac{1}{2}\right)v_F\vec{q}\cdot\vec{\gamma}\right]\alpha^2\ln\left (\frac{\mu}{|\vec{q}|}\right ). \nonumber \\
\end{eqnarray}
Plugging these results into the first-order equation that follows from Eq.\ \eqref{CSeqnSigma},
\begin{equation}
\left(v_F\gamma_{v_F}^{(1)}\partial_{v_F}+\gamma_\psi^{(1)}\right)\frac{i}{\slashed q}+\mu\partial_\mu\frac{i}{\slashed q}\Sigma_1(q)\frac{i}{\slashed q}=0,\label{CSeqnSigma1}
\end{equation}
and using the identity,
\begin{equation}
\partial_{v_F}\frac{i}{\slashed q}=\frac{i}{\slashed q}i\vec{q}\cdot\vec{\gamma}\frac{i}{\slashed q},\label{vFderividentity}
\end{equation}
yields the following leading-order behavior of $\gamma_{v_F}$ and $\gamma_\psi$:
\begin{equation}
\gamma_{v_F}^{(1)}=-\frac{2\alpha}{3\pi},\qquad \gamma_\psi^{(1)}=0.\label{vF1andgammapsi1}
\end{equation}

We now turn our attention to the effects of the two-loop corrections to the electron self-energy. Collecting all the second-order terms in the Callan-Symanzik equation gives
\begin{eqnarray}
&&\frac{i}{\slashed q}\mu\partial_\mu\Sigma_2\frac{i}{\slashed q}+\frac{i}{\slashed q}\mu\partial_\mu\Sigma_1\frac{i}{\slashed q}\Sigma_1\frac{i}{\slashed q}+\frac{i}{\slashed q}\Sigma_1\frac{i}{\slashed q}\mu\partial_\mu\Sigma_1\frac{i}{\slashed q} \cr
&&+\beta_\alpha^{(2)}\frac{i}{\slashed q}\partial_\alpha\Sigma_1\frac{i}{\slashed q}+v_F\gamma_{v_F}^{(2)}\partial_{v_F}\frac{i}{\slashed q}+v_F\gamma_{v_F}^{(1)}\partial_{v_F}\left[\frac{i}{\slashed q}\Sigma_1\frac{i}{\slashed q}\right] \cr
&&+2\gamma_\psi^{(2)}\frac{i}{\slashed q}=0.\label{CSeqnSigma2}
\end{eqnarray}
Using the identity from Eq.~(\ref{vFderividentity}) and collecting all the finite terms, we find
\begin{eqnarray}
&&\alpha^2\left(\frac{15+N}{3\pi^2}-\frac{1}{2}\right)\frac{i}{\slashed q}iq_0\gamma^0\frac{i}{\slashed q}+\alpha^2\left(\frac{45+N}{9\pi^2}-\frac{1}{2}\right) \cr
&&\times\frac{i}{\slashed q}iv_F\vec{q}\cdot\vec{\gamma}\frac{i}{\slashed q} +\gamma_{v_F}^{(2)}\frac{i}{\slashed q}iv_F\vec{q}\cdot\vec{\gamma}\frac{i}{\slashed q}+2\gamma_\psi^{(2)}\frac{i}{\slashed q}=0.
\end{eqnarray}
Writing
\begin{equation}
\frac{i}{\slashed q}=-\frac{i}{\slashed q}iq_0\gamma^0\frac{i}{\slashed q}-\frac{i}{\slashed q}iv_F\vec{q}\cdot\vec{\gamma}\frac{i}{\slashed q},
\end{equation}
and collecting like terms, we obtain the following two relations:
\begin{eqnarray}
\alpha^2\left(\frac{15+N}{3\pi^2}-\frac{1}{2}\right)-2\gamma_\psi^{(2)}&=&0,\nonumber\\
\alpha^2\left(\frac{45+N}{9\pi^2}-\frac{1}{2}\right)+\gamma_{v_F}^{(2)}-2\gamma_\psi^{(2)}&=&0,
\end{eqnarray}
which imply
\begin{eqnarray}
\gamma_\psi^{(2)}&=&\alpha^2\left(\frac{15+N}{6\pi^2}-\frac{1}{4}\right),\nonumber\\
\gamma_{v_F}^{(2)}&=&\frac{2N\alpha^2}{9\pi^2}.
\end{eqnarray}

\subsection{Electron-photon three-point function}

To complete the analysis, we must also obtain $\gamma_\varphi$, the scaling function for the photon field strength. We can compute this from the Callan-Symanzik equation for the electron-photon three-point function. Summing up the various contributions to the second-order vertex function, we find the renormalized result
\begin{equation}
{\cal V}^{(2)}(p,q)=i\gamma^0\left(\frac{N}{3\pi^2}+\frac{10-\pi^2}{2\pi^2}\right)\alpha^2\ln\frac{\mu}{|\vec{q}|}+\hbox{finite},\label{twoloopvertexrenorm}
\end{equation}
while we found in Sec.~\ref{Sec:Renormalization} that the first-order vertex function has no divergence, ${\cal V}^{(1)}(p,q)=\hbox{finite}$, and is thus independent of the renormalization scale. The second-order Callan-Symanzik equation reads as
\begin{eqnarray}
&&\left(\beta_\alpha^{(2)}\partial_\alpha+v_F\gamma_{v_F}^{(1)}\partial_{v_F}+\gamma^{(1)}_\varphi+2\gamma_\psi^{(1)}\right)\Gamma(p,q) \cr && +\mu\partial_\mu\bigg[V_p\Pi^{(0)}(p)\Gamma(p,q)+V_p\frac{i}{\slashed q}{\cal V}^{(1)}(p,q)\frac{i}{\slashed p+\slashed q} \cr && +\Gamma(p,q)\Sigma_1(p+q)\frac{i}{\slashed p+\slashed q}+\frac{i}{\slashed q}\Sigma_1(q)\Gamma(p,q)\bigg]=0, \cr &&\label{CSeqnV2}
\end{eqnarray}
where $V_p=4\pi\alpha v_F/|\vec{p}|^2$, and where we have defined
\begin{equation}
\Gamma(p,q)\equiv V_p\frac{i}{\slashed q}i\gamma^0\frac{i}{\slashed p+\slashed q}.
\end{equation}
Equation \eqref{CSeqnV2} can be drastically simplified by using the Callan-Symanzik equations quoted earlier, namely, Eqs.~\eqref{CSeqnSigma1} and \eqref{CSeqnPi2}, yielding
\begin{equation}
\left(-\gamma_\varphi^{(1)}-2\gamma_\psi^{(1)}\right)\Gamma(p,q)=0,
\end{equation}
which implies
\begin{equation}
\gamma_\varphi^{(1)}=-2\gamma_\psi^{(1)}=0.
\end{equation}
Inputting this result and the result for $\gamma_{v_F}^{(1)}$ from Eq.~\eqref{vF1andgammapsi1} into Eq.\ \eqref{betaalpha2} then produces an explicit expression for the leading-order
beta function for $\alpha$:
\begin{equation}
\beta_\alpha^{(2)}=\frac{2(N+1)}{3\pi}\alpha^2.
\end{equation}

To obtain $\gamma_\varphi^{(2)}$, we need to consider the Callan-Symanzik equation at third order in $\alpha$. The full form of this equation is very complicated; however, we may make use of Eqs.~(\ref{CSeqnPi3}) and (\ref{CSeqnSigma2}) to simplify it considerably, with the result
\begin{eqnarray}
&&\left(-\gamma_\varphi^{(2)}-2\gamma_\psi^{(2)}\right)i\gamma^0+\left(\beta_\alpha^{(2)}\partial_\alpha+v_F\gamma_{v_F}^{(1)}\partial_{v_F}\right){\cal V}^{(1)}(p,q)\cr && +\mu\partial_\mu{\cal V}^{(2)}(p,q)=0.
\end{eqnarray}
Although we have not directly computed ${\cal V}^{(1)}(p,q)$, we can see from this equation that it must be independent of momentum since there is no other momentum dependence in the equation. This would mean that ${\cal V}^{(1)}(p,q)={\cal V}^{(1)}(0,0)$. However, we have shown in Eq.~\eqref{eq:vertex} that the latter vanishes, hence, ${\cal V}^{(1)}(p,q)=0$. Using Eq.~\eqref{twoloopvertexrenorm}, we then have
\begin{equation}
\gamma^{(2)}_\varphi=-2\gamma_\psi^{(2)}+\left(\frac{N}{3\pi^2}+\frac{10-\pi^2}{2\pi^2}\right)\alpha^2=0.
\end{equation}
Since we now have $\gamma_\varphi^{(2)}$ and $\gamma_\psi^{(2)}$, we can go back and determine $\beta_\alpha^{(3)}$ from Eq.~\eqref{betaalpha3}:
\begin{eqnarray}
\beta_\alpha^{(3)}&=&-\alpha\gamma_{v_F}^{(2)}-2\alpha\gamma_\varphi^{(2)}+\frac{\alpha^3}{2\pi^2}N\left(C-\frac{32}{27}\right) \cr
&=&-\frac{(44-27C)N\alpha^3}{54\pi^2}.
\end{eqnarray}
We thus find that we obtain all the same results as we did in Sec.~\ref{Sec:Renormalization} of the main text.

\section{Drude-Boltzmann form of the conductivity}\label{sec:drude}

\begin{center}
\begin{figure}[t]
\includegraphics[width=\columnwidth]{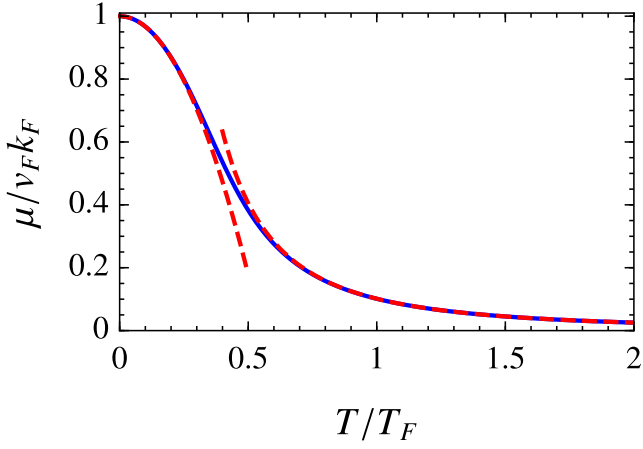}
\caption{Chemical potential vs. temperature as obtained by solving Eq.~\eqref{eq:condmu}. Dashed lines indicate the low- and high-temperature limits~\eqref{eq:muT}.}
\label{fig:muvsT}
\end{figure}
\end{center}

In this appendix, we derive an expression for the Drude part of the conductivity without renormalization effects using kinetic theory. Denote the distribution function momentum ${\vec q}$ at a node $a$ with chirality $s$ by $f_{as}({\vec k}, t)$. In the absence of an external field, the noninteracting distribution is given by the Fermi-Dirac distribution $f_{as}^0({\vec q}) = [\exp[(s v_F |{\vec q}| - \mu)/k_BT]+1]^{-1}$. The full distribution in the presence of an external electric field exerting a force ${\vec F} = \hbar \tfrac{\partial {\vec k}}{\partial t} = e {\vec E}$ on the electrons solves the Boltzmann equation
\begin{align}
\biggl[\frac{\partial}{\partial t} + {\vec F} \cdot \frac{\partial}{\partial {\vec q}}\biggr] f_{as}({\vec q},t) &= - \frac{f_{as}({\vec q},t) - f_{as}^0({\vec q},t)}{\tau_{\vec q}} ,
\end{align}
where we assumed a relaxation time ansatz for the collision integral. In linear response, the distribution function can be expanded as
\begin{align}
f_{as}({\vec q}, \omega) &= 2 \pi \delta(\omega) f_{as}^0({\vec q}) + g_{as}({\vec q}, \omega) ,
\end{align}
where $g({\vec q}, \omega)$ is determined from the Boltzmann equation:
\begin{align}
\biggl[i \omega + \frac{1}{\tau_{\vec q}}\biggr] g_{as}({\vec q}, \omega) &= - \frac{e {\vec E}}{\hbar} \cdot \frac{\partial}{\partial {\vec q}} f_{as}^0({\vec q})
\end{align}
Hence,
\begin{align}
g_{as}({\vec q}, \omega) &= \biggl[i \omega + \frac{1}{\tau_{\vec q}}\biggr]^{-1} \biggl(- \frac{e {\vec E}}{\hbar} \cdot \frac{\partial}{\partial {\vec q}} f_{as}^0({\vec q})\biggr) ,
\end{align}
and the current is (assuming a constant scattering time)
\begin{align}
&{\vec j}(\omega) = v_F e \sum_{as} \int \frac{d^3\vec{q}}{(2\pi)^3} \frac{{\vec q}}{|{\vec q}|} f_{as}({\vec q}, \omega)
\nonumber \\
&= \frac{1}{i \omega + 1/\tau} \times \frac{v_F^2 e^2 g}{3 \pi^2 (\hbar v_F)^3} \int_0^\infty d\varepsilon \, \varepsilon^2 \, \biggl(- \frac{\partial f^0(\varepsilon)}{\partial \varepsilon}\biggr) \cdot  {\vec E} ,
\end{align}
The conductivity is
\begin{align}
\sigma(\omega) &= \frac{1}{i \omega + 1/\tau} \times \frac{v_F^2 e^2 g}{3 \pi^2 (\hbar v_F)^3} \int_0^\infty d\varepsilon \, \varepsilon^2 \, \biggl(- \frac{\partial f^0(\varepsilon)}{\partial \varepsilon}\biggr) .
\end{align}
This yields the dielectric function
\begin{align}
\varepsilon(\omega) &= \frac{4 \pi i}{\omega} \sigma(\omega) = \frac{\omega_p^2}{\omega^2 - i \omega \gamma} ,
\end{align}
where $\gamma = 1/\tau$ and $\omega_p^2$ denotes the Drude weight
\begin{align}
\omega_d^2= 4 \pi e^2 \frac{2 v_F^2}{3} \frac{D_F}{\varepsilon_F^2} \int_0^\infty d\varepsilon \, \varepsilon^2 \, \biggl(- \frac{\partial f^0(\varepsilon)}{\partial \varepsilon} \biggr) . \label{eq:drudeweight}
\end{align}
Here, $D_F = N \varepsilon_F^2/[\pi^2 (\hbar v_F)^3]$ is the density of states. Aside from the temperature dependence of the energy average, there is an additional temperature dependence of the chemical potential, which starts at $\mu = \varepsilon_F$ at zero temperature and then falls off to zero at high temperature. The extrinsic chemical potential $\mu(T)$ at finite temperature is fixed by the relation
\begin{align}
2 N \int \frac{d^3\vec{k}}{(2\pi)^3} n_+(\vec{k}) + 2 N \int \frac{d^3\vec{k}}{(2\pi)^3} [n_-(\vec{k}) - 1] &= n , \label{eq:def}
\end{align}
where $n_s$ are the Fermi-Dirac distributions with energy $\varepsilon_s(k) = s \hbar v_F k$ and $n=g k_F^3/6 \pi^2$ is the zero temperature carrier density. Equation \eqref{eq:def} reduces to
\begin{align}
\frac{\mu}{\varepsilon_F} \biggl[ \pi^2 \biggl(\frac{T}{T_F}\biggr)^2 + \biggl(\frac{\mu}{\varepsilon_F}\biggr)^2\biggr] = 1 . \label{eq:condmu}
\end{align}
The asymptotic behavior of $\mu$ is
\begin{align}
\frac{\mu}{\varepsilon_F} &=
\begin{cases}
1 - \dfrac{\pi^2}{3} \biggl(\dfrac{T}{T_F}\biggr)^2, & T\ll T_F, \\[2ex]
\dfrac{1}{\pi^2} \biggl(\dfrac{T_F}{T}\biggr)^2, & T \gg T_F.
\end{cases} \label{eq:muT}
\end{align}
This behavior is illustrated in Fig.~\ref{fig:muvsT}.

The temperature dependence of the Drude weight~\eqref{eq:drudeweight} is determined both by the energy average and the temperature dependence of the chemical potential. At small and high temperature, we obtain
\begin{align}
\frac{1}{\varepsilon_F^2} \int_0^\infty d\varepsilon \, \varepsilon^2 \, \biggl(- \frac{\partial f^0(\varepsilon)}{\partial \varepsilon} \biggr) &= \begin{cases}
1 - \dfrac{\pi^2}{3} \biggl(\dfrac{T}{T_F}\biggr)^2 & T \ll T_F \\[2ex]
\dfrac{\pi^2}{6} \biggl(\dfrac{T}{T_F}\biggr)^2 & T \gg T_F
\end{cases} . \label{eq:drudeT}
\end{align}
In the Born approximation, the transport scattering time is given by
\begin{align}
&- \frac{f_a({\vec k}) - f_a^0({\vec k})}{\tau_{\vec k}} \nonumber \\
&= - \frac{2 \pi n_i}{\hbar} \int \frac{d^3\vec{k}'}{(2\pi)^3} \biggl\{ |\langle {\vec k} | V | {\vec k'} \rangle |^2 f_a({\vec k}) (1-f_a({\vec k'})) \nonumber \\
&\qquad  - |\langle {\vec k'} | V | {\vec k} \rangle |^2 f_a({\vec k'}) (1-f_a({\vec k})) \biggr\} \delta(\varepsilon_{\vec k} - \varepsilon_{\vec k'}) \nonumber \\
&= - \frac{2 \pi n_i}{\hbar} g_a({\bf k}) \int \frac{d^3\vec{k}'}{(2\pi)^3} \delta(\varepsilon_{\vec k} - \varepsilon_{\vec k'}) |\langle {\vec k} | V | {\vec k'} \rangle |^2 \bigl( 1 - \cos \theta \bigr)
\end{align}
and hence
\begin{align}
\frac{1}{\tau_{\vec k}} &= \frac{2 \pi n_i}{\hbar} \int \frac{d^3\vec{k}'}{(2\pi)^3} \delta(\varepsilon_{\vec k} - \varepsilon_{\vec k'}) |\langle {\vec k} | V | {\vec k}' \rangle |^2 \bigl( 1 - \cos \theta \bigr) .
\end{align}
Note that this transport scattering time is different from the quasiparticle lifetime derived from the imaginary part of the self-energy to this order in the interaction by the additional factor~\cite{DasSarmaPRB2015} of $(1 - \cos \theta)$. In the present case, we assumed that $\tau_{\vec q}$ is energy-independent.

\bibliography{bibweyl}

\begin{thebibliography}{88}
\expandafter\ifx\csname natexlab\endcsname\relax\def\natexlab#1{#1}\fi
\expandafter\ifx\csname bibnamefont\endcsname\relax
  \def\bibnamefont#1{#1}\fi
\expandafter\ifx\csname bibfnamefont\endcsname\relax
  \def\bibfnamefont#1{#1}\fi
\expandafter\ifx\csname citenamefont\endcsname\relax
  \def\citenamefont#1{#1}\fi
\expandafter\ifx\csname url\endcsname\relax
  \def\url#1{\texttt{#1}}\fi
\expandafter\ifx\csname urlprefix\endcsname\relax\def\urlprefix{URL }\fi
\providecommand{\bibinfo}[2]{#2}
\providecommand{\eprint}[2][]{\url{#2}}

\bibitem[{\citenamefont{Herring}(1937)}]{HerringPR1937}
\bibinfo{author}{\bibfnamefont{C.}~\bibnamefont{Herring}},
  \bibinfo{journal}{Phys. Rev.} \textbf{\bibinfo{volume}{52}},
  \bibinfo{pages}{365} (\bibinfo{year}{1937}).

\bibitem[{\citenamefont{Abrikosov and
  Beneslavski\u{i}}(1970)}]{AbrikosovJETP1971}
\bibinfo{author}{\bibfnamefont{A.~A.} \bibnamefont{Abrikosov}}
  \bibnamefont{and} \bibinfo{author}{\bibfnamefont{S.~D.}
  \bibnamefont{Beneslavski\u{i}}}, \bibinfo{journal}{Zh. Eksp. Teor. Fiz.}
  \textbf{\bibinfo{volume}{59}}, \bibinfo{pages}{1280} (\bibinfo{year}{1970}),
  \bibinfo{note}{[Sov. Phys. JETP {\bf 32}, 699 (1971)]}.

\bibitem[{\citenamefont{Wan et~al.}(2011)\citenamefont{Wan, Turner, Vishwanath,
  and Savrasov}}]{WanPRB2011}
\bibinfo{author}{\bibfnamefont{X.}~\bibnamefont{Wan}},
  \bibinfo{author}{\bibfnamefont{A.~M.} \bibnamefont{Turner}},
  \bibinfo{author}{\bibfnamefont{A.}~\bibnamefont{Vishwanath}},
  \bibnamefont{and} \bibinfo{author}{\bibfnamefont{S.~Y.}
  \bibnamefont{Savrasov}}, \bibinfo{journal}{Phys. Rev. B}
  \textbf{\bibinfo{volume}{83}}, \bibinfo{pages}{205101}
  (\bibinfo{year}{2011}).

\bibitem[{\citenamefont{Burkov and Balents}(2011)}]{BurkovPRL2011}
\bibinfo{author}{\bibfnamefont{A.~A.} \bibnamefont{Burkov}} \bibnamefont{and}
  \bibinfo{author}{\bibfnamefont{L.}~\bibnamefont{Balents}},
  \bibinfo{journal}{Phys. Rev. Lett.} \textbf{\bibinfo{volume}{107}},
  \bibinfo{pages}{127205} (\bibinfo{year}{2011}).

\bibitem[{\citenamefont{Young et~al.}(2012)\citenamefont{Young, Zaheer, Teo,
  Kane, Mele, and Rappe}}]{YoungPRL2012}
\bibinfo{author}{\bibfnamefont{S.~M.} \bibnamefont{Young}},
  \bibinfo{author}{\bibfnamefont{S.}~\bibnamefont{Zaheer}},
  \bibinfo{author}{\bibfnamefont{J.~C.~Y.} \bibnamefont{Teo}},
  \bibinfo{author}{\bibfnamefont{C.~L.} \bibnamefont{Kane}},
  \bibinfo{author}{\bibfnamefont{E.~J.} \bibnamefont{Mele}}, \bibnamefont{and}
  \bibinfo{author}{\bibfnamefont{A.~M.} \bibnamefont{Rappe}},
  \bibinfo{journal}{Phys. Rev. Lett.} \textbf{\bibinfo{volume}{108}},
  \bibinfo{pages}{140405} (\bibinfo{year}{2012}).

\bibitem[{\citenamefont{Zyuzin et~al.}(2012)\citenamefont{Zyuzin, Wu, and
  Burkov}}]{ZyuzinPRB2012a}
\bibinfo{author}{\bibfnamefont{A.~A.} \bibnamefont{Zyuzin}},
  \bibinfo{author}{\bibfnamefont{S.}~\bibnamefont{Wu}}, \bibnamefont{and}
  \bibinfo{author}{\bibfnamefont{A.~A.} \bibnamefont{Burkov}},
  \bibinfo{journal}{Phys. Rev. B} \textbf{\bibinfo{volume}{85}},
  \bibinfo{pages}{165110} (\bibinfo{year}{2012}).

\bibitem[{\citenamefont{Yang and Nagaosa}(2014)}]{YangNatCommun2014}
\bibinfo{author}{\bibfnamefont{B.-J.} \bibnamefont{Yang}} \bibnamefont{and}
  \bibinfo{author}{\bibfnamefont{N.}~\bibnamefont{Nagaosa}},
  \bibinfo{journal}{Nat. Commun.} \textbf{\bibinfo{volume}{5}},
  \bibinfo{pages}{4898} (\bibinfo{year}{2014}).

\bibitem[{\citenamefont{Weng et~al.}(2015)\citenamefont{Weng, Fang, Fang,
  Bernevig, and Dai}}]{WengPRX2015}
\bibinfo{author}{\bibfnamefont{H.}~\bibnamefont{Weng}},
  \bibinfo{author}{\bibfnamefont{C.}~\bibnamefont{Fang}},
  \bibinfo{author}{\bibfnamefont{Z.}~\bibnamefont{Fang}},
  \bibinfo{author}{\bibfnamefont{B.~A.} \bibnamefont{Bernevig}},
  \bibnamefont{and} \bibinfo{author}{\bibfnamefont{X.}~\bibnamefont{Dai}},
  \bibinfo{journal}{Phys. Rev. X} \textbf{\bibinfo{volume}{5}},
  \bibinfo{pages}{011029} (\bibinfo{year}{2015}).

\bibitem[{\citenamefont{{Chiu} and {Schnyder}}(2015)}]{ChiuArXiv2015}
\bibinfo{author}{\bibfnamefont{C.-K.} \bibnamefont{{Chiu}}} \bibnamefont{and}
  \bibinfo{author}{\bibfnamefont{A.~P.} \bibnamefont{{Schnyder}}},
  \bibinfo{journal}{ArXiv e-prints}  (\bibinfo{year}{2015}),
  \eprint{1501.06820}.

\bibitem[{\citenamefont{{Huang}
  et~al.}(2015{\natexlab{a}})\citenamefont{{Huang}, {Xu}, {Belopolski}, {Lee},
  {Chang}, {Wang}, {Alidoust}, {Bian}, {Neupane}, {Bansil}
  et~al.}}]{HuangArXiv2015a}
\bibinfo{author}{\bibfnamefont{S.-M.} \bibnamefont{{Huang}}},
  \bibinfo{author}{\bibfnamefont{S.-Y.} \bibnamefont{{Xu}}},
  \bibinfo{author}{\bibfnamefont{I.}~\bibnamefont{{Belopolski}}},
  \bibinfo{author}{\bibfnamefont{C.-C.} \bibnamefont{{Lee}}},
  \bibinfo{author}{\bibfnamefont{G.}~\bibnamefont{{Chang}}},
  \bibinfo{author}{\bibfnamefont{B.}~\bibnamefont{{Wang}}},
  \bibinfo{author}{\bibfnamefont{N.}~\bibnamefont{{Alidoust}}},
  \bibinfo{author}{\bibfnamefont{G.}~\bibnamefont{{Bian}}},
  \bibinfo{author}{\bibfnamefont{M.}~\bibnamefont{{Neupane}}},
  \bibinfo{author}{\bibfnamefont{A.}~\bibnamefont{{Bansil}}},
  \bibnamefont{et~al.}, \bibinfo{journal}{ArXiv e-prints}
  (\bibinfo{year}{2015}{\natexlab{a}}), \eprint{1501.00755}.

\bibitem[{\citenamefont{{Huang}
  et~al.}(2015{\natexlab{b}})\citenamefont{{Huang}, {Xu}, {Belopolski}, {Lee},
  {Chang}, {Wang}, {Alidoust}, {Neupane}, {Zheng}, {Sanchez}
  et~al.}}]{HuangArXiv2015b}
\bibinfo{author}{\bibfnamefont{S.-M.} \bibnamefont{{Huang}}},
  \bibinfo{author}{\bibfnamefont{S.-Y.} \bibnamefont{{Xu}}},
  \bibinfo{author}{\bibfnamefont{I.}~\bibnamefont{{Belopolski}}},
  \bibinfo{author}{\bibfnamefont{C.-C.} \bibnamefont{{Lee}}},
  \bibinfo{author}{\bibfnamefont{G.}~\bibnamefont{{Chang}}},
  \bibinfo{author}{\bibfnamefont{B.}~\bibnamefont{{Wang}}},
  \bibinfo{author}{\bibfnamefont{N.}~\bibnamefont{{Alidoust}}},
  \bibinfo{author}{\bibfnamefont{M.}~\bibnamefont{{Neupane}}},
  \bibinfo{author}{\bibfnamefont{H.}~\bibnamefont{{Zheng}}},
  \bibinfo{author}{\bibfnamefont{D.}~\bibnamefont{{Sanchez}}},
  \bibnamefont{et~al.}, \bibinfo{journal}{ArXiv e-prints}
  (\bibinfo{year}{2015}{\natexlab{b}}), \eprint{1503.05868}.

\bibitem[{\citenamefont{Yang et~al.}(2011)\citenamefont{Yang, Lu, and
  Ran}}]{YangPRB2011}
\bibinfo{author}{\bibfnamefont{K.-Y.} \bibnamefont{Yang}},
  \bibinfo{author}{\bibfnamefont{Y.-M.} \bibnamefont{Lu}}, \bibnamefont{and}
  \bibinfo{author}{\bibfnamefont{Y.}~\bibnamefont{Ran}},
  \bibinfo{journal}{Phys. Rev. B} \textbf{\bibinfo{volume}{84}},
  \bibinfo{pages}{075129} (\bibinfo{year}{2011}).

\bibitem[{\citenamefont{Hosur et~al.}(2012)\citenamefont{Hosur, Parameswaran,
  and Vishwanath}}]{HosurPRL2012}
\bibinfo{author}{\bibfnamefont{P.}~\bibnamefont{Hosur}},
  \bibinfo{author}{\bibfnamefont{S.~A.} \bibnamefont{Parameswaran}},
  \bibnamefont{and}
  \bibinfo{author}{\bibfnamefont{A.}~\bibnamefont{Vishwanath}},
  \bibinfo{journal}{Phys. Rev. Lett.} \textbf{\bibinfo{volume}{108}},
  \bibinfo{pages}{046602} (\bibinfo{year}{2012}).

\bibitem[{\citenamefont{Zyuzin and Burkov}(2012)}]{ZyuzinPRB2012b}
\bibinfo{author}{\bibfnamefont{A.~A.} \bibnamefont{Zyuzin}} \bibnamefont{and}
  \bibinfo{author}{\bibfnamefont{A.~A.} \bibnamefont{Burkov}},
  \bibinfo{journal}{Phys. Rev. B} \textbf{\bibinfo{volume}{86}},
  \bibinfo{pages}{115133} (\bibinfo{year}{2012}).

\bibitem[{\citenamefont{Rosenstein and Lewkowicz}(2013)}]{RosensteinPRB2013}
\bibinfo{author}{\bibfnamefont{B.}~\bibnamefont{Rosenstein}} \bibnamefont{and}
  \bibinfo{author}{\bibfnamefont{M.}~\bibnamefont{Lewkowicz}},
  \bibinfo{journal}{Phys. Rev. B} \textbf{\bibinfo{volume}{88}},
  \bibinfo{pages}{045108} (\bibinfo{year}{2013}).

\bibitem[{\citenamefont{Moon et~al.}(2013)\citenamefont{Moon, Xu, Kim, and
  Balents}}]{MoonPRL2013}
\bibinfo{author}{\bibfnamefont{E.-G.} \bibnamefont{Moon}},
  \bibinfo{author}{\bibfnamefont{C.}~\bibnamefont{Xu}},
  \bibinfo{author}{\bibfnamefont{Y.~B.} \bibnamefont{Kim}}, \bibnamefont{and}
  \bibinfo{author}{\bibfnamefont{L.}~\bibnamefont{Balents}},
  \bibinfo{journal}{Phys. Rev. Lett.} \textbf{\bibinfo{volume}{111}},
  \bibinfo{pages}{206401} (\bibinfo{year}{2013}).

\bibitem[{\citenamefont{Biswas and Ryu}(2014)}]{BiswasPRB2014}
\bibinfo{author}{\bibfnamefont{R.~R.} \bibnamefont{Biswas}} \bibnamefont{and}
  \bibinfo{author}{\bibfnamefont{S.}~\bibnamefont{Ryu}},
  \bibinfo{journal}{Phys. Rev. B} \textbf{\bibinfo{volume}{89}},
  \bibinfo{pages}{014205} (\bibinfo{year}{2014}).

\bibitem[{\citenamefont{Parameswaran et~al.}(2014)\citenamefont{Parameswaran,
  Grover, Abanin, Pesin, and Vishwanath}}]{ParameswaranPRX2014}
\bibinfo{author}{\bibfnamefont{S.~A.} \bibnamefont{Parameswaran}},
  \bibinfo{author}{\bibfnamefont{T.}~\bibnamefont{Grover}},
  \bibinfo{author}{\bibfnamefont{D.~A.} \bibnamefont{Abanin}},
  \bibinfo{author}{\bibfnamefont{D.~A.} \bibnamefont{Pesin}}, \bibnamefont{and}
  \bibinfo{author}{\bibfnamefont{A.}~\bibnamefont{Vishwanath}},
  \bibinfo{journal}{Phys. Rev. X} \textbf{\bibinfo{volume}{4}},
  \bibinfo{pages}{031035} (\bibinfo{year}{2014}).

\bibitem[{\citenamefont{{Ziegler}}(2015)}]{ZieglerArXiv2015}
\bibinfo{author}{\bibfnamefont{K.}~\bibnamefont{{Ziegler}}},
  \bibinfo{journal}{ArXiv e-prints}  (\bibinfo{year}{2015}),
  \eprint{1501.00268}.

\bibitem[{\citenamefont{Potter et~al.}(2014)\citenamefont{Potter, Kimchi, and
  Vishwanath}}]{PotterNatCommun2014}
\bibinfo{author}{\bibfnamefont{A.~C.} \bibnamefont{Potter}},
  \bibinfo{author}{\bibfnamefont{I.}~\bibnamefont{Kimchi}}, \bibnamefont{and}
  \bibinfo{author}{\bibfnamefont{A.}~\bibnamefont{Vishwanath}},
  \bibinfo{journal}{Nat. Commun.} \textbf{\bibinfo{volume}{5}},
  \bibinfo{pages}{5161} (\bibinfo{year}{2014}).

\bibitem[{\citenamefont{Ganeshan and Das~Sarma}(2015)}]{GaneshanPRB2015}
\bibinfo{author}{\bibfnamefont{S.}~\bibnamefont{Ganeshan}} \bibnamefont{and}
  \bibinfo{author}{\bibfnamefont{S.}~\bibnamefont{Das~Sarma}},
  \bibinfo{journal}{Phys. Rev. B} \textbf{\bibinfo{volume}{91}},
  \bibinfo{pages}{125438} (\bibinfo{year}{2015}).

\bibitem[{\citenamefont{{Xu} et~al.}(2013)\citenamefont{{Xu}, {Liu},
  {Kushwaha}, {Chang}, {Krizan}, {Sankar}, {Polley}, {Adell},
  {Balasubramanian}, {Miyamoto} et~al.}}]{XuArXiv2013}
\bibinfo{author}{\bibfnamefont{S.-Y.} \bibnamefont{{Xu}}},
  \bibinfo{author}{\bibfnamefont{C.}~\bibnamefont{{Liu}}},
  \bibinfo{author}{\bibfnamefont{S.~K.} \bibnamefont{{Kushwaha}}},
  \bibinfo{author}{\bibfnamefont{T.-R.} \bibnamefont{{Chang}}},
  \bibinfo{author}{\bibfnamefont{J.~W.} \bibnamefont{{Krizan}}},
  \bibinfo{author}{\bibfnamefont{R.}~\bibnamefont{{Sankar}}},
  \bibinfo{author}{\bibfnamefont{C.~M.} \bibnamefont{{Polley}}},
  \bibinfo{author}{\bibfnamefont{J.}~\bibnamefont{{Adell}}},
  \bibinfo{author}{\bibfnamefont{T.}~\bibnamefont{{Balasubramanian}}},
  \bibinfo{author}{\bibfnamefont{K.}~\bibnamefont{{Miyamoto}}},
  \bibnamefont{et~al.}, \bibinfo{journal}{ArXiv e-prints}
  (\bibinfo{year}{2013}), \eprint{1312.7624}.

\bibitem[{\citenamefont{Liu et~al.}(2014{\natexlab{a}})\citenamefont{Liu, Zhou,
  Zhang, Wang, Weng, Prabhakaran, Mo, Shen, Fang, Dai et~al.}}]{LiuScience2014}
\bibinfo{author}{\bibfnamefont{Z.~K.} \bibnamefont{Liu}},
  \bibinfo{author}{\bibfnamefont{B.}~\bibnamefont{Zhou}},
  \bibinfo{author}{\bibfnamefont{Y.}~\bibnamefont{Zhang}},
  \bibinfo{author}{\bibfnamefont{Z.~J.} \bibnamefont{Wang}},
  \bibinfo{author}{\bibfnamefont{H.~M.} \bibnamefont{Weng}},
  \bibinfo{author}{\bibfnamefont{D.}~\bibnamefont{Prabhakaran}},
  \bibinfo{author}{\bibfnamefont{S.-K.} \bibnamefont{Mo}},
  \bibinfo{author}{\bibfnamefont{Z.~X.} \bibnamefont{Shen}},
  \bibinfo{author}{\bibfnamefont{Z.}~\bibnamefont{Fang}},
  \bibinfo{author}{\bibfnamefont{X.}~\bibnamefont{Dai}}, \bibnamefont{et~al.},
  \bibinfo{journal}{Science} \textbf{\bibinfo{volume}{343}},
  \bibinfo{pages}{864} (\bibinfo{year}{2014}{\natexlab{a}}).

\bibitem[{\citenamefont{Neupane et~al.}(2014)\citenamefont{Neupane, Xu, Sankar,
  Alidoust, Bian, Liu, Belopolski, Chang, Jeng, Lin
  et~al.}}]{NeupaneNatCommun2014}
\bibinfo{author}{\bibfnamefont{M.}~\bibnamefont{Neupane}},
  \bibinfo{author}{\bibfnamefont{S.-Y.} \bibnamefont{Xu}},
  \bibinfo{author}{\bibfnamefont{R.}~\bibnamefont{Sankar}},
  \bibinfo{author}{\bibfnamefont{N.}~\bibnamefont{Alidoust}},
  \bibinfo{author}{\bibfnamefont{G.}~\bibnamefont{Bian}},
  \bibinfo{author}{\bibfnamefont{C.}~\bibnamefont{Liu}},
  \bibinfo{author}{\bibfnamefont{I.}~\bibnamefont{Belopolski}},
  \bibinfo{author}{\bibfnamefont{T.-R.} \bibnamefont{Chang}},
  \bibinfo{author}{\bibfnamefont{H.-T.} \bibnamefont{Jeng}},
  \bibinfo{author}{\bibfnamefont{H.}~\bibnamefont{Lin}}, \bibnamefont{et~al.},
  \bibinfo{journal}{Nat. Commun.} \textbf{\bibinfo{volume}{5}},
  \bibinfo{pages}{3786} (\bibinfo{year}{2014}).

\bibitem[{\citenamefont{Liu et~al.}(2014{\natexlab{b}})\citenamefont{Liu,
  Jiang, Zhou, Wang, Zhang, Weng, Prabhakaran, Mo, Peng, Dudin
  et~al.}}]{LiuNatMater2014}
\bibinfo{author}{\bibfnamefont{Z.~K.} \bibnamefont{Liu}},
  \bibinfo{author}{\bibfnamefont{J.}~\bibnamefont{Jiang}},
  \bibinfo{author}{\bibfnamefont{B.}~\bibnamefont{Zhou}},
  \bibinfo{author}{\bibfnamefont{Z.~J.} \bibnamefont{Wang}},
  \bibinfo{author}{\bibfnamefont{Y.}~\bibnamefont{Zhang}},
  \bibinfo{author}{\bibfnamefont{H.~M.} \bibnamefont{Weng}},
  \bibinfo{author}{\bibfnamefont{D.}~\bibnamefont{Prabhakaran}},
  \bibinfo{author}{\bibfnamefont{S.-K.} \bibnamefont{Mo}},
  \bibinfo{author}{\bibfnamefont{H.}~\bibnamefont{Peng}},
  \bibinfo{author}{\bibfnamefont{P.}~\bibnamefont{Dudin}},
  \bibnamefont{et~al.}, \bibinfo{journal}{Nat. Mater.}
  \textbf{\bibinfo{volume}{13}}, \bibinfo{pages}{677}
  (\bibinfo{year}{2014}{\natexlab{b}}).

\bibitem[{\citenamefont{Borisenko et~al.}(2014)\citenamefont{Borisenko, Gibson,
  Evtushinsky, Zabolotnyy, B\"uchner, and Cava}}]{BorisenkoPRL2014}
\bibinfo{author}{\bibfnamefont{S.}~\bibnamefont{Borisenko}},
  \bibinfo{author}{\bibfnamefont{Q.}~\bibnamefont{Gibson}},
  \bibinfo{author}{\bibfnamefont{D.}~\bibnamefont{Evtushinsky}},
  \bibinfo{author}{\bibfnamefont{V.}~\bibnamefont{Zabolotnyy}},
  \bibinfo{author}{\bibfnamefont{B.}~\bibnamefont{B\"uchner}},
  \bibnamefont{and} \bibinfo{author}{\bibfnamefont{R.~J.} \bibnamefont{Cava}},
  \bibinfo{journal}{Phys. Rev. Lett.} \textbf{\bibinfo{volume}{113}},
  \bibinfo{pages}{027603} (\bibinfo{year}{2014}).

\bibitem[{\citenamefont{{Xu} et~al.}(2015)\citenamefont{{Xu}, {Liu},
  {Belopolski}, {Kushwaha}, {Sankar}, {Krizan}, {Chang}, {Polley}, {Adell},
  {Balasubramanian} et~al.}}]{XuArXiv2015c}
\bibinfo{author}{\bibfnamefont{S.-Y.} \bibnamefont{{Xu}}},
  \bibinfo{author}{\bibfnamefont{C.}~\bibnamefont{{Liu}}},
  \bibinfo{author}{\bibfnamefont{I.}~\bibnamefont{{Belopolski}}},
  \bibinfo{author}{\bibfnamefont{S.~K.} \bibnamefont{{Kushwaha}}},
  \bibinfo{author}{\bibfnamefont{R.}~\bibnamefont{{Sankar}}},
  \bibinfo{author}{\bibfnamefont{J.~W.} \bibnamefont{{Krizan}}},
  \bibinfo{author}{\bibfnamefont{T.-R.} \bibnamefont{{Chang}}},
  \bibinfo{author}{\bibfnamefont{C.~M.} \bibnamefont{{Polley}}},
  \bibinfo{author}{\bibfnamefont{J.}~\bibnamefont{{Adell}}},
  \bibinfo{author}{\bibfnamefont{T.}~\bibnamefont{{Balasubramanian}}},
  \bibnamefont{et~al.}, \bibinfo{journal}{Phys. Rev. B}
  \textbf{\bibinfo{volume}{92}}, \bibinfo{pages}{075115}
  (\bibinfo{year}{2015}).

\bibitem[{\citenamefont{Xu et~al.}(2015)\citenamefont{Xu, Liu, Kushwaha,
  Sankar, Krizan, Belopolski, Neupane, Bian, Alidoust, Chang
  et~al.}}]{XuScience2015}
\bibinfo{author}{\bibfnamefont{S.-Y.} \bibnamefont{Xu}},
  \bibinfo{author}{\bibfnamefont{C.}~\bibnamefont{Liu}},
  \bibinfo{author}{\bibfnamefont{S.~K.} \bibnamefont{Kushwaha}},
  \bibinfo{author}{\bibfnamefont{R.}~\bibnamefont{Sankar}},
  \bibinfo{author}{\bibfnamefont{J.~W.} \bibnamefont{Krizan}},
  \bibinfo{author}{\bibfnamefont{I.}~\bibnamefont{Belopolski}},
  \bibinfo{author}{\bibfnamefont{M.}~\bibnamefont{Neupane}},
  \bibinfo{author}{\bibfnamefont{G.}~\bibnamefont{Bian}},
  \bibinfo{author}{\bibfnamefont{N.}~\bibnamefont{Alidoust}},
  \bibinfo{author}{\bibfnamefont{T.-R.} \bibnamefont{Chang}},
  \bibnamefont{et~al.}, \bibinfo{journal}{Science}
  \textbf{\bibinfo{volume}{347}}, \bibinfo{pages}{294} (\bibinfo{year}{2015}).

\bibitem[{\citenamefont{Banerjee and Pickett}(2012)}]{BanerjeePRB2012}
\bibinfo{author}{\bibfnamefont{S.}~\bibnamefont{Banerjee}} \bibnamefont{and}
  \bibinfo{author}{\bibfnamefont{W.~E.} \bibnamefont{Pickett}},
  \bibinfo{journal}{Phys. Rev. B} \textbf{\bibinfo{volume}{86}},
  \bibinfo{pages}{075124} (\bibinfo{year}{2012}).

\bibitem[{\citenamefont{Jeon et~al.}(2014)\citenamefont{Jeon, Zhou, Gyenis,
  Feldman, Kimchi, Potter, Gibson, Cava, Vishwanath, and
  Yazdani}}]{JeonNatMater2014}
\bibinfo{author}{\bibfnamefont{S.}~\bibnamefont{Jeon}},
  \bibinfo{author}{\bibfnamefont{B.~B.} \bibnamefont{Zhou}},
  \bibinfo{author}{\bibfnamefont{A.}~\bibnamefont{Gyenis}},
  \bibinfo{author}{\bibfnamefont{B.~E.} \bibnamefont{Feldman}},
  \bibinfo{author}{\bibfnamefont{I.}~\bibnamefont{Kimchi}},
  \bibinfo{author}{\bibfnamefont{A.~C.} \bibnamefont{Potter}},
  \bibinfo{author}{\bibfnamefont{Q.~D.} \bibnamefont{Gibson}},
  \bibinfo{author}{\bibfnamefont{R.~J.} \bibnamefont{Cava}},
  \bibinfo{author}{\bibfnamefont{A.}~\bibnamefont{Vishwanath}},
  \bibnamefont{and} \bibinfo{author}{\bibfnamefont{A.}~\bibnamefont{Yazdani}},
  \bibinfo{journal}{Nat. Mater.} \textbf{\bibinfo{volume}{13}},
  \bibinfo{pages}{851} (\bibinfo{year}{2014}).

\bibitem[{\citenamefont{Liang et~al.}(2015)\citenamefont{Liang, Gibson, Ali,
  Liu, Cava, and Ong}}]{LiangNatMater2015}
\bibinfo{author}{\bibfnamefont{T.}~\bibnamefont{Liang}},
  \bibinfo{author}{\bibfnamefont{Q.}~\bibnamefont{Gibson}},
  \bibinfo{author}{\bibfnamefont{M.~N.} \bibnamefont{Ali}},
  \bibinfo{author}{\bibfnamefont{M.}~\bibnamefont{Liu}},
  \bibinfo{author}{\bibfnamefont{R.~J.} \bibnamefont{Cava}}, \bibnamefont{and}
  \bibinfo{author}{\bibfnamefont{N.~P.} \bibnamefont{Ong}},
  \bibinfo{journal}{Nat. Mater.} \textbf{\bibinfo{volume}{14}},
  \bibinfo{pages}{280} (\bibinfo{year}{2015}).

\bibitem[{\citenamefont{Narayanan et~al.}(2015)\citenamefont{Narayanan, Watson,
  Blake, Bruyant, Drigo, Chen, Prabhakaran, Yan, Felser, Kong
  et~al.}}]{NarayananPRL2015}
\bibinfo{author}{\bibfnamefont{A.}~\bibnamefont{Narayanan}},
  \bibinfo{author}{\bibfnamefont{M.~D.} \bibnamefont{Watson}},
  \bibinfo{author}{\bibfnamefont{S.~F.} \bibnamefont{Blake}},
  \bibinfo{author}{\bibfnamefont{N.}~\bibnamefont{Bruyant}},
  \bibinfo{author}{\bibfnamefont{L.}~\bibnamefont{Drigo}},
  \bibinfo{author}{\bibfnamefont{Y.~L.} \bibnamefont{Chen}},
  \bibinfo{author}{\bibfnamefont{D.}~\bibnamefont{Prabhakaran}},
  \bibinfo{author}{\bibfnamefont{B.}~\bibnamefont{Yan}},
  \bibinfo{author}{\bibfnamefont{C.}~\bibnamefont{Felser}},
  \bibinfo{author}{\bibfnamefont{T.}~\bibnamefont{Kong}}, \bibnamefont{et~al.},
  \bibinfo{journal}{Phys. Rev. Lett.} \textbf{\bibinfo{volume}{114}},
  \bibinfo{pages}{117201} (\bibinfo{year}{2015}).

\bibitem[{\citenamefont{Kushwaha et~al.}(2015)\citenamefont{Kushwaha, Krizan,
  Feldman, Gyenis, Randeria, Xiong, Xu, Alidoust, Belopolski, Liang
  et~al.}}]{Kushwaha_APLMat15}
\bibinfo{author}{\bibfnamefont{S.~K.} \bibnamefont{Kushwaha}},
  \bibinfo{author}{\bibfnamefont{J.~W.} \bibnamefont{Krizan}},
  \bibinfo{author}{\bibfnamefont{B.~E.} \bibnamefont{Feldman}},
  \bibinfo{author}{\bibfnamefont{A.}~\bibnamefont{Gyenis}},
  \bibinfo{author}{\bibfnamefont{M.~T.} \bibnamefont{Randeria}},
  \bibinfo{author}{\bibfnamefont{J.}~\bibnamefont{Xiong}},
  \bibinfo{author}{\bibfnamefont{S.-Y.} \bibnamefont{Xu}},
  \bibinfo{author}{\bibfnamefont{N.}~\bibnamefont{Alidoust}},
  \bibinfo{author}{\bibfnamefont{I.}~\bibnamefont{Belopolski}},
  \bibinfo{author}{\bibfnamefont{T.}~\bibnamefont{Liang}},
  \bibnamefont{et~al.}, \bibinfo{journal}{APL Mat.}
  \textbf{\bibinfo{volume}{3}}, \bibinfo{pages}{041504} (\bibinfo{year}{2015}).

\bibitem[{\citenamefont{Taguchi and Tanaka}(2015)}]{TaguchiPRB2015}
\bibinfo{author}{\bibfnamefont{K.}~\bibnamefont{Taguchi}} \bibnamefont{and}
  \bibinfo{author}{\bibfnamefont{Y.}~\bibnamefont{Tanaka}},
  \bibinfo{journal}{Phys. Rev. B} \textbf{\bibinfo{volume}{91}},
  \bibinfo{pages}{054422} (\bibinfo{year}{2015}).

\bibitem[{\citenamefont{{Feng} et~al.}(2015)\citenamefont{{Feng}, {Pang}, {Wu},
  {Wang}, {Weng}, {Li}, {Dai}, {Fang}, {Shi}, and {Lu}}}]{FengArXiv2014}
\bibinfo{author}{\bibfnamefont{J.}~\bibnamefont{{Feng}}},
  \bibinfo{author}{\bibfnamefont{Y.}~\bibnamefont{{Pang}}},
  \bibinfo{author}{\bibfnamefont{D.}~\bibnamefont{{Wu}}},
  \bibinfo{author}{\bibfnamefont{Z.}~\bibnamefont{{Wang}}},
  \bibinfo{author}{\bibfnamefont{H.}~\bibnamefont{{Weng}}},
  \bibinfo{author}{\bibfnamefont{J.}~\bibnamefont{{Li}}},
  \bibinfo{author}{\bibfnamefont{X.}~\bibnamefont{{Dai}}},
  \bibinfo{author}{\bibfnamefont{Z.}~\bibnamefont{{Fang}}},
  \bibinfo{author}{\bibfnamefont{Y.}~\bibnamefont{{Shi}}}, \bibnamefont{and}
  \bibinfo{author}{\bibfnamefont{L.}~\bibnamefont{{Lu}}},
  \bibinfo{journal}{Phys. Rev. B} \textbf{\bibinfo{volume}{92}},
  \bibinfo{pages}{081306} (\bibinfo{year}{2015}).

\bibitem[{\citenamefont{Das~Sarma et~al.}(2015)\citenamefont{Das~Sarma, Hwang,
  and Min}}]{DasSarmaPRB2015}
\bibinfo{author}{\bibfnamefont{S.}~\bibnamefont{Das~Sarma}},
  \bibinfo{author}{\bibfnamefont{E.~H.} \bibnamefont{Hwang}}, \bibnamefont{and}
  \bibinfo{author}{\bibfnamefont{H.}~\bibnamefont{Min}},
  \bibinfo{journal}{Phys. Rev. B} \textbf{\bibinfo{volume}{91}},
  \bibinfo{pages}{035201} (\bibinfo{year}{2015}).

\bibitem[{\citenamefont{{Xiong} et~al.}(2015)\citenamefont{{Xiong}, {Kushwaha},
  {Krizan}, {Liang}, {Cava}, and {Ong}}}]{XiongArXiv2015}
\bibinfo{author}{\bibfnamefont{J.}~\bibnamefont{{Xiong}}},
  \bibinfo{author}{\bibfnamefont{S.}~\bibnamefont{{Kushwaha}}},
  \bibinfo{author}{\bibfnamefont{J.}~\bibnamefont{{Krizan}}},
  \bibinfo{author}{\bibfnamefont{T.}~\bibnamefont{{Liang}}},
  \bibinfo{author}{\bibfnamefont{R.~J.} \bibnamefont{{Cava}}},
  \bibnamefont{and} \bibinfo{author}{\bibfnamefont{N.~P.} \bibnamefont{{Ong}}},
  \bibinfo{journal}{ArXiv e-prints}  (\bibinfo{year}{2015}),
  \eprint{1502.06266}.

\bibitem[{\citenamefont{{Das Sarma} and Hwang}(2015)}]{DasSarmaPRB2015_2}
\bibinfo{author}{\bibfnamefont{S.}~\bibnamefont{{Das Sarma}}} \bibnamefont{and}
  \bibinfo{author}{\bibfnamefont{E.~H.} \bibnamefont{Hwang}},
  \bibinfo{journal}{Phys. Rev. B} \textbf{\bibinfo{volume}{91}},
  \bibinfo{pages}{195104} (\bibinfo{year}{2015}).

\bibitem[{\citenamefont{Goswami and Chakravarty}(2011)}]{GoswamiPRL2011}
\bibinfo{author}{\bibfnamefont{P.}~\bibnamefont{Goswami}} \bibnamefont{and}
  \bibinfo{author}{\bibfnamefont{S.}~\bibnamefont{Chakravarty}},
  \bibinfo{journal}{Phys. Rev. Lett.} \textbf{\bibinfo{volume}{107}},
  \bibinfo{pages}{196803} (\bibinfo{year}{2011}).

\bibitem[{\citenamefont{Witczak-Krempa and Kim}(2012)}]{WitczakKrempaPRB2012}
\bibinfo{author}{\bibfnamefont{W.}~\bibnamefont{Witczak-Krempa}}
  \bibnamefont{and} \bibinfo{author}{\bibfnamefont{Y.~B.} \bibnamefont{Kim}},
  \bibinfo{journal}{Phys. Rev. B} \textbf{\bibinfo{volume}{85}},
  \bibinfo{pages}{045124} (\bibinfo{year}{2012}),
  \urlprefix\url{http://link.aps.org/doi/10.1103/PhysRevB.85.045124}.

\bibitem[{\citenamefont{Isobe and Nagaosa}(2012)}]{IsobePRB2012}
\bibinfo{author}{\bibfnamefont{H.}~\bibnamefont{Isobe}} \bibnamefont{and}
  \bibinfo{author}{\bibfnamefont{N.}~\bibnamefont{Nagaosa}},
  \bibinfo{journal}{Phys. Rev. B} \textbf{\bibinfo{volume}{86}},
  \bibinfo{pages}{165127} (\bibinfo{year}{2012}).

\bibitem[{\citenamefont{Wang and Zhang}(2013)}]{WangPRB2013b}
\bibinfo{author}{\bibfnamefont{Z.}~\bibnamefont{Wang}} \bibnamefont{and}
  \bibinfo{author}{\bibfnamefont{S.-C.} \bibnamefont{Zhang}},
  \bibinfo{journal}{Phys. Rev. B} \textbf{\bibinfo{volume}{87}},
  \bibinfo{pages}{161107} (\bibinfo{year}{2013}).

\bibitem[{\citenamefont{Goswami and Tewari}(2013)}]{GoswamiPRB2013}
\bibinfo{author}{\bibfnamefont{P.}~\bibnamefont{Goswami}} \bibnamefont{and}
  \bibinfo{author}{\bibfnamefont{S.}~\bibnamefont{Tewari}},
  \bibinfo{journal}{Phys. Rev. B} \textbf{\bibinfo{volume}{88}},
  \bibinfo{pages}{245107} (\bibinfo{year}{2013}).

\bibitem[{\citenamefont{Gonz\'alez}(2014)}]{GonzalezPRB2014}
\bibinfo{author}{\bibfnamefont{J.}~\bibnamefont{Gonz\'alez}},
  \bibinfo{journal}{Phys. Rev. B} \textbf{\bibinfo{volume}{90}},
  \bibinfo{pages}{121107} (\bibinfo{year}{2014}).

\bibitem[{\citenamefont{Zhang et~al.}(2015)\citenamefont{Zhang, Wu, Schoop,
  Ali, Shi, Ni, Gibson, Jiang, Sidorov, Yi et~al.}}]{ZhangPRB2015}
\bibinfo{author}{\bibfnamefont{S.}~\bibnamefont{Zhang}},
  \bibinfo{author}{\bibfnamefont{Q.}~\bibnamefont{Wu}},
  \bibinfo{author}{\bibfnamefont{L.}~\bibnamefont{Schoop}},
  \bibinfo{author}{\bibfnamefont{M.~N.} \bibnamefont{Ali}},
  \bibinfo{author}{\bibfnamefont{Y.}~\bibnamefont{Shi}},
  \bibinfo{author}{\bibfnamefont{N.}~\bibnamefont{Ni}},
  \bibinfo{author}{\bibfnamefont{Q.}~\bibnamefont{Gibson}},
  \bibinfo{author}{\bibfnamefont{S.}~\bibnamefont{Jiang}},
  \bibinfo{author}{\bibfnamefont{V.}~\bibnamefont{Sidorov}},
  \bibinfo{author}{\bibfnamefont{W.}~\bibnamefont{Yi}}, \bibnamefont{et~al.},
  \bibinfo{journal}{Phys. Rev. B} \textbf{\bibinfo{volume}{91}},
  \bibinfo{pages}{165133} (\bibinfo{year}{2015}).

\bibitem[{\citenamefont{{Schoop} et~al.}(2014)\citenamefont{{Schoop}, {Xie},
  {Chen}, {Gibson}, {Lapidus}, {Kimchi}, {Hirschberger}, {Haldolaarachchige},
  {Ali}, {Belvin} et~al.}}]{SchoopArXiv2014}
\bibinfo{author}{\bibfnamefont{L.~M.} \bibnamefont{{Schoop}}},
  \bibinfo{author}{\bibfnamefont{L.~S.} \bibnamefont{{Xie}}},
  \bibinfo{author}{\bibfnamefont{R.}~\bibnamefont{{Chen}}},
  \bibinfo{author}{\bibfnamefont{Q.~D.} \bibnamefont{{Gibson}}},
  \bibinfo{author}{\bibfnamefont{S.~H.} \bibnamefont{{Lapidus}}},
  \bibinfo{author}{\bibfnamefont{I.}~\bibnamefont{{Kimchi}}},
  \bibinfo{author}{\bibfnamefont{M.}~\bibnamefont{{Hirschberger}}},
  \bibinfo{author}{\bibfnamefont{N.}~\bibnamefont{{Haldolaarachchige}}},
  \bibinfo{author}{\bibfnamefont{M.~N.} \bibnamefont{{Ali}}},
  \bibinfo{author}{\bibfnamefont{C.~A.} \bibnamefont{{Belvin}}},
  \bibnamefont{et~al.}, \bibinfo{journal}{ArXiv e-prints}
  (\bibinfo{year}{2014}), \eprint{1412.2767}.

\bibitem[{\citenamefont{Yang et~al.}(2014)\citenamefont{Yang, Moon, Isobe, and
  Nagaosa}}]{YangNatPhys2014}
\bibinfo{author}{\bibfnamefont{B.-J.} \bibnamefont{Yang}},
  \bibinfo{author}{\bibfnamefont{E.-G.} \bibnamefont{Moon}},
  \bibinfo{author}{\bibfnamefont{H.}~\bibnamefont{Isobe}}, \bibnamefont{and}
  \bibinfo{author}{\bibfnamefont{N.}~\bibnamefont{Nagaosa}},
  \bibinfo{journal}{Nat. Phys.} \textbf{\bibinfo{volume}{10}},
  \bibinfo{pages}{774} (\bibinfo{year}{2014}).

\bibitem[{\citenamefont{{Pixley} et~al.}(2015)\citenamefont{{Pixley},
  {Goswami}, and {Das Sarma}}}]{PixleyArXiv2015}
\bibinfo{author}{\bibfnamefont{J.~H.} \bibnamefont{{Pixley}}},
  \bibinfo{author}{\bibfnamefont{P.}~\bibnamefont{{Goswami}}},
  \bibnamefont{and} \bibinfo{author}{\bibfnamefont{S.}~\bibnamefont{{Das
  Sarma}}}, \bibinfo{journal}{ArXiv e-prints}  (\bibinfo{year}{2015}),
  \eprint{1502.07778}.

\bibitem[{\citenamefont{Roy and {Das Sarma}}(2014)}]{RoyPRB2014}
\bibinfo{author}{\bibfnamefont{B.}~\bibnamefont{Roy}} \bibnamefont{and}
  \bibinfo{author}{\bibfnamefont{S.}~\bibnamefont{{Das Sarma}}},
  \bibinfo{journal}{Phys. Rev. B} \textbf{\bibinfo{volume}{90}},
  \bibinfo{pages}{241112} (\bibinfo{year}{2014}).

\bibitem[{\citenamefont{{Lv} et~al.}(2015{\natexlab{a}})\citenamefont{{Lv},
  {Weng}, {Fu}, {Wang}, {Miao}, {Ma}, {Richard}, {Huang}, {Zhao}, {Chen}
  et~al.}}]{LvArXiv2015a}
\bibinfo{author}{\bibfnamefont{B.~Q.} \bibnamefont{{Lv}}},
  \bibinfo{author}{\bibfnamefont{H.~M.} \bibnamefont{{Weng}}},
  \bibinfo{author}{\bibfnamefont{B.~B.} \bibnamefont{{Fu}}},
  \bibinfo{author}{\bibfnamefont{X.~P.} \bibnamefont{{Wang}}},
  \bibinfo{author}{\bibfnamefont{H.}~\bibnamefont{{Miao}}},
  \bibinfo{author}{\bibfnamefont{J.}~\bibnamefont{{Ma}}},
  \bibinfo{author}{\bibfnamefont{P.}~\bibnamefont{{Richard}}},
  \bibinfo{author}{\bibfnamefont{X.~C.} \bibnamefont{{Huang}}},
  \bibinfo{author}{\bibfnamefont{L.~X.} \bibnamefont{{Zhao}}},
  \bibinfo{author}{\bibfnamefont{G.~F.} \bibnamefont{{Chen}}},
  \bibnamefont{et~al.}, \bibinfo{journal}{ArXiv e-prints}
  (\bibinfo{year}{2015}{\natexlab{a}}), \eprint{1502.04684}.

\bibitem[{\citenamefont{{Lv} et~al.}(2015{\natexlab{b}})\citenamefont{{Lv},
  {Xu}, {Weng}, {Ma}, {Richard}, {Huang}, {Zhao}, {Chen}, {Matt}, {Bisti}
  et~al.}}]{LvArXiv2015b}
\bibinfo{author}{\bibfnamefont{B.~Q.} \bibnamefont{{Lv}}},
  \bibinfo{author}{\bibfnamefont{N.}~\bibnamefont{{Xu}}},
  \bibinfo{author}{\bibfnamefont{H.~M.} \bibnamefont{{Weng}}},
  \bibinfo{author}{\bibfnamefont{J.~Z.} \bibnamefont{{Ma}}},
  \bibinfo{author}{\bibfnamefont{P.}~\bibnamefont{{Richard}}},
  \bibinfo{author}{\bibfnamefont{X.~C.} \bibnamefont{{Huang}}},
  \bibinfo{author}{\bibfnamefont{L.~X.} \bibnamefont{{Zhao}}},
  \bibinfo{author}{\bibfnamefont{G.~F.} \bibnamefont{{Chen}}},
  \bibinfo{author}{\bibfnamefont{C.}~\bibnamefont{{Matt}}},
  \bibinfo{author}{\bibfnamefont{F.}~\bibnamefont{{Bisti}}},
  \bibnamefont{et~al.}, \bibinfo{journal}{Nat. Phys.}
  \textbf{\bibinfo{volume}{11}}, \bibinfo{pages}{724}
  (\bibinfo{year}{2015}{\natexlab{b}}).

\bibitem[{\citenamefont{{Xu} et~al.}(2015)\citenamefont{{Xu}, {Alidoust},
  {Belopolski}, {Zhang}, {Bian}, {Chang}, {Zheng}, {Strokov}, {Sanchez},
  {Chang} et~al.}}]{XuArxiv2015a}
\bibinfo{author}{\bibfnamefont{S.-Y.} \bibnamefont{{Xu}}},
  \bibinfo{author}{\bibfnamefont{N.}~\bibnamefont{{Alidoust}}},
  \bibinfo{author}{\bibfnamefont{I.}~\bibnamefont{{Belopolski}}},
  \bibinfo{author}{\bibfnamefont{C.}~\bibnamefont{{Zhang}}},
  \bibinfo{author}{\bibfnamefont{G.}~\bibnamefont{{Bian}}},
  \bibinfo{author}{\bibfnamefont{T.-R.} \bibnamefont{{Chang}}},
  \bibinfo{author}{\bibfnamefont{H.}~\bibnamefont{{Zheng}}},
  \bibinfo{author}{\bibfnamefont{V.}~\bibnamefont{{Strokov}}},
  \bibinfo{author}{\bibfnamefont{D.~S.} \bibnamefont{{Sanchez}}},
  \bibinfo{author}{\bibfnamefont{G.}~\bibnamefont{{Chang}}},
  \bibnamefont{et~al.}, \bibinfo{journal}{Nat. Phys.}
  \textbf{\bibinfo{volume}{11}}, \bibinfo{pages}{748} (\bibinfo{year}{2015}).

\bibitem[{\citenamefont{Wang et~al.}(2013)\citenamefont{Wang, Weng, Wu, Dai,
  and Fang}}]{WangPRB2013a}
\bibinfo{author}{\bibfnamefont{Z.}~\bibnamefont{Wang}},
  \bibinfo{author}{\bibfnamefont{H.}~\bibnamefont{Weng}},
  \bibinfo{author}{\bibfnamefont{Q.}~\bibnamefont{Wu}},
  \bibinfo{author}{\bibfnamefont{X.}~\bibnamefont{Dai}}, \bibnamefont{and}
  \bibinfo{author}{\bibfnamefont{Z.}~\bibnamefont{Fang}},
  \bibinfo{journal}{Phys. Rev. B} \textbf{\bibinfo{volume}{88}},
  \bibinfo{pages}{125427} (\bibinfo{year}{2013}).

\bibitem[{\citenamefont{{Zhang}
  et~al.}(2015{\natexlab{a}})\citenamefont{{Zhang}, {Yuan}, {Xu}, {Lin},
  {Tong}, {Zahid Hasan}, {Wang}, {Zhang}, and {Jia}}}]{ZhangArXiv2015a}
\bibinfo{author}{\bibfnamefont{C.}~\bibnamefont{{Zhang}}},
  \bibinfo{author}{\bibfnamefont{Z.}~\bibnamefont{{Yuan}}},
  \bibinfo{author}{\bibfnamefont{S.}~\bibnamefont{{Xu}}},
  \bibinfo{author}{\bibfnamefont{Z.}~\bibnamefont{{Lin}}},
  \bibinfo{author}{\bibfnamefont{B.}~\bibnamefont{{Tong}}},
  \bibinfo{author}{\bibfnamefont{M.}~\bibnamefont{{Zahid Hasan}}},
  \bibinfo{author}{\bibfnamefont{J.}~\bibnamefont{{Wang}}},
  \bibinfo{author}{\bibfnamefont{C.}~\bibnamefont{{Zhang}}}, \bibnamefont{and}
  \bibinfo{author}{\bibfnamefont{S.}~\bibnamefont{{Jia}}},
  \bibinfo{journal}{ArXiv e-prints}  (\bibinfo{year}{2015}{\natexlab{a}}),
  \eprint{1502.00251}.

\bibitem[{\citenamefont{{Zhang}
  et~al.}(2015{\natexlab{b}})\citenamefont{{Zhang}, {Xu}, {Belopolski}, {Yuan},
  {Lin}, {Tong}, {Alidoust}, {Lee}, {Huang}, {Lin} et~al.}}]{ZhangArXiv2015b}
\bibinfo{author}{\bibfnamefont{C.}~\bibnamefont{{Zhang}}},
  \bibinfo{author}{\bibfnamefont{S.-Y.} \bibnamefont{{Xu}}},
  \bibinfo{author}{\bibfnamefont{I.}~\bibnamefont{{Belopolski}}},
  \bibinfo{author}{\bibfnamefont{Z.}~\bibnamefont{{Yuan}}},
  \bibinfo{author}{\bibfnamefont{Z.}~\bibnamefont{{Lin}}},
  \bibinfo{author}{\bibfnamefont{B.}~\bibnamefont{{Tong}}},
  \bibinfo{author}{\bibfnamefont{N.}~\bibnamefont{{Alidoust}}},
  \bibinfo{author}{\bibfnamefont{C.-C.} \bibnamefont{{Lee}}},
  \bibinfo{author}{\bibfnamefont{S.-M.} \bibnamefont{{Huang}}},
  \bibinfo{author}{\bibfnamefont{H.}~\bibnamefont{{Lin}}},
  \bibnamefont{et~al.}, \bibinfo{journal}{ArXiv e-prints}
  (\bibinfo{year}{2015}{\natexlab{b}}), \eprint{1503.02630}.

\bibitem[{\citenamefont{Elias et~al.}(2011)\citenamefont{Elias, Gorbachev,
  Mayorov, Morozov, Zhukov, Blake, Ponomarenko, Grigorieva, Novoselov, Guinea
  et~al.}}]{EliasNatPhys2011}
\bibinfo{author}{\bibfnamefont{D.~C.} \bibnamefont{Elias}},
  \bibinfo{author}{\bibfnamefont{R.~V.} \bibnamefont{Gorbachev}},
  \bibinfo{author}{\bibfnamefont{A.~S.} \bibnamefont{Mayorov}},
  \bibinfo{author}{\bibfnamefont{S.~V.} \bibnamefont{Morozov}},
  \bibinfo{author}{\bibfnamefont{A.~A.} \bibnamefont{Zhukov}},
  \bibinfo{author}{\bibfnamefont{P.}~\bibnamefont{Blake}},
  \bibinfo{author}{\bibfnamefont{L.~A.} \bibnamefont{Ponomarenko}},
  \bibinfo{author}{\bibfnamefont{I.~V.} \bibnamefont{Grigorieva}},
  \bibinfo{author}{\bibfnamefont{K.~S.} \bibnamefont{Novoselov}},
  \bibinfo{author}{\bibfnamefont{F.}~\bibnamefont{Guinea}},
  \bibnamefont{et~al.}, \bibinfo{journal}{Nat. Phys.}
  \textbf{\bibinfo{volume}{7}}, \bibinfo{pages}{701} (\bibinfo{year}{2011}).

\bibitem[{\citenamefont{Li et~al.}(2008)\citenamefont{Li, Henriksen, Jiang,
  Hao, Martin, Kim, Stormer, and Basov}}]{LiNatPhys2008}
\bibinfo{author}{\bibfnamefont{Z.~Q.} \bibnamefont{Li}},
  \bibinfo{author}{\bibfnamefont{E.~A.} \bibnamefont{Henriksen}},
  \bibinfo{author}{\bibfnamefont{Z.}~\bibnamefont{Jiang}},
  \bibinfo{author}{\bibfnamefont{Z.}~\bibnamefont{Hao}},
  \bibinfo{author}{\bibfnamefont{M.~C.} \bibnamefont{Martin}},
  \bibinfo{author}{\bibfnamefont{P.}~\bibnamefont{Kim}},
  \bibinfo{author}{\bibfnamefont{H.~L.} \bibnamefont{Stormer}},
  \bibnamefont{and} \bibinfo{author}{\bibfnamefont{D.~N.} \bibnamefont{Basov}},
  \bibinfo{journal}{Nat. Phys.} \textbf{\bibinfo{volume}{4}},
  \bibinfo{pages}{532} (\bibinfo{year}{2008}).

\bibitem[{\citenamefont{Chae et~al.}(2012)\citenamefont{Chae, Jung, Young,
  Dean, Wang, Gao, Watanabe, Taniguchi, Hone, Shepard et~al.}}]{ChaePRL2012}
\bibinfo{author}{\bibfnamefont{J.}~\bibnamefont{Chae}},
  \bibinfo{author}{\bibfnamefont{S.}~\bibnamefont{Jung}},
  \bibinfo{author}{\bibfnamefont{A.~F.} \bibnamefont{Young}},
  \bibinfo{author}{\bibfnamefont{C.~R.} \bibnamefont{Dean}},
  \bibinfo{author}{\bibfnamefont{L.}~\bibnamefont{Wang}},
  \bibinfo{author}{\bibfnamefont{Y.}~\bibnamefont{Gao}},
  \bibinfo{author}{\bibfnamefont{K.}~\bibnamefont{Watanabe}},
  \bibinfo{author}{\bibfnamefont{T.}~\bibnamefont{Taniguchi}},
  \bibinfo{author}{\bibfnamefont{J.}~\bibnamefont{Hone}},
  \bibinfo{author}{\bibfnamefont{K.~L.} \bibnamefont{Shepard}},
  \bibnamefont{et~al.}, \bibinfo{journal}{Phys. Rev. Lett.}
  \textbf{\bibinfo{volume}{109}}, \bibinfo{pages}{116802}
  (\bibinfo{year}{2012}).

\bibitem[{\citenamefont{Siegel et~al.}(2011)\citenamefont{Siegel, Park, Hwang,
  Deslippe, Fedorov, Louie, and Lanzara}}]{SiegelPNAS2011}
\bibinfo{author}{\bibfnamefont{D.~A.} \bibnamefont{Siegel}},
  \bibinfo{author}{\bibfnamefont{C.-H.} \bibnamefont{Park}},
  \bibinfo{author}{\bibfnamefont{C.}~\bibnamefont{Hwang}},
  \bibinfo{author}{\bibfnamefont{J.}~\bibnamefont{Deslippe}},
  \bibinfo{author}{\bibfnamefont{A.~V.} \bibnamefont{Fedorov}},
  \bibinfo{author}{\bibfnamefont{S.~G.} \bibnamefont{Louie}}, \bibnamefont{and}
  \bibinfo{author}{\bibfnamefont{A.}~\bibnamefont{Lanzara}},
  \bibinfo{journal}{Proc. Nat. Acad. Sci. USA} \textbf{\bibinfo{volume}{108}},
  \bibinfo{pages}{11365} (\bibinfo{year}{2011}).

\bibitem[{\citenamefont{Yu et~al.}(2013)\citenamefont{Yu, Jalil, Belle,
  Mayorov, Blake, Schedin, Morozov, Ponomarenko, Chiappini, Wiedmann
  et~al.}}]{YuPNAS2013}
\bibinfo{author}{\bibfnamefont{G.~L.} \bibnamefont{Yu}},
  \bibinfo{author}{\bibfnamefont{R.}~\bibnamefont{Jalil}},
  \bibinfo{author}{\bibfnamefont{B.}~\bibnamefont{Belle}},
  \bibinfo{author}{\bibfnamefont{A.~S.} \bibnamefont{Mayorov}},
  \bibinfo{author}{\bibfnamefont{P.}~\bibnamefont{Blake}},
  \bibinfo{author}{\bibfnamefont{F.}~\bibnamefont{Schedin}},
  \bibinfo{author}{\bibfnamefont{S.~V.} \bibnamefont{Morozov}},
  \bibinfo{author}{\bibfnamefont{L.~A.} \bibnamefont{Ponomarenko}},
  \bibinfo{author}{\bibfnamefont{F.}~\bibnamefont{Chiappini}},
  \bibinfo{author}{\bibfnamefont{S.}~\bibnamefont{Wiedmann}},
  \bibnamefont{et~al.}, \bibinfo{journal}{Proc. Nat. Acad. Sci. USA}
  \textbf{\bibinfo{volume}{110}}, \bibinfo{pages}{3282} (\bibinfo{year}{2013}).

\bibitem[{\citenamefont{Lv and Zhang}(2013)}]{LvIJMPB2013}
\bibinfo{author}{\bibfnamefont{M.}~\bibnamefont{Lv}} \bibnamefont{and}
  \bibinfo{author}{\bibfnamefont{S.-C.} \bibnamefont{Zhang}},
  \bibinfo{journal}{Int. J. Mod. Phys. B} \textbf{\bibinfo{volume}{27}},
  \bibinfo{pages}{1250177} (\bibinfo{year}{2013}).

\bibitem[{\citenamefont{{Hofmann} et~al.}(2014)\citenamefont{{Hofmann},
  {Barnes}, and {Das Sarma}}}]{HofmannArXiv2014}
\bibinfo{author}{\bibfnamefont{J.}~\bibnamefont{{Hofmann}}},
  \bibinfo{author}{\bibfnamefont{E.}~\bibnamefont{{Barnes}}}, \bibnamefont{and}
  \bibinfo{author}{\bibfnamefont{S.}~\bibnamefont{{Das Sarma}}},
  \bibinfo{journal}{ArXiv e-prints}  (\bibinfo{year}{2014}),
  \eprint{1410.1547}.

\bibitem[{\citenamefont{{Hofmann} and {Das Sarma}}(2015)}]{HofmannArXiv2015}
\bibinfo{author}{\bibfnamefont{J.}~\bibnamefont{{Hofmann}}} \bibnamefont{and}
  \bibinfo{author}{\bibfnamefont{S.}~\bibnamefont{{Das Sarma}}},
  \bibinfo{journal}{ArXiv e-prints}  (\bibinfo{year}{2015}),
  \eprint{1501.04636}.

\bibitem[{\citenamefont{Isobe and Nagaosa}(2013)}]{IsobePRB2013}
\bibinfo{author}{\bibfnamefont{H.}~\bibnamefont{Isobe}} \bibnamefont{and}
  \bibinfo{author}{\bibfnamefont{N.}~\bibnamefont{Nagaosa}},
  \bibinfo{journal}{Phys. Rev. B} \textbf{\bibinfo{volume}{87}},
  \bibinfo{pages}{205138} (\bibinfo{year}{2013}).

\bibitem[{\citenamefont{Barnes et~al.}(2014)\citenamefont{Barnes, Hwang,
  Throckmorton, and Das~Sarma}}]{BarnesPRB2014}
\bibinfo{author}{\bibfnamefont{E.}~\bibnamefont{Barnes}},
  \bibinfo{author}{\bibfnamefont{E.~H.} \bibnamefont{Hwang}},
  \bibinfo{author}{\bibfnamefont{R.~E.} \bibnamefont{Throckmorton}},
  \bibnamefont{and}
  \bibinfo{author}{\bibfnamefont{S.}~\bibnamefont{Das~Sarma}},
  \bibinfo{journal}{Phys. Rev. B} \textbf{\bibinfo{volume}{89}},
  \bibinfo{pages}{235431} (\bibinfo{year}{2014}).

\bibitem[{\citenamefont{Hofmann et~al.}(2014)\citenamefont{Hofmann, Barnes, and
  Das~Sarma}}]{HofmannPRL2014}
\bibinfo{author}{\bibfnamefont{J.}~\bibnamefont{Hofmann}},
  \bibinfo{author}{\bibfnamefont{E.}~\bibnamefont{Barnes}}, \bibnamefont{and}
  \bibinfo{author}{\bibfnamefont{S.}~\bibnamefont{Das~Sarma}},
  \bibinfo{journal}{Phys. Rev. Lett.} \textbf{\bibinfo{volume}{113}},
  \bibinfo{pages}{105502} (\bibinfo{year}{2014}).

\bibitem[{\citenamefont{Zivitz and Stevenson}(1974)}]{ZivitzPRB1974}
\bibinfo{author}{\bibfnamefont{M.}~\bibnamefont{Zivitz}} \bibnamefont{and}
  \bibinfo{author}{\bibfnamefont{J.~R.} \bibnamefont{Stevenson}},
  \bibinfo{journal}{Phys. Rev. B} \textbf{\bibinfo{volume}{10}},
  \bibinfo{pages}{2457} (\bibinfo{year}{1974}).

\bibitem[{\citenamefont{Jay-Gerin et~al.}(1977)\citenamefont{Jay-Gerin, Aubin,
  and Caron}}]{JayGerinSSC1977}
\bibinfo{author}{\bibfnamefont{J.-P.} \bibnamefont{Jay-Gerin}},
  \bibinfo{author}{\bibfnamefont{M.}~\bibnamefont{Aubin}}, \bibnamefont{and}
  \bibinfo{author}{\bibfnamefont{L.}~\bibnamefont{Caron}},
  \bibinfo{journal}{Solid State Communications} \textbf{\bibinfo{volume}{21}},
  \bibinfo{pages}{771} (\bibinfo{year}{1977}).

\bibitem[{\citenamefont{{Das Sarma} and Hwang}(2013)}]{DasSarmaPRB2013}
\bibinfo{author}{\bibfnamefont{S.}~\bibnamefont{{Das Sarma}}} \bibnamefont{and}
  \bibinfo{author}{\bibfnamefont{E.~H.} \bibnamefont{Hwang}},
  \bibinfo{journal}{Phys. Rev. B} \textbf{\bibinfo{volume}{87}},
  \bibinfo{pages}{035415} (\bibinfo{year}{2013}).

\bibitem[{\citenamefont{{Das Sarma} and Li}(2013)}]{DasSarmaPRB2013_2}
\bibinfo{author}{\bibfnamefont{S.}~\bibnamefont{{Das Sarma}}} \bibnamefont{and}
  \bibinfo{author}{\bibfnamefont{Q.}~\bibnamefont{Li}}, \bibinfo{journal}{Phys.
  Rev. B} \textbf{\bibinfo{volume}{87}}, \bibinfo{pages}{235418}
  (\bibinfo{year}{2013}).

\bibitem[{\citenamefont{Sodemann and Fogler}(2012)}]{SodemannPRB2012}
\bibinfo{author}{\bibfnamefont{I.}~\bibnamefont{Sodemann}} \bibnamefont{and}
  \bibinfo{author}{\bibfnamefont{M.~M.} \bibnamefont{Fogler}},
  \bibinfo{journal}{Phys. Rev. B} \textbf{\bibinfo{volume}{86}},
  \bibinfo{pages}{115408} (\bibinfo{year}{2012}).

\bibitem[{\citenamefont{Peskin and Schroeder}(1995)}]{PeskinSchroederBook}
\bibinfo{author}{\bibfnamefont{M.~E.} \bibnamefont{Peskin}} \bibnamefont{and}
  \bibinfo{author}{\bibfnamefont{D.~V.} \bibnamefont{Schroeder}},
  \emph{\bibinfo{title}{An Introduction to Quantum Field Theory}}
  (\bibinfo{publisher}{Westview Press}, \bibinfo{year}{1995}).

\bibitem[{\citenamefont{Jost and Luttinger}(1950)}]{JostHPA1950}
\bibinfo{author}{\bibfnamefont{R.}~\bibnamefont{Jost}} \bibnamefont{and}
  \bibinfo{author}{\bibfnamefont{J.~M.} \bibnamefont{Luttinger}},
  \bibinfo{journal}{Helvetica Physica Acta} \textbf{\bibinfo{volume}{23}},
  \bibinfo{pages}{201} (\bibinfo{year}{1950}).

\bibitem[{\citenamefont{Itzykson and Zuber}(2005)}]{ItzyksonBook}
\bibinfo{author}{\bibfnamefont{C.}~\bibnamefont{Itzykson}} \bibnamefont{and}
  \bibinfo{author}{\bibfnamefont{J.-B.} \bibnamefont{Zuber}},
  \emph{\bibinfo{title}{Quantum Field Theory}} (\bibinfo{publisher}{Dover
  Publications}, \bibinfo{year}{2005}).

\bibitem[{\citenamefont{Nozieres}(1964)}]{Nozieres1964}
\bibinfo{author}{\bibfnamefont{P.}~\bibnamefont{Nozieres}},
  \emph{\bibinfo{title}{Theory of Interacting Fermi Systems}}
  (\bibinfo{publisher}{W. A. Benjamin, Inc.}, \bibinfo{year}{1964}).

\bibitem[{\citenamefont{Abrikosov et~al.}(1975)\citenamefont{Abrikosov, Gorkov,
  and Dzyaloshinski}}]{Abrikosov1975}
\bibinfo{author}{\bibfnamefont{A.~A.} \bibnamefont{Abrikosov}},
  \bibinfo{author}{\bibfnamefont{L.~P.} \bibnamefont{Gorkov}},
  \bibnamefont{and} \bibinfo{author}{\bibfnamefont{I.~E.}
  \bibnamefont{Dzyaloshinski}}, \emph{\bibinfo{title}{Methods of Quantum Field
  Theory in Statistical Physics}} (\bibinfo{publisher}{Dover Pulbications},
  \bibinfo{year}{1975}).

\bibitem[{\citenamefont{Mahan}(1981)}]{Mahan1981}
\bibinfo{author}{\bibfnamefont{G.~D.} \bibnamefont{Mahan}},
  \emph{\bibinfo{title}{Many-Particle Physics}} (\bibinfo{publisher}{Plenum
  Press, New York}, \bibinfo{year}{1981}).

\bibitem[{\citenamefont{Fetter and Walecka}(2003)}]{Fetter2003}
\bibinfo{author}{\bibfnamefont{A.~L.} \bibnamefont{Fetter}} \bibnamefont{and}
  \bibinfo{author}{\bibfnamefont{J.~D.} \bibnamefont{Walecka}},
  \emph{\bibinfo{title}{Quantum Theory of Many-Partice Systems}}
  (\bibinfo{publisher}{Dover Pulbications}, \bibinfo{year}{2003}).

\bibitem[{\citenamefont{Shankar}(1994)}]{ShankarRMP1994}
\bibinfo{author}{\bibfnamefont{R.}~\bibnamefont{Shankar}},
  \bibinfo{journal}{Rev. Mod. Phys.} \textbf{\bibinfo{volume}{66}},
  \bibinfo{pages}{129} (\bibinfo{year}{1994}).

\bibitem[{\citenamefont{Polchinski}(1999)}]{PolchinskiArXiv1999}
\bibinfo{author}{\bibfnamefont{J.}~\bibnamefont{Polchinski}},
  \bibinfo{journal}{ArXiv e-prints}  (\bibinfo{year}{1999}),
  \eprint{hep-th/9210046v2}.

\bibitem[{\citenamefont{Kohn and Luttinger}(1965)}]{KohnPRL1965}
\bibinfo{author}{\bibfnamefont{W.}~\bibnamefont{Kohn}} \bibnamefont{and}
  \bibinfo{author}{\bibfnamefont{J.~M.} \bibnamefont{Luttinger}},
  \bibinfo{journal}{Phys. Rev. Lett.} \textbf{\bibinfo{volume}{15}},
  \bibinfo{pages}{524} (\bibinfo{year}{1965}).

\bibitem[{\citenamefont{Zhang and {Das Sarma}}(2005)}]{ZhangPRB2005}
\bibinfo{author}{\bibfnamefont{Y.}~\bibnamefont{Zhang}} \bibnamefont{and}
  \bibinfo{author}{\bibfnamefont{S.}~\bibnamefont{{Das Sarma}}},
  \bibinfo{journal}{Phys. Rev. B} \textbf{\bibinfo{volume}{71}},
  \bibinfo{pages}{045322} (\bibinfo{year}{2005}).

\bibitem[{\citenamefont{Dyson}(1952)}]{DysonPR1952}
\bibinfo{author}{\bibfnamefont{F.~J.} \bibnamefont{Dyson}},
  \bibinfo{journal}{Phys. Rev.} \textbf{\bibinfo{volume}{85}},
  \bibinfo{pages}{631} (\bibinfo{year}{1952}).

\bibitem[{\citenamefont{{Kolomeisky}}(2014)}]{KolomeiskyArXiv2014}
\bibinfo{author}{\bibfnamefont{E.~B.} \bibnamefont{{Kolomeisky}}},
  \bibinfo{journal}{ArXiv e-prints}  (\bibinfo{year}{2014}),
  \eprint{1409.3765}.

\bibitem[{\citenamefont{G\"ockeler et~al.}(1998)\citenamefont{G\"ockeler,
  Horsley, Linke, Rakow, Schierholz, and St\"uben}}]{GockelerPRL1998}
\bibinfo{author}{\bibfnamefont{M.}~\bibnamefont{G\"ockeler}},
  \bibinfo{author}{\bibfnamefont{R.}~\bibnamefont{Horsley}},
  \bibinfo{author}{\bibfnamefont{V.}~\bibnamefont{Linke}},
  \bibinfo{author}{\bibfnamefont{P.}~\bibnamefont{Rakow}},
  \bibinfo{author}{\bibfnamefont{G.}~\bibnamefont{Schierholz}},
  \bibnamefont{and} \bibinfo{author}{\bibfnamefont{H.}~\bibnamefont{St\"uben}},
  \bibinfo{journal}{Phys. Rev. Lett.} \textbf{\bibinfo{volume}{80}},
  \bibinfo{pages}{4119} (\bibinfo{year}{1998}).

\bibitem[{\citenamefont{Ashby and Carbotte}(2014)}]{AshbyPRB2014}
\bibinfo{author}{\bibfnamefont{P.~E.~C.} \bibnamefont{Ashby}} \bibnamefont{and}
  \bibinfo{author}{\bibfnamefont{J.~P.} \bibnamefont{Carbotte}},
  \bibinfo{journal}{Phys. Rev. B} \textbf{\bibinfo{volume}{89}},
  \bibinfo{pages}{245121} (\bibinfo{year}{2014}).

\bibitem[{\citenamefont{Rosenstein et~al.}(2014)\citenamefont{Rosenstein, Kao,
  and Lewkowicz}}]{RosensteinPRB2014}
\bibinfo{author}{\bibfnamefont{B.}~\bibnamefont{Rosenstein}},
  \bibinfo{author}{\bibfnamefont{H.~C.} \bibnamefont{Kao}}, \bibnamefont{and}
  \bibinfo{author}{\bibfnamefont{M.}~\bibnamefont{Lewkowicz}},
  \bibinfo{journal}{Phys. Rev. B} \textbf{\bibinfo{volume}{90}},
  \bibinfo{pages}{045137} (\bibinfo{year}{2014}).

\bibitem[{\citenamefont{{Rosenstein} et~al.}(2015)\citenamefont{{Rosenstein},
  {Kao}, and {Lewkowicz}}}]{RosensteinArXiv2015}
\bibinfo{author}{\bibfnamefont{B.}~\bibnamefont{{Rosenstein}}},
  \bibinfo{author}{\bibfnamefont{H.~C.} \bibnamefont{{Kao}}}, \bibnamefont{and}
  \bibinfo{author}{\bibfnamefont{M.}~\bibnamefont{{Lewkowicz}}},
  \bibinfo{journal}{ArXiv e-prints}  (\bibinfo{year}{2015}),
  \eprint{1508.01604}.

\end{thebibliography}

\end{document}